\documentclass[11pt,a4paper]{article}
\pdfoutput=1
\usepackage{jheppub}
\usepackage{caption,subcaption}
\usepackage{float,subfloat}
\usepackage{graphicx,amsmath,amsthm,dcolumn,hyperref,verbatim,amsfonts,tikz,amssymb,mathtools,tikz-cd,upgreek,pgfplots}
\usepackage[capitalize]{cleveref}
\pgfplotsset{compat=1.5}

\title{\huge Musings on SVD and pseudo entanglement entropies }

\author[a,b]{Pawe{\l} Caputa,}
\author[a]{Souradeep Purkayastha,}
\author[a]{Abhigyan Saha,}
\author[a,c]{Piotr Su{\l}kowski}

\newcommand{\be}{\begin{equation}}
\newcommand{\ee}{\end{equation}}
\newcommand{\bea}{\begin{eqnarray}}
\newcommand{\eea}{\end{eqnarray}}

\newcommand{\nn}{\nonumber}
\newcommand{\Tr}{\text{Tr}}

\newcommand{\ket}[1]{\left| #1 \right>}
\newcommand{\bra}[1]{\left< #1 \right|}
\newcommand{\ketbra}[1]{\left|#1\right>\left<#1\right|}

\newcommand{\ba}{\begin{eqnarray}}
\newcommand{\ea}{\end{eqnarray}}

\newcommand{\dd}{\mathrm{d}}
\newcommand{\cH}{{\mathcal{H}}}

\newcommand{\cK}{{\mathcal{K}}}
\newcommand{\cL}{{\mathcal{L}}}

\newcommand{\cS}{{\mathcal{S}}}
\newcommand{\cT}{{\mathcal{T}}}

\newcommand{\cP}{{\mathcal{P}}}

\preprint{CALT-2024-029, YITP-24-97}

\affiliation[a]{Faculty of Physics, University of Warsaw, Pasteura 5, 02-093 Warsaw, Poland}
\affiliation[b]{Yukawa Institute for Theoretical Physics, Kyoto University, Kitashirakawa Oiwakecho, Sakyo-ku, Kyoto 606-8502, Japan}
\affiliation[c]
{Walter Burke Institute for Theoretical Physics, California Institute of Technology, Pasadena, CA 91125}

\abstract{Pseudo-entropy and SVD entropy are generalizations of the entanglement entropy that involve post-selection. In this work we analyze their properties as measures on the spaces of quantum states and argue that their excess provides useful characterization of a difference between two (i.e. pre-selected and post-selected) states, which shares certain features and in certain cases can be identified as a metric. In particular, when applied to link complement states that are associated to topological links via Chern--Simons theory, these generalized entropies and their excess provide a novel quantification of a difference between corresponding links. We discuss the dependence of such entropy measures on the level of Chern--Simons theory and determine their asymptotic values for certain link states. We find that imaginary part of the pseudo-entropy is sensitive to, and can diagnose chirality of knots. We also consider properties of entropy measures for simpler quantum mechanical systems, such as generalized SU(2) and SU(1,1) coherent states, and tripartite GHZ and W states.  }

\begin{document}

\maketitle

\newpage
%%%%%%%%%%%%%%%%%%%%%%%%%%%
%%%%%%%%%%%%%%%%%%%%%%%%%%%
\section{Introduction}\label{sec_introduction}
%%%%%%%%%%%%%%%%%%%%%%%%%%%
%%%%%%%%%%%%%%%%%%%%%%%%%%%
In recent years intriguing connections between high energy physics and quantum information theory have been revealed. One link between these research areas is provided by the notion of entanglement entropy and its generalizations (see e.g. review \cite{Horodecki:2009zz}). Apart from providing means to describe complex systems, other motivations to study various incarnations of entropy include their geometric interpretation via AdS/CFT correspondence \cite{Ryu:2006bv,Takayanagi:2017knl,Dutta:2019gen,Dong:2016fnf}, the potential to characterize topological properties of various systems \cite{Kitaev:2005dm,levin2006detecting} and topological field theories in particular, the capability to describe the process of post-selection \cite{aharonov1988result,Horowitz:2003he}, applications of these ideas in condensed matter physics \cite{PasqualeCalabrese_2009}, etc.

Generalizations of the entanglement entropy $S_{\mathrm{E}}^\phi$ (of a state $|\phi\rangle$) of our primary interest in this work are pseudo-entropy denoted $S^{\phi|\psi}_{\mathrm{P}}$ \cite{Nakata:2020luh}, and SVD entropy denoted $S^{\phi|\psi}_{\mathrm{SVD}}$ \cite{Parzygnat:2023avh}. Recall that entanglement entropy characterizes entanglement between two subsets of a Hilbert space; they are often taken to be associated to two subregions of the spatial domain on which a system under consideration is defined. Pseudo-entropy, which arises naturally from the AdS/CFT perspective, is a generalization of the entanglement entropy that involves post-selection and depends on two states, the initial one $|\phi\rangle$ and the final (post-selected) one $|\psi\rangle$, as indicated in the notation above. Pseudo-entropy takes complex values and can be also larger than the logarithm of the dimension of the Hilbert space, which obscures its quantum-information interpretation. To remedy these issues, the SVD entropy has been introduced in \cite{Parzygnat:2023avh}. SVD entropy also depends on the initial and post-selected state, however it takes real values, which in addition do not exceed the logarithm of the dimension of the Hilbert space. Moreover, it admits an elegant operational meaning as a number of Bell pairs in the intermediate states between $\ket{\phi}$ and $\ket{\psi}$. 

In this work we argue that pseudo-entropy, SVD entropy, and their excess are useful in quantifying a difference between (pre-selected and post-selected) quantum states. In particular, for link states which are associated to topological links via Chern--Simons theory -- and which are of our main interest -- these generalized entropies provide a novel quantification of a difference between corresponding links. We analyze the dependence of such measures on the level of Chern--Simons theory, and in particular determine their asymptotic values for large level. Note that these results (as well as classes of links under our consideration) extend and generalize earlier analysis of the entanglement and pseudo-entropy for link states in  \cite{Balasubramanian_2017,Balasubramanian:2018por,SO3}. Furthermore, as a warm up, we also study these concepts for simpler quantum mechanical systems, involving generalized $\mathrm{SU}(2)$ and $\mathrm{SU}(1,1)$  coherent states, as well as tripartite GHZ and W states. The systems that we analyze are characterized by increasing dimension of Hilbert spaces and increasing number of components; for link states the dimension of the Hilbert space is determined by the level of Chern--Simons theory, while the number of components is equal to the number of components of a link. These quantities may take arbitrary values; in particular, we analyze the limit of infinite level, which is also of interest in other contexts, such as the volume conjecture \cite{kashaev1996,murakami1999,SO3}.

While we provide precise definitions of pseudo-entropy $S^{\phi|\psi}_{\mathrm{P}}$ and SVD entropy $S^{\phi|\psi}_{\mathrm{SVD}}$ in section \ref{sec:definitions}, we note here that their excess is defined respectively as
\be
\Delta S^{\phi|\psi}_{\mathrm{P}}=\text{Re}\big(S^{\phi|\psi}_{\mathrm{P}}\big)-\frac{ S_{\mathrm{E}}^\phi+S_{\mathrm{E}}^\psi }{2}, \qquad \quad \Delta S^{\phi|\psi}_{\mathrm{SVD}}=S^{\phi|\psi}_{\mathrm{SVD}}-\frac{S_{\mathrm{E}}^\phi+S_{\mathrm{E}}^\psi}{2}, \label{intro-excess}
\ee
where $S_{\mathrm{E}}^\phi$ is the entanglement entropy of a state $|\phi\rangle$. These excess functions have interesting properties. For example, it was conjectured in \cite{Mollabashi:2020yie,Mollabashi:2021xsd} that the pseudo-entropy excess is non-positive or positive if the two states are respectively in the same or different quantum phases. In this work we analyze link states and other states from this perspective and argue that they can be associated to the same or different phases, depending on particular choice of parameters characterizing a given system we consider. 

Furthermore, our main observation is that the entropy excess (\ref{intro-excess}) satisfies certain -- and in some cases all, depending on features of a given quantum system -- axioms of the metric. In quantum information theory certain metrics have been introduced before (however some of them only for pure states), which are referred to as Fisher metric, Fubini-Study metric, Bures metric or Helstrom metric, and which provide a notion of distance on a space of quantum states. We analyze for which systems and under which conditions the absolute value of an excess function, i.e. either $|\Delta S^{\phi|\psi}_{\mathrm{P}}|$ or $|\Delta S^{\phi|\psi}_{\mathrm{SVD}}|$, has analogous interpretation and thus provides a proper notion of a distance between quantum states. The absolute value of either pseudo-entropy's or SVD entropy's excess is clearly non-negative, equal to zero for $|\phi\rangle=|\psi\rangle$, and symmetric (with respect to the interchange of $|\phi\rangle$ and $|\psi\rangle$), which are a subset of the axioms of a metric. In what follows we analyze for which systems of our interest the triangle inequality holds (the space of states is called semi-metric if this inequality is violated), and when the separation axiom (meaning that the distance cannot vanish for different states) holds (the distance function is referred to as pseudo-metric when this axiom is violated). 

A prototype example of a metric structure that we find is SVD entropy excess for two-component link states in U(1) Chern--Simons theory. Consider two two-component links with linking numbers $l_1$ and $l_2$ respectively. We show that for the corresponding pre-selected and post-selected link states, in U(1) theory at level $k$, the SVD entropy takes the form
\begin{equation}
 S_{\mathrm{SVD}} = \log \Big( \frac{k}{\gcd(k, l_1 l_2)} \Big) 
\end{equation}
whenever the greatest common divisor (commonly denoted by $\gcd$) $\gcd(k,l_1l_2) \neq np^2$ for $n,p\in \mathbb{N}$ (when $\gcd(k,l_1l_2) = np^2$, the expression is more complicated). It then follows that the absolute value of the SVD entropy excess takes the form
\begin{equation}
\left|\Delta S_{\mathrm{SVD}}\right| =  \frac{1}{2} \log \Big( \frac{(\gcd(k, l_1 l_2))^2}{\gcd(k, l_1) \cdot \gcd(k, l_2)} \Big). 
\end{equation}
We show that for this expression (and also more generally, for $\gcd(k,l_1l_2) = np^2$) the triangle inequality holds and thus $|\Delta S_{SVD}|$ provides a pseudo-metric on the space of two-component links (it is a pseudo-metric, as in U(1) theory the entropy measures depend only on linking numbers, so the distance between two different links with the same linking number vanishes). Motivated by this example, we discuss for what other systems, including link states in Chern--Simons theory with non-abelian gauge group as well as quantum mechanical examples, the metric interpretation holds -- this turns out to be the case for some specific ranges of parameters specifying quantum states in a given system. We stress that whenever the SVD entropy excess can be interpreted as a metric on the space of link states, it also provides a measure on the space of links that may be of interest from the knot theory perspective.

Apart from the metric interpretation, we also identify other properties of entropy measures. On one hand, we find classes of links states for which the SVD entropy take values between the entanglement entropies of pre-selected and post-selected states, or exceeds the value of one of these entanglement entropies. While the former case can be explained in terms of Bell pairs exchanged between the two states under consideration, the latter phenomenon is more surprising. Furthermore, we find that the imaginary part of the pseudo-entropy, whose quantum information interpretation has been not so clear, detects chirality of link states associated to topological links. 
 
\bigskip

The paper is organised as follows. In~\cref{sec:definitions} we introduce entropy measures of our interest: entanglement entropy, pseudo-entropy and SVD entropy, and their excess. In  \cref{sec:link_complement} we review basics of Chern--Simons theory, knot invariants, and introduce the link states that are of our main interest in what follows. In  \cref{sec:toymodels} we analyze quantum mechanical examples involving generalized $\mathrm{SU}(2)$ and $\mathrm{SU}(1,1)$ coherent states, as well as tripartite GHZ and W states. In  \cref{sec-links} we determine entropy measures and discuss their properties for various classes of link complement states: two-component links in U(1) Chern--Simons theory, connected sums $\mathcal{K}\# 2^2_1$ and $(p,pn)$ torus links in non-abelian Chern--Simons theory, and other examples involving in particular Borromean links. In  \cref{subsec:largek} we determine asymptotic values of entropy measures for large $k$ for various link states, and in  \cref{sec_chirality} we show that imaginary part of pseudo-entropy detects chirality of link states. 
%%%%%%%%%%%%%%%%%%%%%%%%%%%
%%%%%%%%%%%%%%%%%%%%%%%%%%%
\section{Review of entropy measures}
\label{sec:definitions}
%%%%%%%%%%%%%%%%%%%%%%%%%%%
%%%%%%%%%%%%%%%%%%%%%%%%%%%
In this section we introduce von Neumann entanglement entropy, pseudo-entropy \cite{Nakata:2020luh}, SVD entropy \cite{Parzygnat:2023avh}, and the entropy excess. In the following sections we will employ these quantities to characterize entanglement structure of quantum states in various models. 
%%%%%%%%%%%%%%%%%%%%%%%%%%%%%%%%%%%%%%%%%%%%%%%%%%%
%%%%%%%%%%%%%%%%%%%%%%%%%%%%%%%%%%%%%%%%%%%%%%%%%%%
\subsection{Entanglement entropy, pseudo-entropy and SVD entropy}
\label{subsec:defn_QI}
%%%%%%%%%%%%%%%%%%%%%%%%%%%%%%%%%%%%%%%%%%%%%%%%%%%
%%%%%%%%%%%%%%%%%%%%%%%%%%%%%%%%%%%%%%%%%%%%%%%%%%%
To set up the stage, consider a Hilbert space $\mathcal{H}$ that admits a decomposition into two parts\footnote{A generalisation to multiple parts is analogous.}, $A$ and its complement $B$
\be
\mathcal{H}=\cH_A\otimes \cH_B.\label{HAB}
\ee
We denote dimensions of $\cH_A$ and $\cH_B$ by $d_{A}$ and $d_{B}$ respectively. In what follows we study both finite and infinite dimensional spaces. Next, we pick a pure quantum state $\ket{\psi}$ in $\mathcal{H}$ and define the (normalized) reduced density matrix of $A$ by tracing over $B$
\be
\rho_A=\Tr_B(\rho),\qquad \Tr(\rho_A)=1.
\ee
To characterize the entanglement between $A$ and $B$, we will study von Neumann entanglement entropy (denoted by the subscript E) of $\rho_A$
\be
S^\psi_{\mathrm{E}}=S(\rho_A)\equiv-\Tr(\rho_A\log \rho_A)=-\sum_i p_i\log p_i,   \label{SE}
\ee
where $p_i$ are eigenvalues of $\rho_A$. From the Schmidt decomposition of the pure state $\ket{\psi}$ in the Hilbert spaces $\mathcal{H}_A\otimes \mathcal{H}_B$ we have $S(\rho_A)=S(\rho_B)$ where $\rho_B=\Tr_A(\rho)$. 

Next, we introduce two interesting generalisations of entanglement entropy. The first one is the pseudo-entropy \cite{Nakata:2020luh}. Its
 definition requires two pure states $\ket{\phi}$ and $\ket{\psi}$ in $\mathcal{H}$ \eqref{HAB} satisfying $\langle \phi |\psi \rangle \ne 0$. In what follows we sometimes refer to them as the reference or pre-selected state and the target or post-selected state respectively. Then, we define a transition matrix for these two states
\be
\tau^{\phi|\psi}=\frac{\ket{\phi}\bra{\psi}}{\langle \psi|\phi\rangle}. \label{tau-phi-psi}
\ee
Such objects are very natural not only in quantum information but also in physical studies of post-selection or weak values and quantum measurements \cite{aharonov1988result}. 
By analogy with the reduced density matrix, we have the reduced transition matrix for $A$
\be
\tau^{\phi|\psi}_A=\Tr_{B}(\tau^{\phi|\psi}).\label{RedTrM}
\ee
Since these transition matrices are not Hermitian, they will generally have complex eigenvalues (see \cite{Nakata:2020luh} for some classification), but one can still define a complex extension of the von Neumann entropy, referred to as the pseudo-entropy (that we denote by subscript P)  \cite{Nakata:2020luh}
\be
S^{\phi|\psi}_{\mathrm{P}}=-\Tr_{A}(\tau^{\phi|\psi}_A\log \tau^{\phi|\psi}_A). \label{def:psent_general}
\ee
Pseudo-entropy has several interesting properties and we only mention a few. Firstly, for the specific case of $\ket{\phi} =  \ket{\psi}$, \eqref{def:psent_general} reduces to the entanglement entropy (\ref{SE}). It vanishes if the the states are product states $\ket{\varphi_1}_A\ket{\varphi_2}_B$. Swapping the two states is equivalent to complex conjugation of the pseudo-entropy. Its real part has various interesting properties, including operational meaning for some classes of states and playing the role of an order parameter for different quantum phases \cite{Mollabashi:2020yie,Mollabashi:2021xsd}. While imaginary part remains mysterious, in this work we reveal some of its properties. Similar to von Neumann entropy, pseudo-entropy is symmetric under exchanging $A$ with its complement $B$. Nevertheless, it is still not clear and very interesting open problem to determine conditions for obeying (perhaps saturation) or violation of the famous entropy inequalities. For example violations of sub-additivity were discussed in \cite{Nakata:2020luh}. Further important developments on pseudo-entropy can be found e.g. in \cite{Nishioka:2021cxe,Doi:2024nty,Kawamoto:2023nki,He:2024jog,Omidi:2023env,Narayan:2023ebn,Chen:2023gnh,Doi:2023zaf,Ishiyama:2022odv,Mukherjee:2022jac,Miyaji:2021lcq,Goto:2021kln}.

Another generalization of entanglement entropy that involves post-selection is Singular Value Decomposition entropy (SVD entropy for short) recently defined in \cite{Parzygnat:2023avh}. To define it we introduce analogous quantities as for the pseudo-entropy, up to the reduced transition matrix $\tau^{\phi|\psi}_A$ in \eqref{RedTrM}. Then we perform the SVD decomposition
\be
\tau^{\phi|\psi}_A=U\Lambda V^\dagger,
\ee
with unitary matrices $U$ and $V$ and diagonal matrix with real and non-negative eigenvalues
\be
\Lambda=\text{diag}(\lambda_1,\cdots,\lambda_{d_A}).
\ee
In general, these eigenvalues are not normalized, so we normalize them by introducing
\be
\hat{\lambda}_i=\frac{\lambda_i}{\sum_j\lambda_j},\qquad \sum^{d_A}_{i=1}\hat{\lambda}_i=1.
\ee
Note that it is useful to interpret them as eigenvalues of the following density matrix constructed from the transition matrix \eqref{RedTrM} 
\be
\rho^{\phi|\psi}_A=\frac{\sqrt{\left(\tau^{\phi|\psi}_A\right)^\dagger\tau^{\phi|\psi}_A}}{\Tr\left(\sqrt{\left(\tau^{\phi|\psi}_A\right)^\dagger\tau^{\phi|\psi}_A}\right)}. \label{def:svde}
\ee
From this data, we finally define the SVD entropy as 
\be
S^{\phi|\psi}_{\mathrm{SVD}}=-\Tr(\rho^{\phi|\psi}_A\log\rho^{\phi|\psi}_A)=-\sum_i \hat{\lambda}_i\log(\hat{\lambda}_i).
\ee
This quantity is manifestly real and  has several interesting properties. It is positive and bounded  \cite{Parzygnat:2023avh}
\be
0\le S^{\phi|\psi}_{\mathrm{SVD}}\le \log d_A,
\ee
and also vanishes if any of the states is a product $\ket{\psi}=\ket{\varphi}_A\ket{\varphi}_B$. Formally, it can be defined for states that have a vanishing inner product (which cancels in the computation with $\rho^{\phi|\psi}_A$). However, in contrast to the previous two quantities above, it is not symmetric under swapping $A$ and $B$, i.e. $S(\rho^{\phi|\psi}_A)\neq S(\rho^{\phi|\psi}_B)$. In fact, one can show that application of a unitary operator on $B$, that we trace over, changes $S(\rho^{\phi|\psi}_A)$. In general, SVD entropy violates Araki-Lieb inequality and (strong) sub-additivity. Nevertheless, it admits a very elegant operational meaning as a number of Bell pairs in the intermediate states between (arbitrary) $\ket{\phi}$ and $\ket{\psi}$. See \cite{Guo:2024edr,He:2023syy,Kanda:2023jyi,He:2023wko,Guo:2023aio,Singh:2024pbo} for further progress on this quantity.

Furthermore, following \cite{Mollabashi:2020yie,Mollabashi:2021xsd}, we define the excess of the entropy measures introduced above, i.e. the pseudo-entropy excess
\be
\Delta S^{\phi|\psi}_{\mathrm{P}}=\text{Re}\big(S^{\phi|\psi}_{\mathrm{P}}\big)-\frac{1}{2}\big(S_{\mathrm{E}}(\rho^\phi_A)+S_{\mathrm{E}}(\rho^\psi_A)\big), \label{defn:entropy_excess}
\ee
and analogously the excess of the SVD entropy 
\be
\Delta S^{\phi|\psi}_{\mathrm{SVD}}=S^{\phi|\psi}_{\mathrm{SVD}}-\frac{1}{2}\big(S_{\mathrm{E}}(\rho^\phi_A)+S_{\mathrm{E}}(\rho^\psi_A)\big). \label{defn:SVDentropy_excess}
\ee
The entropy excess was conjectured to be a useful order parameter for detecting or distinguishing quantum phases in $\ket{\phi}$ and $\ket{\psi}$ (see also \cite{Shinmyo_2024}). In particular, this excess was observed to be non-positive when the two states belong to the same quantum phase, while its positivity was correlated with different phases of the two states under consideration. We will examine this property for our quantum mechanical as well as link complement states below. 
%%%%%%%%%%%%%%%%%%%%%%%%%%%%%%%%%%%%%%%%%
\subsection{Entropy measures for two-component states}\label{sec:general 2 component}
%%%%%%%%%%%%%%%%%%%%%%%%%%%%%%%%%%%%%%%%%
Before we proceed with specific models, let us analyze a general class of quantum states in a product Hilbert space $\mathcal{H}=\mathcal{H}_A \otimes \mathcal{H}_B$ of the form
\be
\ket{\psi_{i}}=\sum^{d-1}_{n=0}c^{(i)}_n\ket{n}_A\otimes\ket{n}_B,     \label{psi-2-component}
\ee
with equal dimensions of the two components $d=\dim \mathcal{H}_A=\dim \mathcal{H}_B$ and complex coefficients $c^{(i)}_n$. These coefficients can be normalized as $\sum^{d-1}_{n=0}|c^{(i)}_n|^2=1$, however we do not necessarily impose this condition, as the normalization cancels in the transition matrix
\be
\tau^{1|2}=\frac{\ket{\psi_1}\bra{\psi_2}}{\langle\psi_{2}|\psi_{1}\rangle}.
\ee
We denote the overlap of our two states by
\be f^{(1|2)}\equiv\langle\psi_{2}|\psi_{1}\rangle=\sum^{d-1}_{n=0}c^{(1)}_n\bar{c}^{(2)}_n,  \label{f-overlap}
\ee
and compute the reduced transition matrix by tracing over $\mathcal{H}_B$
\be
\tau^{1|2}_A=\frac{1}{f^{(1|2)}}\sum^{d-1}_{n=0}c^{(1)}_n\bar{c}^{(2)}_n\ket{n}_A\bra{n}_A.
\ee
This matrix is already diagonal and has complex eigenvalues. Moreover, its normalized singular  values are encoded in the density matrix \eqref{def:svde} that becomes
\be
\rho^{1|2}_A=\frac{1}{\tilde{f}^{(1|2)}}\sum^{d-1}_{n=0}|c^{(1)}_n\bar{c}^{(2)}_n|\ket{n}_A\bra{n}_A,\qquad \Tr(\rho^{1|2}_A)=1,
\ee
where we denoted the real normalization by
\be
\tilde{f}^{(1|2)}\equiv\sum^{d-1}_{n=0}|c^{(1)}_n\bar{c}^{(2)}_n|. 
   \label{f-tilde-overlap}
\ee
This way we derive the singular values
\be
\hat{\lambda}_n=\frac{|c^{(1)}_n\bar{c}^{(2)}_n|}{\tilde{f}^{(1|2)}},\qquad \sum^{d-1}_{n=0}\hat{\lambda}_n=1.
\ee
Based on the above formulas, we obtain the pseudo-entropy 
\begin{align}
\begin{split}
S_{\mathrm{P}}^{1|2}&=-\frac{1}{f^{(1|2)}}\sum^{d-1}_{n=0}c^{(1)}_n\bar{c}^{(2)}_n\log\left(\frac{c^{(1)}_n\bar{c}^{(2)}_n}{f^{(1|2)}}\right) = \\
&=\log(f^{(1|2)})-\frac{1}{f^{(1|2)}}\sum^{d-1}_{n=0}c^{(1)}_n\bar{c}^{(2)}_n\log\left(c^{(1)}_n\bar{c}^{(2)}_n\right),\label{PEGen2Q}
\end{split}
\end{align}
as well as the SVD entropy
\begin{align}
\begin{split}
S_{\mathrm{SVD}}^{1|2}&=-\frac{1}{\tilde{f}^{(1|2)}}\sum^{d-1}_{n=0}|c^{(1)}_n\bar{c}^{(2)}_n|\log\left(\frac{|c^{(1)}_n\bar{c}^{(2)}_n|}{\tilde{f}^{(1|2)}}\right) = \\
&=\log(\tilde{f}^{(1|2)})-\frac{1}{\tilde{f}^{(1|2)}}\sum^{d-1}_{n=0}|c^{(1)}_n\bar{c}^{(2)}_n|\log\left(|c^{(1)}_n\bar{c}^{(2)}_n|\right). \label{SSVD-2-component}
\end{split}
\end{align}
In particular, we present explicit formulas for two qubits with $d=2$ in appendix \ref{App:2-qubits}. We take advantage of all these formulas in what follows. 
%%%%%%%%%%%%%%%%%%%%%%%%%%%%%%%%%%%%%%%%
%%%%%%%%%%%%%%%%%%%%%%%%%%%%%%%%%%%%%%%%
\section{Chern--Simons theory and link complement states}\label{sec:link_complement}
%%%%%%%%%%%%%%%%%%%%%%%%%%%%%%%%%%%%%%%%
%%%%%%%%%%%%%%%%%%%%%%%%%%%%%%%%%%%%%%%%
The main objects that we will examine using the quantum-information tools introduced above will be the link complement states, which are defined using formalism of Chern--Simons theory. Here we briefly review their construction and refer to~\cite{witten1989quantum,Cabra_Rossini_1997,kohno2002conformal,Kaul:1993hb,Kaul:1998ye} and \cite{Balasubramanian_2017,Balasubramanian:2018por,Leigh:2021trp,Camilo:2019bbl,Fliss:2020yrd} for more details and applications.

Chern--Simons theory is a 3-dimensional topological quantum field theory defined by the action
\begin{equation}
S = \frac{k}{4\pi} \int_M \Tr \Big(A \wedge \dd A+ \frac{2}{3} A \wedge A \wedge A  \Big), \label{eq:cs_action}
\end{equation}
where $A=A_\mu dx^\mu$ is a gauge field, the coupling $k$ (that takes integer values) is called the level, and $M$ is a 3-manifold on which the theory is defined. Various expectation values in this theory are naturally expressed in terms of a parameter
\begin{equation}
q=\exp \left( \frac{2 \pi i}{k+\gamma}\right), \label{def:q}
\end{equation}
where $\gamma$ is the dual Coxeter number of the gauge group under consideration; in particular $\gamma=N$ in $\mathrm{SU}(N)$ theory. In what follows we also use the $q$-number, $q$-factorial and $q$-Pochhammer symbol, defined respectively as 
\begin{equation}
    [x] = \frac{q^{x/2}-q^{-x/2}}{q^{1/2}-q^{-1/2}}, \qquad
    [x]! = [x][x-1]\cdots[1], \qquad
    (z;q)_k = \prod_{j=0}^{k-1}(1-zq^j).
    \label{eq: q-nfp}
\end{equation}

An important role in Chern--Simons theory is played by modular matrices $\cS$ and $\cT$. In the $\mathrm{SU}(2)$ case these matrices are related to the quantum representation of the modular group $\mathrm{PSL}(2,\mathbb{Z})$ at level $k$; they take the form
\begin{align}
\begin{split}
\cS_{lm} &= \sqrt{\frac{2}{k+2}} \sin \Big(\frac{(l+1)(m+1) \pi }{k+2}\Big)
= \frac{q^{\frac{(l+1)(m+1)}{2}}-q^{-\frac{(l+1)(m+1)}{2}}}{i\sqrt{2(k+2)}}, \\ \cT_{lm} &= \delta_{lm} q^{\frac{l(l+2)}{4}}, 
\label{def:S_T_matrices}
\end{split}
\end{align}
where $0 \le l,m \le k$ label integrable representations of $\mathrm{SU}(2)$. The above matrices satisfy the relations ~\cite{francesco1997conformal,kohno2002conformal}
\begin{equation}
\cS^{2}=1, \;\qquad (\cS\cT)^{3}=q^{\frac{3k}{8}}. \label{eq:SandTidentities}
\end{equation}

\begin{figure}[t!]
    \centering
    \includegraphics[width=0.4\linewidth,height=3cm]{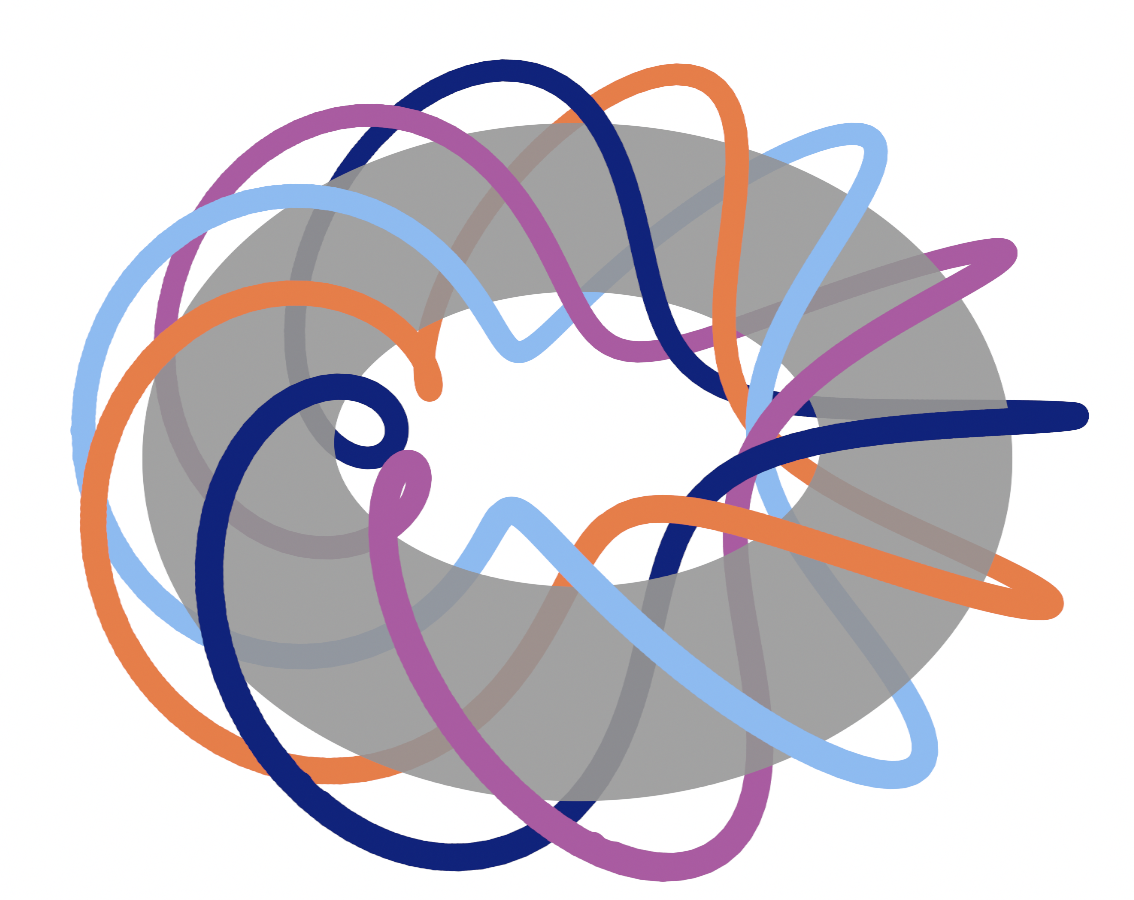}
    \caption{Representative example of a generic torus link $\mathrm{T}(p,q)$ with $p=4$ and $q=12$.} \label{fig-torus}
\end{figure}

\begin{figure}[b!]
    \centering
        \includegraphics[width=0.6\linewidth, height = 4cm]{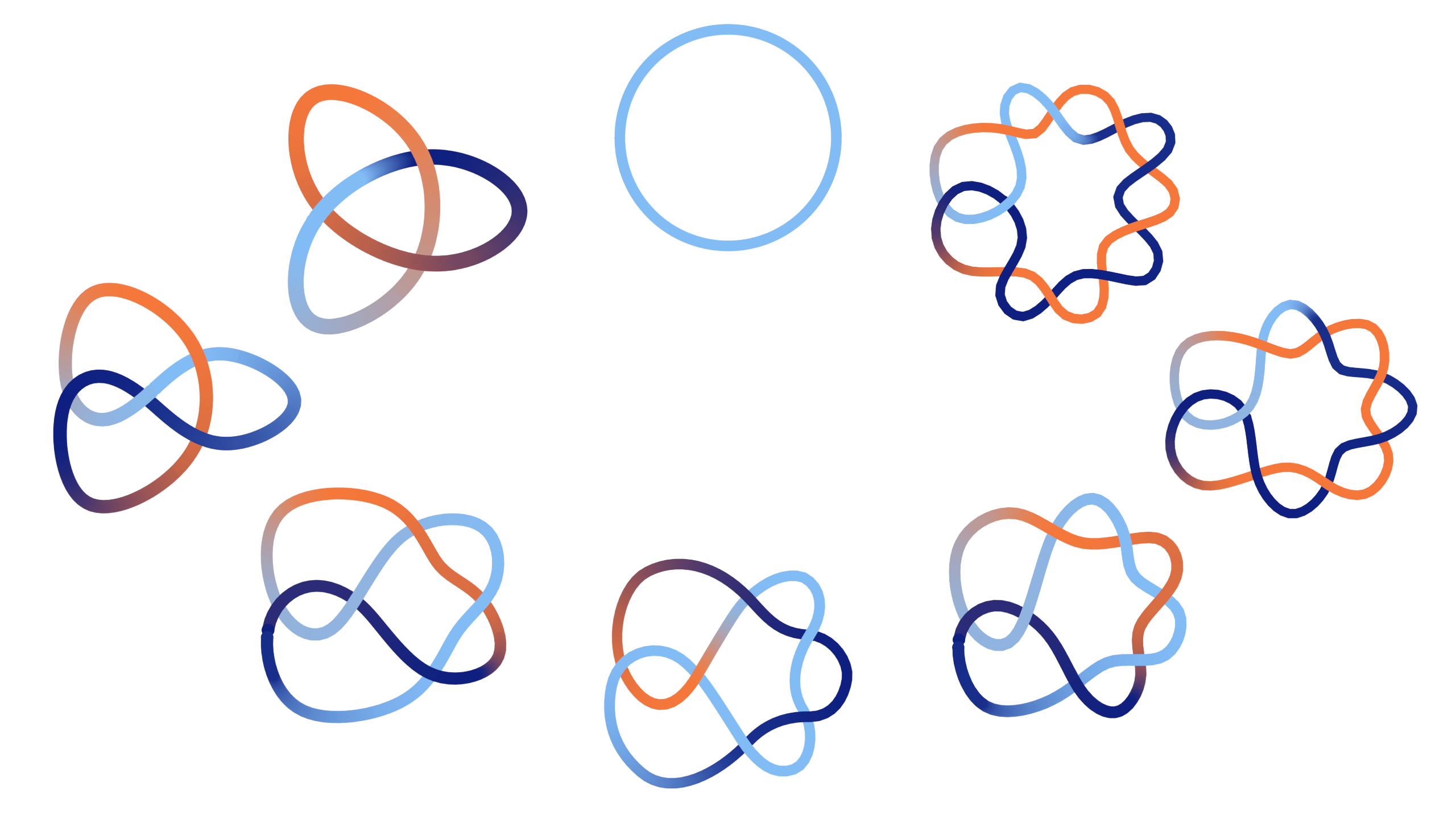}
    \caption{Examples of twist knots $\mathcal{K}_p$ for $p=0$ (unknot), 1 (trefoil) , $-1$ (figure-8), $2,-2,3,-3,4$ (in anti-clockwise order, starting from the unknot at the top).}
    \label{twistknots pic}
\end{figure}

\begin{figure}[t!]
    \centering
    \begin{subfigure}[b]{0.32\textwidth}
        \centering
        \includegraphics[width=0.9\linewidth, height = 3cm]{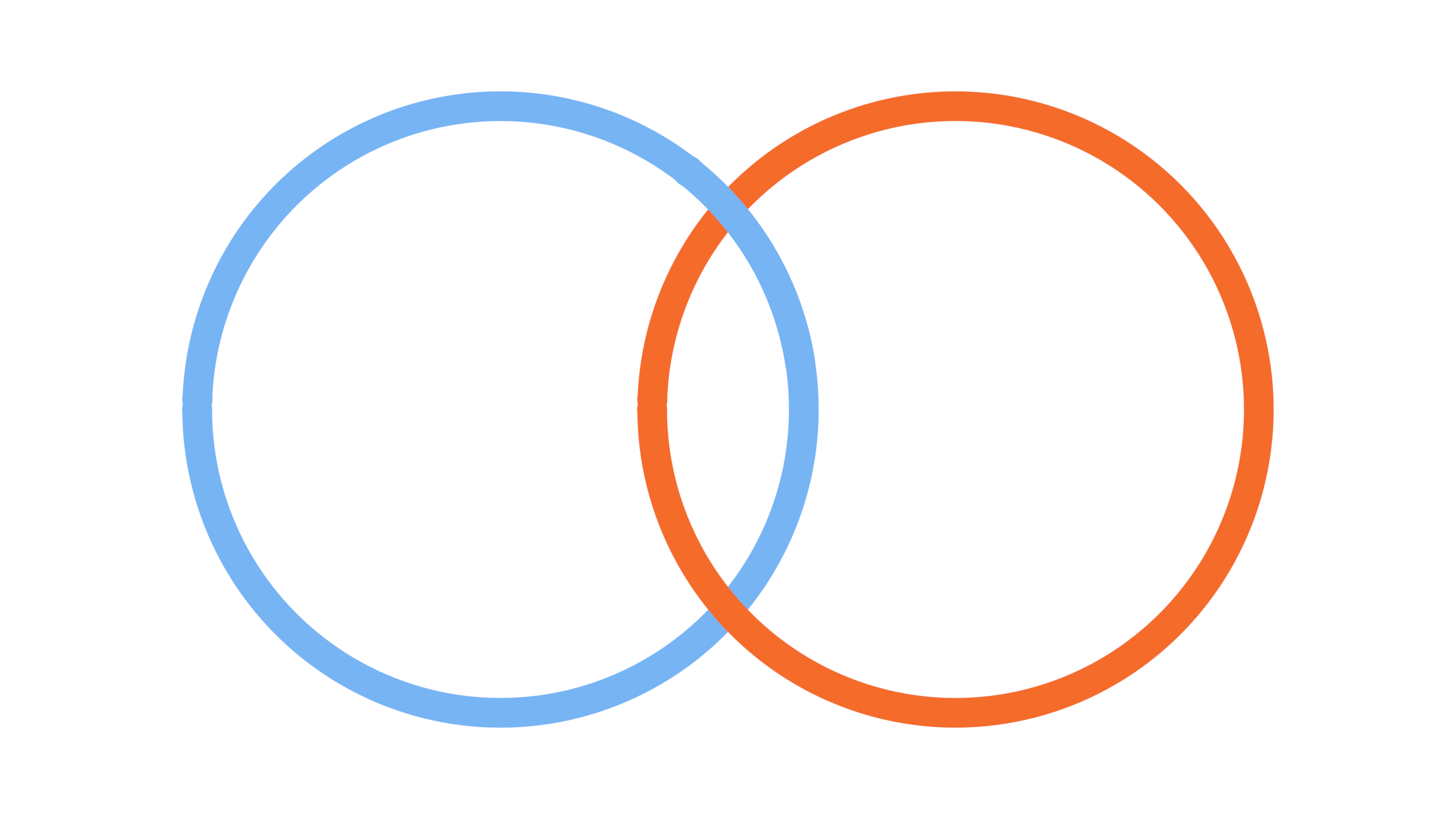}
%        \caption{Unknot $0_1$}
        \label{hopf picture}
    \end{subfigure}
    \hfill
    \begin{subfigure}[b]{0.32\textwidth}
        \centering
        \includegraphics[width=0.9\linewidth, height = 3cm]{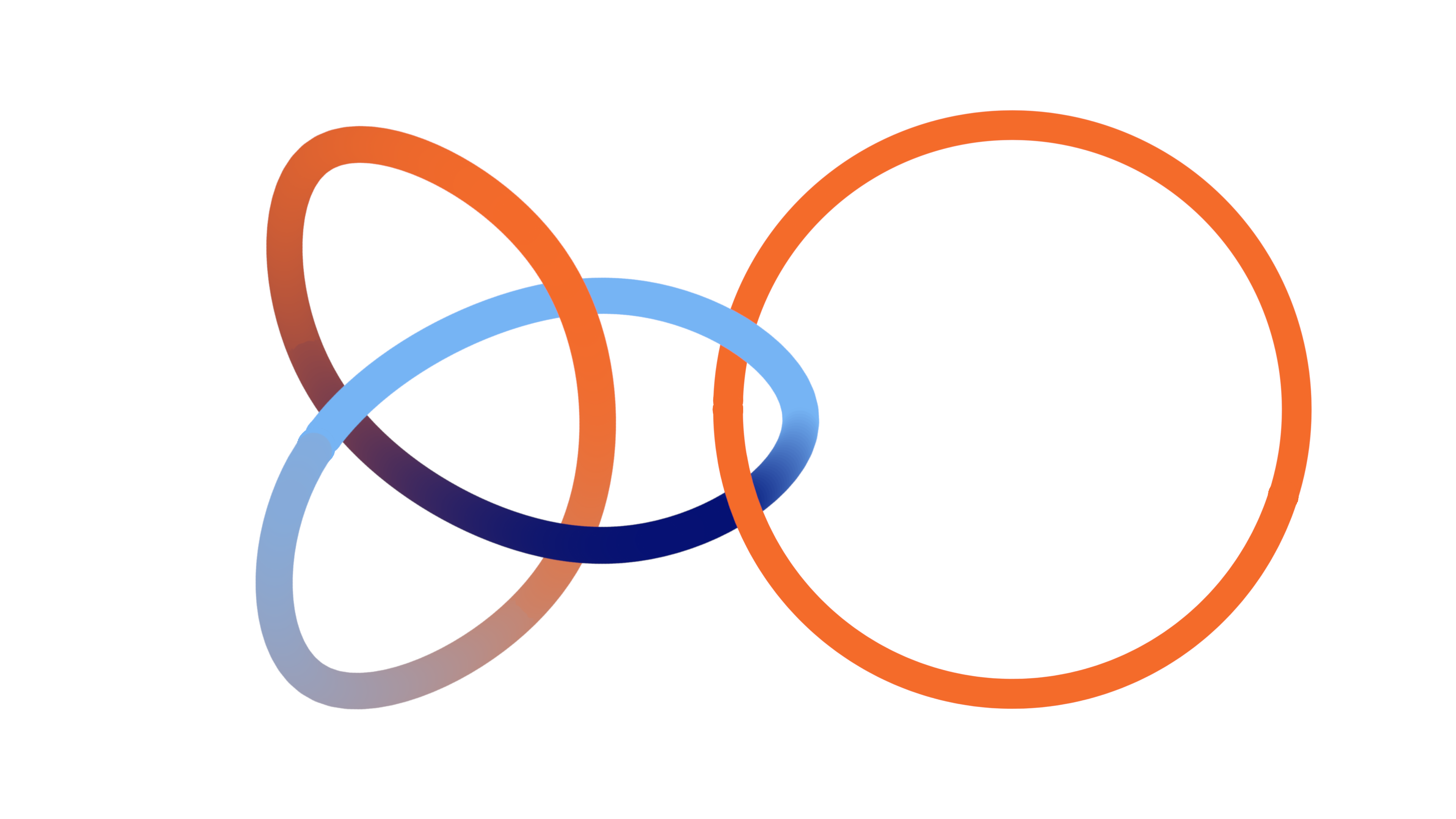}
%        \caption{Trefoil knot $3_1$}
    \end{subfigure}
        \hfill
    \begin{subfigure}[b]{0.32\textwidth}
        \centering
        \includegraphics[width=0.9\linewidth, height = 3cm]{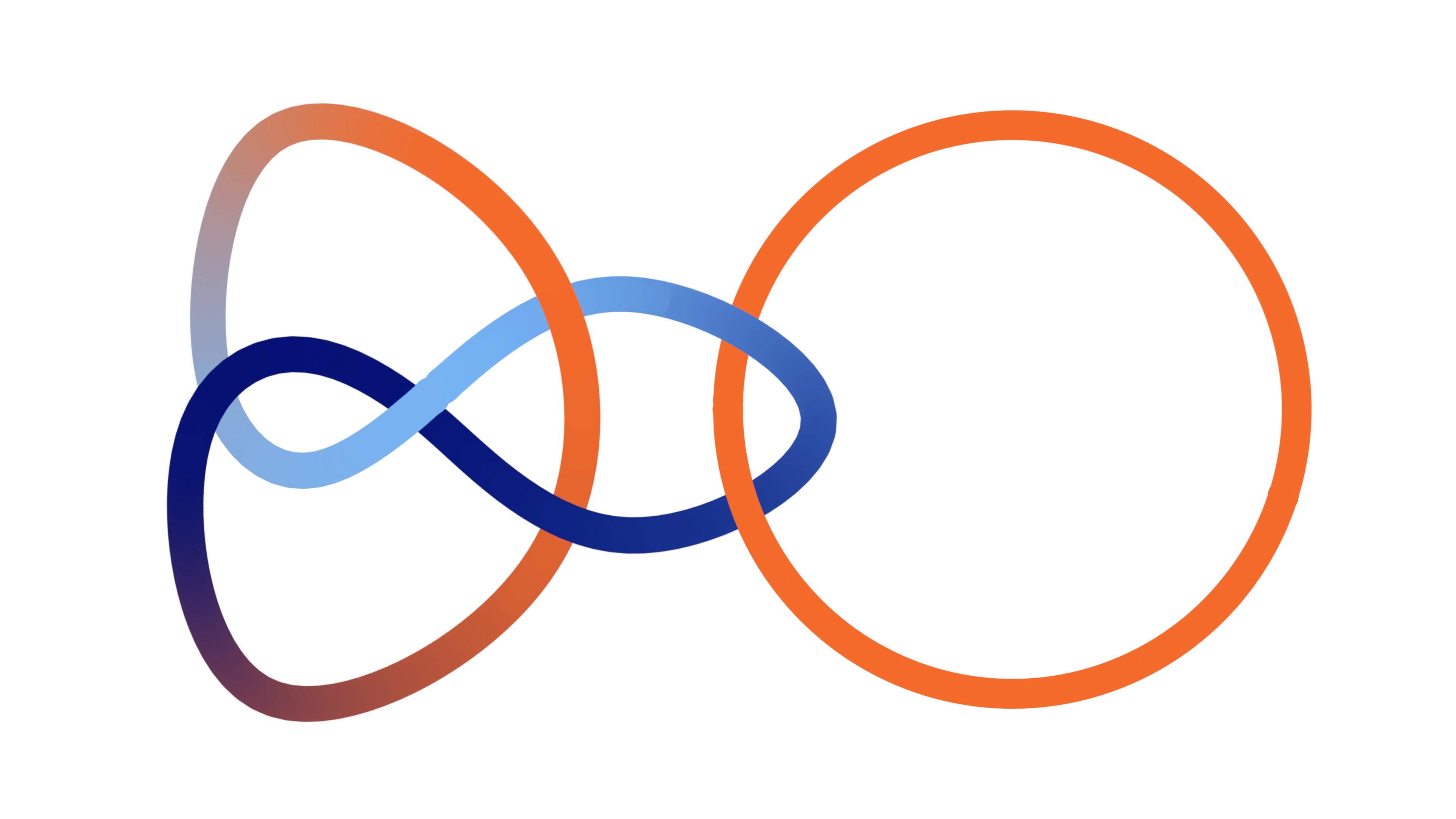}
 %       \caption{Figure-8 knot $4_1$}
    \end{subfigure}
    \caption{Links of the form $\mathcal{K}\#{2^{2}_{1}}$ for $\mathcal{K}=0_1 $\textrm{\ (unknot)}$, 3_1 \textrm{ (trefoil knot)}$ and $4_1$ (figure-8 knot)  respectively.}
    \label{K221links}
\end{figure}

Interesting observables in Chern--Simons theory -- which are also building blocks of the link states that we are going to consider -- are expectation values of Wilson loops associated to knots $\cK$ and ($n$-component) links $\cL^n= \bigsqcup_{i=1}^{n} \cK^{i}$ (i.e. disjoint unions of knots $\cK^{1},\ldots,\cK^{n}$). The simplest knot, i.e. unentangled loop, is called the unknot, denoted $0_1$. One infinite family of knots and links that we consider are the torus knots (for relatively prime $p$ and $q$) and torus links (for $p$ and $q$ not relatively prime) $\mathrm{T}(p,q)$, i.e. those that can be formed by winding a piece of rope respectively $p$ and $q$ times along two cycles of a torus, see fig. \ref{fig-torus}. A $\mathrm{T}(p,q)$ torus link has $\gcd(p,q)$ components; any two of them weave around one another with linking number $\frac{pq}{\gcd(p,q)^{2}}$. The simplest non-trivial torus knot is the trefoil knot $\mathrm{T}(2,3)$ also denoted $3_1$, while the simplest torus link is the Hopf-link $\mathrm{T}(2,2)$, also denoted $2^2_1$, which is made of two interlacing unknots. Of our interest are also twist knots $\mathcal{K}_p$, see fig. \ref{twistknots pic}, which are an infinite family of knots constructed by taking a loop, making respectively $2p-1$ half-twists for positive $p$ or $|2p|$ half-twists for negative $p$, and linking its ends together. These include the unknot $\mathcal{K}_0=0_1$, trefoil knot $\mathcal{K}_1 = 3_1$, figure-8 knot $\mathcal{K}_{-1} = 4_1$, as well as $\mathcal{K}_2 = 5_2, \mathcal{K}_{-2} = 6_1, \mathcal{K}_3 = 7_2, \mathcal{K}_{-3} = 8_1, \mathcal{K}_{4} = 9_2$, etc.  Another class of links that we consider are connected sums of the form $\mathcal{K}\#2^2_1$, which take the form of the Hopf-link whose one component is replaced by the knot $\mathcal{K}$; in our considerations we choose
$\mathcal{K}$ to be a twist knot, see fig. \ref{K221links}. 
In particular, $0_1\#2^2_1 = 2^2_1$ is the Hopf-link.


For a link $\cL^n$ (and in particular for a knot, for $n=1$), the expectation values of Wilson loops in Chern--Simons theory reproduce colored (knot and) link invariants and take the form 
\begin{equation}
C^{\cL^n}_{m_{1},\ldots,m_{n}}=\Big\langle \prod_{i=1}^{n} W_{R_{m_{i}}}(\cK^{i}) \Big\rangle_{S^{3}},\qquad W_{R_{m_{i}}}(\cK^{i})= \Tr_{m_{j_{i}}} \cP \exp \Big( i \oint_{\cK^{i}} A \Big), \label{CLn}
\end{equation}
where $W_{R_{m_{i}}}(\cK^{i})$ involves an integral of the gauge field along $i$-th component of a link $\cK^i$, $\cP$ denotes the path ordering, and $R_{m_{i}}$ (also referred to as the color) for a given level $k$ is an integrable representation of the gauge group~\cite{GEPNER1986493}. For $\textrm{SU}(N)$ gauge group, the labels $m_i$ of integrable representations are given by $\frac{(k+N-1)!}{(N-1)!k!}$ Young diagrams that fit into the rectangle of size $k\times (N-1)$; for $\textrm{SU}(2)$ they can be identified as integers $m_i= 0,\ldots,k$ that label symmetric representations $S^{m_i}$. Furthermore, for $\textrm{SU}(N)$ gauge group, link invariants (\ref{CLn}) are polynomials (or rational functions, depending on normalization) in $q$ and $a=q^N$, referred to as colored HOMFLY-PT polynomials; for $\textrm{SU}(2)$ and the specialization $a=q^2$ they reduce to colored Jones polynomials. The polynomials that we consider in what follows are normalized so that for the unknot they are equal to 1. We denote HOMFLY-PT polynomials of a knot $\mathcal{K}$ colored by $m$-th symmetric representation $S^m$ by $P^{\mathcal{K}}_m(a,q)$, while colored Jones polynomials by $V^{\mathcal{K}}_m(q) \equiv P^{\mathcal{K}}_m(q^2,q)$. When we refer to an arbitrary knot, or it is clear to which knot we refer to, we ignore the knot label, and we also often skip the representation label when a knot polynomial is uncolored, i.e. when it is colored by the fundamental representation; e.g. $V(q)\equiv V_1(q)$.
Let us provide some examples of knot invariants colored by symmetric representations $S^m$ (i.e. those represented by Young diagrams that consist of one row of length $m$), which we also use in what follows. The  colored HOMFLY-PT polynomials of twist knots $\mathcal{K}_{p}$ take the form \cite{nawata2012super} 
\begin{equation}
\begin{aligned}
P^{\mathcal{K}_{p}}_m\left(a, q\right)= \sum_{k=0}^{\infty} \sum_{\ell=0}^k &q^k \frac{\left(aq^{-1} ; q\right)_k}{(q ; q)_k}\left(q^{1-m} ; q\right)_k\left(a q^{m-1} ; q\right)_k \\
& \times(-1)^{\ell} a^{p \ell} q^{(p+1 / 2) \ell(\ell-1)} \frac{1-a q^{2 \ell-1}}{\left(a q^{\ell-1} ; q\right)_{k+1}} \frac{(q; q)_k}{(q; q)_\ell(q; q)_{k-\ell}}.
\end{aligned}  
\label{eqn: twist knots master formula}
\end{equation}
For $N=2$ (i.e. $a=q^2$) they reduce to colored Jones polynomials~\cite{CJP-twist}
\begin{equation}
    V_m^{\mathcal{K}_p} = \sum_{k=0}^{\infty} \sum_{\ell=0}^{k} q^{k} (q^{1-m}; q)_{k}(q^{m+1}; q)_{k} (-1)^{\ell}q^{\ell(p+1)+p(\ell-1)/2}(1 - q^{2\ell+1}) \frac{(q;q)_{k}}{(q;q)_{\ell+k+1}(q;q)_{k-\ell}},
    \label{eq: twist knots}
\end{equation} 
and in particular, for
 trefoil $3_1=\mathcal{K}_1$ and figure-8 knot $4_1=\mathcal{K}_{-1}$, they take the form \cite{nawata2012super}\footnote{Equivalent formulas can be found in \cite{Habiro}.}
\begin{equation}
\begin{aligned}
V^{3_1}_m  &= \sum_{k=0}^{\infty} q^k \left(q^{1-m}; q\right)_k \left(q^{1+m}; q\right)_k, \\
V^{4_1}_m  &=  \sum_{k=0}^{\infty} (-1)^k q^{-\frac{k(k+1)}{2}} \left(q^{1-m}; q\right)_k \left(q^{1+m}; q\right)_k.  
\end{aligned}
\label{eqn: 3_1 and 4_1 cjp}
\end{equation}

Colored Jones polynomials for torus knots $\mathrm{T}(P,Q)$ (with relatively prime $P$ and $Q$) read \cite{hikami2004difference}
\begin{equation}
        V^{\mathrm{T}(P,Q)}_m = \frac{q^{-PQ(1-m^2)/4}}{q^{-m/2} - q^{m/2}} \sum_{r=-(m-1)/2}^{(m-1)/2} \Big( q^{-PQ r^2+(P+Q)r-1/2} - q^{-PQ r^2+(P-Q)r+1/2} \Big).
    \label{torusknots}
\end{equation}
For a Whitehead link shown in \cref{whiteheadpics}, with its two components colored respectively by $S^m$ and $S^n$ symmetric representations, Jones polynomial takes the form (\ref{Whitehead-Jones}). For the three component Borromean link shown in \cref{borromean link}  the colored Jones polynomial takes the form (\ref{borro eqn}). Other examples of colored polynomials for various knots and links, and also their generalizations to super-polynomials (i.e. deformations of HOMFLY-PT polynomials that depend on an additional parameter $t$ and capture some information about homological invariants) can be found in \cite{Fuji:2012pi,Gukov:2015gmm}.

The link states of our interest are associated to manifolds $M_n=S^{3} \setminus N(\cL^n)$, which are obtained by removing a tubular neighbourhood $N(\cL^n)$ of a link $\cL^n$ from $S^3$ (a tubular neighbourhood is obtained by thickening each component of a link to a solid torus). The boundary of $M_n$ takes the form of $n$ copies of a torus, $\partial M_n=\cup^n_{i=1}T^2$. The path integral in Chern--Simons theory with gauge group $G$ and level $k$ produces then a link state, i.e. quantum state in the $n$-fold tensor product Hilbert space $\mathcal{H}^{\otimes n}=\mathcal{H}(T^2,G,k)^{\otimes n}$, which can be expanded as
\begin{equation}
\ket{\cL^n}= \sum_{m_{1},\ldots,m_{n}} C^{\cL^n}_{m_{1},\ldots,m_{n}} \ket{m_{1},\ldots,m_{n}}, \label{link-state}
\end{equation}
where $\ket{m}$ are basis elements of the torus Hilbert space $\mathcal{H}$ labeled by integrable representations $m$ of $G$, and the coefficients $C^{\cL^n}_{m_{1},\ldots,m_{n}}$ are the link invariants (\ref{CLn}). Note that these states are not normalized, which however does not affect our considerations, as any normalization factors cancel in 
(\ref{tau-phi-psi}) and lead to the same values of entropy measures that we analyze.

In what follows we will take advantage of the unitarity of the $\cS$ matrix (\ref{def:S_T_matrices}), which can be used to implement a unitary basis transformation
\begin{equation}
\label{unitary basis}
    \ket{m} \equiv  \sum_n \cS_{mn} \ket{n}.
\end{equation}
In our cases this transformation does not affect the entanglement properties of the link states and can conveniently be used to diagonalize various transition matrices. For example, in $\mathrm{SU}(2)$ Chern--Simons theory at level $k$ we can rewrite link states associated to links of the form $\mathcal{K}\#{2^{2}_{1}}$ as follows \cite{Balasubramanian_2017} 
\begin{equation}
    |\mathcal{K}\#{2^{2}_{1}} \rangle = \sum_{m,n = 0}^{k} \frac{C^{\mathcal{K}}_m}{\cS_{0m}} \cS_{mn}  |m, n \rangle \equiv \sum_{m = 0}^k  \tilde{C}^{\mathcal{K}}_m   |m, m\rangle,
   % \label{K221 link eq}
   \label{eqn:connected_sum_firstinstance}
\end{equation}
where $C^{\mathcal{K}}_m$ and $\tilde{C}^{\mathcal{K}}_m= C^{\mathcal{K}}_m/\cS_{0m}$ are respectively unreduced and reduced colored invariants of the knot $\mathcal{K}$. The link states for $(p,q)$ torus links with $d=\textrm{gcd}(p,q)$ components in $\mathrm{SU}(2)$ Chern--Simons theory can be written as \cite{Leigh:2021trp, Balasubramanian:2018por,Dwivedi_2020,Stevan_2010,Brini_2012}
\begin{align}
\begin{split}
\ket{\mathrm{T}(p,q)} & = \sum_{m=0}^k \sum_{l=0}^k \frac{1}{S_{0m}^{d-1}} S_{m l} V^{T_{p/d,q/d}}_{l} |\underbrace{m,\cdots,m}_{d \, \mathrm{entries}}\rangle = \\
&=\sum_{m=0}^{k} \left(\cS X\left(\frac{p}{d}\right) \cT^{\frac{q}{p}} \cS \right)_{m0} \frac{1}{\cS_{0m}^{d-1}} \ket{m,\ldots,m}, \label{eq:general_torus_link_component_state_firstinstance}
\end{split}
\end{align}
where the matrix $X(t)$, whose components $X_{ab}(t)$ are referred to as Adams coefficients, is defined by~\cite{Dwivedi_2020} 
\begin{equation}
\Tr_{a}\left(U^{t}\right) = \sum_{b=0}^{k} X_{ab}(t) \Tr_{b} \left(U\right), \qquad \textrm{for}\ U \in \mathrm{SU}(2).
 \label{defn:xmatrix}
\end{equation}
%%%%%%%%%%%%%%%%%%%%%%%%%%%
%%%%%%%%%%%%%%%%%%%%%%%%%%%
\section{Quantum mechanical examples}
\label{sec:toymodels}
%%%%%%%%%%%%%%%%%%%%%%%%%%%
%%%%%%%%%%%%%%%%%%%%%%%%%%%
In this section we use entropy measures to characterize states in a few quantum mechanical examples. We consider generalized coherent states for $\mathrm{SU}(2)$ and $\mathrm{SU}(1,1)$ Lie algebras in the two-mode (bipartite) representation \cite{Perelomov:1971bd},
as well as tripartite GHZ and W states. 

Note that the above systems have different numbers of modes and different sizes of a Hilbert space (for each of those modes). In case of bipartite coherent states for $\mathrm{SU}(2)$, a choice of a given representation determines the dimension of the Hilbert space for each mode, which can be any (finite) integer number. On the other hand, for each of the two modes of $\mathrm{SU}(1,1)$ coherent states, the Hilbert space is infinite-dimensional. Finally, we consider tripartite GHZ and W states associated to two-dimensional (qubit) Hilbert space. The link states that we analyze in the next section can be thought of as generalizations of these systems, in the sense they may involve an arbitrary number of components, and the size of the Hilbert space is an arbitrary integer fixed by the choice of the level.

In this section, for the above quantum mechanical systems we evaluate their pseudo-entropy and SVD entropy, and discuss how various states are distinguished by the entropy excess. 

%%%%%%%%%%%%%%%%%%%%%%%%%%%%%%%%%%%%%%%%%%%%
\subsection{$\mathrm{SU}(2)$ coherent states}
\label{sec: coherent states}
%%%%%%%%%%%%%%%%%%%%%%%%%%%%%%%%%%%%%%%%%%%%%
As the first example we consider coherent states associated to  the $\mathrm{SU}(2)$ Lie algebra. Its generators $J_i$, $i=1,2,3$, satisfy commutation relations $[J_i,J_k]=i\epsilon_{ijk}J_k$, which can be written in terms of the ladder operators $J_\pm=J_1\pm iJ_2$ as
\be
[J_3,J_\pm]=\pm J_\pm,\qquad [J_+,J_-]=2J_3.
\ee
The lowest weight states $\ket{-j}$ are defined by
\be
J_3 \ket{-j} = -j \ket{-j}, \qquad J_- \ket{-j} = 0,
\ee
and generalized coherent states are conventionally defined by acting with a displacement operator on $\ket{-j}$ 
\be
\ket{z,j}=e^{\xi J_+-\bar{\xi}J_-}\ket{-j}.
\ee
Using the Baker–Campbell–Hausdorff (BCH) formula for $\mathrm{SU}(2)$ we can expand this state as
\be
\ket{z_i,j}=(1+|z_i|^2)^{-j}\sum^{2j}_{n=0}z^{n}_i\,\sqrt{\frac{\Gamma(2j+1)}{n!\Gamma(2j-n+1)}}\,\ket{n}_A\otimes\ket{n}_B, \label{ketSU2}
\ee
where basis vectors $\ket{n}_i$ have $n$ powers of $J_+$ acting on them\footnote{Formally they can be constructed by using the two-mode representation of the algebra that we associate with $A$ and $B$.}. The Hilbert space $\mathcal{H}=\mathcal{H}_A\otimes \mathcal{H}_B$ is finite dimensional and each component $\mathcal{H}_i$ has dimension $d_i=2j+1$. Moreover, the complex coordinate $z_i$ admits a geometric interpretation as a point on the stereographic projection of the unit sphere\footnote{Relation to coordinate $\xi$ enters through $\xi=\frac{\theta}{2}e^{i\phi}$.}
\be
z_i=\tan\left(\frac{\theta_i}{2}\right)e^{i\phi_i}\equiv t_i e^{i\phi_i},\qquad\theta_i\in [0,\pi],\qquad \phi_i\in [0,2\pi]. \label{zi-tan}
\ee
In our context, we consider the transition matrix between two coherent states labeled by different $z_i$'s 
\be
\tau^{1|2}=\frac{\ket{z_1,j}\bra{z_2,j}}{\langle z_2,j|z_1,j\rangle}.
\ee
Since the coherent states form an over-complete basis, the overlap between them is non-trivial and given by
\be
\langle z_2,j|z_1,j\rangle=\frac{(1+|z_1|^2)^{-j}(1+|z_2|^2)^{-j}}{(1+z_1\bar{z}_2)^{-2j}}.
\ee
This way, after tracing over the second Hilbert space, we obtain the reduced transition matrix
\be
\tau^{1|2}_A=Tr_{\mathcal{H}_B}\big(\tau^{1|2}\big)=(1+z_1\bar{z}_2)^{-2j}\sum^{2j}_{n=0}\frac{\Gamma(2j+1)}{n!\Gamma(2j-n+1)}(z_1\bar{z}_2)^n\ket{n} \bra{n},
\ee
and similarly the density matrix \eqref{def:svde} 
\be
\rho^{1|2}_A=(1+|z_1\bar{z}_2|)^{-2j}\sum^{2j}_{n=0}\frac{\Gamma(2j+1)}{n!\Gamma(2j-n+1)}|z_1\bar{z}_2|^n\ket{n} \bra{n}.
\ee
Clearly the $2j+1$ complex eigenvalues of $\tau^{1|2}_A$ are parametrized by
\be
z_1\bar{z}_2=t_1 t_2e^{i\phi_{12}},\qquad \phi_{12}=\phi_1-\phi_2,
\ee
whereas those of $\rho^{1|2}_A$ are real and parametrized by the absolute value of the expression above which has the complex phase removed
\be
\hat{\lambda}_n=\frac{\Gamma(2j+1)}{n!\Gamma(2j-n+1)}(1+t_1t_2)^{-2j}(t_1t_2)^n,\qquad \sum^{2j}_{n=0}\hat{\lambda}_n=1.
\ee

After closer examination of these eigenvalues, we can compute von Neumann, pseudo and SVD entropies at once. Indeed, $\hat{\lambda}$ simply follows the binomial distribution so it is convenient to introduce 
\be
P_n(X)=\binom{2j}{n}p^n(1-p)^{2j-n},\qquad p=\frac{X}{1+X},
\ee
and compute
\be
S(j,X)=-\sum^{2j}_{n=0}P_n(X)\log(P_n(X)),
\ee
where $X=t^2_i$ for the computation of von Neumann entropies, $X=t_1t_2\exp(i\phi_{12})$ for the pseudo-entropy and $X=t_1t_2$ for the SVD entropy. While getting a closed form for arbitrary $j$ is not possible, we can easily derive the answer for small $j$ or perform the sums and plot numerically. For instance, for the first three $j=m/2$ with $m=1,2,3$ we simply get
\be
S(m/2,X)=m(-p\log(p)-(1-p)\log(1-p))-p(1-p)m\log(m).\label{EntropsSUp}
\ee
For higher $j$ this expression gets corrected by higher polynomials in $p(1-p)$. On the other hand, we can use Stirling's approximation, or equivalently the central limit or de Moivre–Laplace theorem, to derive the asymptotic expression for large $j$. The answer diverges logarithmically with $j$
\be
S(j,X)\simeq\frac{1}{2}\log\left(4\pi e\, p(1-p)j\right).\label{AsymptotjEQ}
\ee

\begin{figure}[h!]
\centering
    \includegraphics[width=8cm]{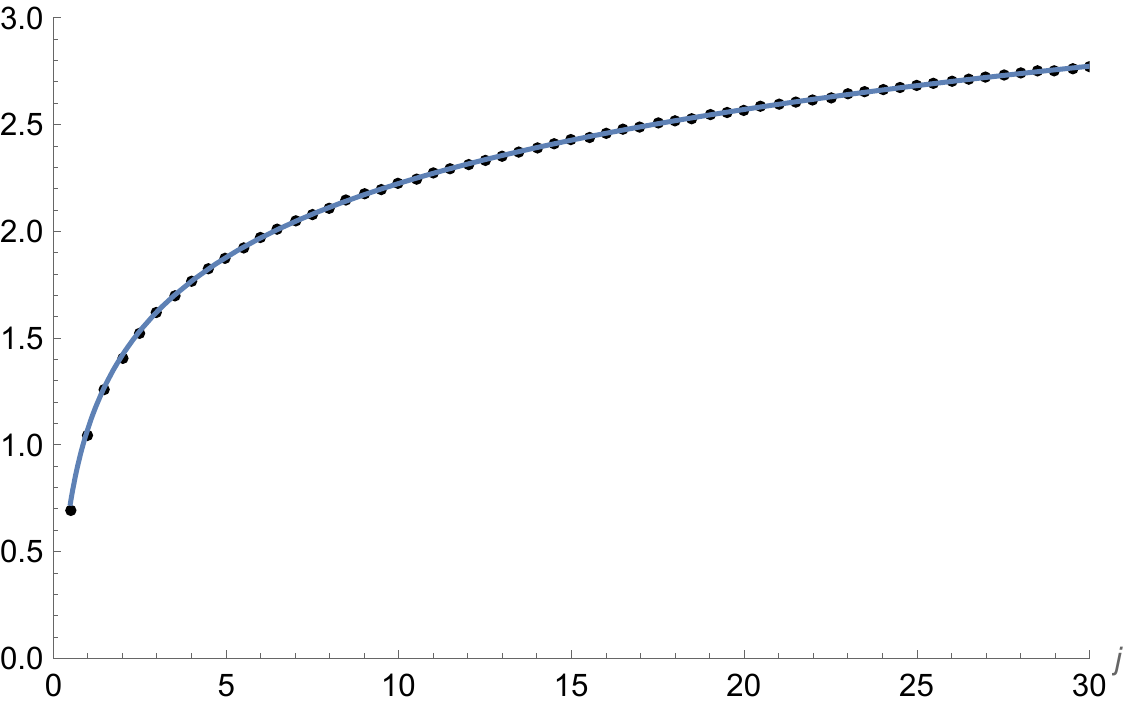}
    \caption{Entanglement entropy as a function of $j$ for $\theta=\pi/2$ (dots) vs asymptotic formula \eqref{AsymptotjEQ} (solid blue curve).}
    \label{Asymptotj}
\end{figure}

Let us now analyze our quantities of interests. Firstly, the von Neumann entropies that are parametrized by $t=\tan(\theta/2)$ have the maximum value (for any $j$) for $\theta=\pi/2$ or $t=1$ and they vanish for $\theta=\{0,\pi\}$. We will use these ``maximally entangled states" as our  target states in the transition matrix. As an example, in Fig. \ref{Asymptotj} we present the von Neumann entropy as a function of $j$ for $\theta=\pi/2$ (equivalent to $p=1/2$) vs. the asymptotic formula \eqref{AsymptotjEQ}.

Next, for concreteness and analytical control, let us focus on the $j=1/2$ example where entanglement entropies are
\be
S^i_{\mathrm{E}}=\log(1+t^2_i)-\frac{2t^2_i\log(t_i)}{1+t^2_i}.
\ee
The pseudo-entropy becomes
\be
S^{1|2}_{\mathrm{P}}=\log\left(1+t_1t_2e^{i\phi_{12}}\right)-\frac{(i\phi_{12}+\log(t_1t_2))t_1t_2e^{i\phi_{12}}}{1+t_1t_2e^{i\phi_{12}}},
\ee
and we can decompose it into real and imaginary parts as
\bea
S^{1|2}_{\mathrm{P}}&=&\frac{1}{2}\log(\Delta_{12})-t_1t_2\frac{(t_1t_2+\cos(\phi_{12}))\log(t_1t_2)-\phi_{12}\sin(\phi_{12})}{\Delta_{12}}\nn\\
&-&i\left[t_1t_2\frac{(t_1t_2+\cos(\phi_{12}))\phi_{12}+\sin(\phi_{12})\log(t_1t_2)}{\Delta_{12}}+\frac{i}{2}\log\left(\frac{1+t_1t_2e^{i\phi_{12}}}{1+t_1t_2e^{-i\phi_{12}}}\right)\right],\nn\\
\eea
where
\be
\Delta_{12}=|1+t_1t_2e^{i\phi_{12}}|^2=1+t^2_1t^2_2+2t_1t_2\cos(\phi_{12}).
\ee
Clearly, the imaginary part of the pseudo-entropy arises due to the non-trivial phases of our two states that have different $\phi_i$ points on the sphere. Moreover, if we flip the phases, $\phi_1\leftrightarrow\phi_2$ the imaginary part changes the sign. We will elaborate on this property in the context of link states in the last section.

Analogously, we can evaluate the SVD entropy that in our $j=1/2$ example can be simply obtained from pseudo-entropy by setting $\phi_{12}=0$
\be
S^{1|2}_{\mathrm{SVD}}=\log(1+t_1t_2)-\frac{t_1t_2\log(t_1t_2)}{1+t_1t_2}.
\ee
Since it has the form \eqref{EntropsSUp} with real $p\in (0,1)$ parametrized by $t_1t_2$, it is simply the entropy of a qubit density matrix with eigenvalues $\{p,1-p\}$ with maximum at $p=1/2$ at which it saturates the upper bound $S_{\mathrm{SVD}}=\log(d_A)=\log(2)$.

\begin{figure}[t!]
 \centering
     \includegraphics[width=7cm]{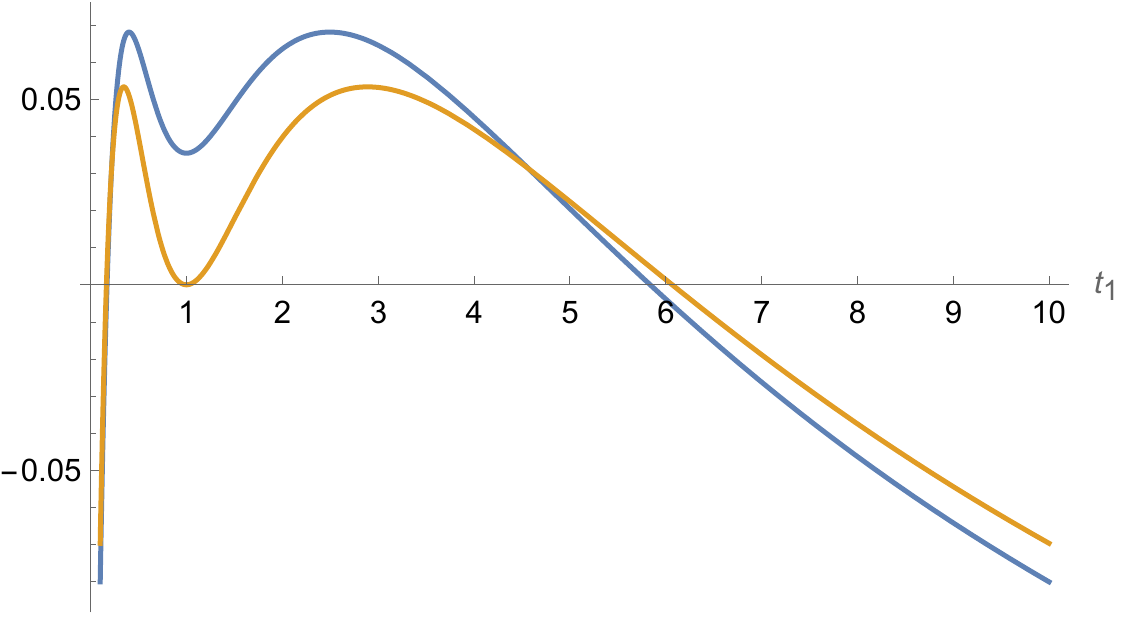} 
     \includegraphics[width=7cm]{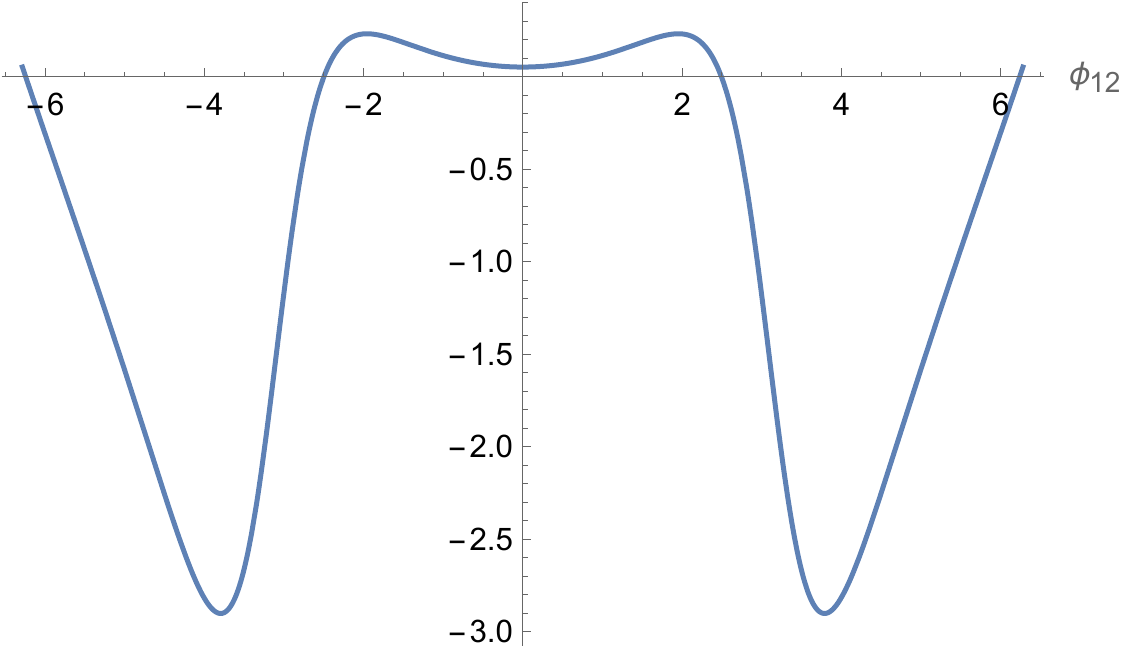} 
\caption{Pseudo-entropy and SVD entropy excess for $t_2=1$. Left: pseudo-entropy excess for $\phi_{12}=\pi/6$ (blue curve) and SVD excess (orange curve). Right: pseudo-entropy excess for $t_2=1$ and $t_1=0.4$. }\label{ExcessSU2}
\end{figure}

Finally, we analyze the excess of both quantities. The pseudo-entropy excess becomes
\bea
\Delta S^{1|2}_{\mathrm{P}}&=&\frac{1}{2}\log\left(\frac{\Delta_{12}}{(1+t^2_1)(1+t^2_2)}\right)+\frac{t^2_1\log(t_1)}{1+t^2_1}+\frac{t^2_2\log(t_2)}{1+t^2_2}\nn\\
& & - t_1t_2\frac{(t_1t_2+\cos(\phi_{12}))\log(t_1t_2)-\phi_{12}\sin(\phi_{12})}{\Delta_{12}},
\eea
whereas the SVD excess is
\be
\Delta S^{1|2}_{\mathrm{SVD}}= \frac{1}{2}\log\left(\frac{(1+t_1t_2)^2}{(1+t^2_1)(1+t^2_2)}\right)+\frac{t^2_1\log(t_1)}{1+t^2_1}+\frac{t^2_2\log(t_2)}{1+t^2_2}-\frac{t_1t_2\log(t_1t_2)}{1+t_1t_2}.
\ee
To plot these quantities it is instructive to take one of the states (post selected) as the maximally entangled state with $t_2=1$. We observe that it is possible to have both, negative and positive values of the excess as functions of $t_1$ as well as of $\phi_{12}$, see fig.\ref{ExcessSU2}.  

It is also interesting to evaluate the pseudo-entropy and SVD entropy excess using the asymptotic formula \eqref{AsymptotjEQ}. Inserting all the terms for pseudo-entropy yields
\be
\Delta S^{1|2}_{\mathrm{P}}=\frac{1}{2}\log\left[\frac{1+t^2_1t^2_2+t^2_1+t^2_2}{1+t^2_1t^2_2+2t_1t_2\cos\phi_{12}}\right].
\ee
Comparing the numerator and denominator we note that this excess is always positive since
\be
t^2_1+t^2_2>2t_1t_2\cos\phi_{12}\qquad \Leftrightarrow \qquad |t_1-t_2 e^{i\phi_{12}}|^2>0.
\ee
Similar steps for the SVD entropy give
\be
\Delta S^{1|2}_{\mathrm{SVD}}=\frac{1}{2}\log\left[\frac{1+t^2_1t^2_2+t^2_1+t^2_2}{1+t^2_1t^2_2+2t_1t_2}\right],
\ee
which is also always positive since $(t_1-t_2)^2>0$. This suggests that the positivity of the excess may be also correlated with the properties and inequalities satisfied by the entropies that we use in this combination. For instance the asymptotic formulas at large $j$ may not fulfill all the properties of the exact expressions. We leave exploring these properties as an important future research direction.
%%%%%%%%%%%%%%%%%%%%%%%%%%
\subsection{$\mathrm{SU}(1,1)$ coherent states}
%%%%%%%%%%%%%%%%%%%%%%%%%%
Next we move to coherent states of the simplest non-compact infinite dimensional Lie group  $\mathrm{SU}(1,1)$. This is the algebra of the group of all $2\times 2$ matrices with unit determinant, that leave invariant the Hermitian form $|z_1|^2-|z_2|^2$. The commutation relations of this algebra can be written as
\be
[L_0,L_{\pm 1}]=\mp L_{\pm 1},\qquad [L_1,L_{-1}]=2L_0,
\ee
where $L_{1}$ plays the role of the lowering  and $L_{-1}$ the rising operator. The highest weight states $\ket{h}$ are defined by
\be 
L_0\ket{h}=h\ket{h}, \qquad L_1\ket{h}=0,
\ee
while coherent states $\ket{z,h}$ are constructed by acting with displacement operator on $\ket{h}$
\be
\ket{z,h}=e^{\xi L_{-1}-\bar{\xi}L_1}\ket{h}.
\ee
Using the BCH relation for this algebra and the two-mode representation of the generators (see e.g. \cite{Perelomov:1971bd}), we can expand the state as
\be
\ket{z_i,h}=(1-|z_i|^2)^h\sum^\infty_{n=0}z^n_i\sqrt{\frac{\Gamma(2h+n)}{n!\Gamma(2h)}}\ket{n}_A\otimes\ket{n}_B.
\ee
This time, the Hilbert space of each bi-partition is infinite dimensional\footnote{A special case of these states with real $z_i$ is the thermofield double state of two harmonic oscillators for which pseudo-entropy was computed in \cite{Nakata:2020luh}.} and thus the summation over $n$ has an infinite range, in contrast to a finite summation in $\mathrm{SU}(2)$ case (\ref{ketSU2}). Moreover, the complex numbers parametrizing the state geometrize the hyperbolic disc\footnote{Here we have the relation to $\xi=\frac{\rho}{2}e^{i\theta}$.}
\be
z_i=\tanh\left(\frac{\rho_i}{2}\right)e^{i\theta_i}\equiv r_i e^{i\theta_i},\qquad |z_i|\leq 1. \label{zi-tanh}
\ee
Analogously to the previous example we define the transition matrix as
\be
\tau^{1|2}=\frac{\ket{z_1,h}\bra{z_2,h}}{\langle z_2,h|z_1,h\rangle},
\ee
where the overlap is now given by
\be
\langle z_2,h|z_1,h\rangle=\left(\frac{(1-|z_1|^2)(1-|z_2|^2)}{(1-z_1\bar{z}_2)^2}\right)^h.
\ee
This way, the reduced transition matrix is also diagonal
\be
\tau^{1|2}_A=(1-z_1\bar{z}_2)^{2h}\sum^\infty_{n=0}\frac{\Gamma(2h+n)}{n!\Gamma(2h)}(z_1\bar{z}_2)^n\ket{n} \bra{n},
\ee
and has infinite number of complex eigenvalues parametrized by
\be
z_1\bar{z}_2=r_1 r_2\, e^{i\theta_{12}},\qquad \theta_{12}\equiv\theta_1-\theta_2.
\ee
On the other hand the density matrix that contains the normalized singular values of this transition matrix becomes
\be
\rho^{1|2}_A=\left(1-|z_1\bar{z}_2|\right)^{2h}\sum^\infty_{n=0}\frac{\Gamma(2h+n)}{n!\Gamma(2h)}|z_1\bar{z}_2|^n\,\ketbra{n}.
\ee
The singular values are obviously real and take the form
\be
\hat{\lambda}_n=\frac{\Gamma(2h+n)}{n!\Gamma(2h)}\left(1-r_1r_2\right)^{2h}(r_1r_2)^n,\qquad \sum^\infty_{n=0}\hat{\lambda}_n=1. 
\ee
Examining this result we see that, similarly to the $\mathrm{SU}(2)$ coherent states, this time we end up with the negative binomial distribution. Analogously, we define
\be
P_n(Y)=\binom{2h+n-1}{n}(1-p)^n p^{2h},\qquad p=1-Y,
\ee
where we use $Y=r^2_i$ for von Neumann entropies, $Y=r_1r_2 e^{i\theta_{12}}$ for the pseudo-entropy and $Y=r_1r_2$ for the SVD entropy.

Again, obtaining the closed expression for higher $h$ is not possible. As an illustration, for $h=1/2$ we get
\be
S(1/2,Y)=-\log(1-Y)-\frac{Y}{1-Y}\log(Y),
\ee
and inserting the explicit expressions we obtain the pseudo-entropy
\bea
S^{1|2}_{\mathrm{P}}&=&-\frac{1}{2}\log(\tilde{\Delta}_{12})+r_1r_2\frac{(r_1r_2-\cos(\theta_{12}))\log(r_1r_2)+\theta_{12}\sin(\theta_{12})}{\tilde{\Delta}_{12}}\nn\\
&+&i\left[r_1r_2\frac{\theta_{12}(r_1r_2-\cos(\theta_{12}))-\sin(\theta_{12})\log(r_1r_2)}{\tilde{\Delta}_{12}}+\frac{i}{2}\log\left(\frac{1-r_1r_2 e^{i\theta_{12}}}{1-r_1r_2 e^{-i\theta_{12}}}\right)\right],\nn\\
\eea
with 
\be
\tilde{\Delta}_{12}=|1-r_1r_2 e^{i\theta_{12}}|^2=1+r^2_1r^2_2-2r_1r_2\cos(\theta_{12}).
\ee 
Similarly, the SVD entropy can be obtained by  setting $\theta_{12}=0$ and in this case reads
\be
S^{1|2}_{\mathrm{SVD}}=-\log(1-r_1r_2)-\frac{r_1r_2}{1-r_1r_2}\log(r_1r_2).
\ee
Recall that $r_i=\tanh(\rho_i/2)$ and we can see that the SVD entropy as well as von Neumann entropies (obtained by setting $r_i$ equal) are  positive but they don't have maximum as for the $\mathrm{SU}(2)$ coherent states. 

From the expressions above we evaluate the excess
\bea
\Delta S^{1|2}_{\mathrm{P}}&=&-\frac{1}{2}\log\left(\frac{\tilde{\Delta}_{12}}{(1-r^2_1)(1-r^2_2)}\right)+\frac{r^2_1\log(r_1)}{1-r^2_1}+\frac{r^2_2\log(r_2)}{1-r^2_2}\nn\\
&+&r_1r_2\frac{(r_1r_2-\cos(\theta_{12}))\log(r_1r_2)+\theta_{12}\sin(\theta_{12})}{\tilde{\Delta}_{12}},
\eea
and similarly
\be
\Delta S^{1|2}_{\mathrm{SVD}}=\frac{r^2_1\log(r_1)}{1-r^2_1}+\frac{r^2_2\log(r_2)}{1-r^2_2}-\frac{r_1r_2\log(r_1r_2)}{1-r_1r_2}-\frac{1}{2}\log\left[\frac{(1-r_1r_2)^2}{(1-r^2_1)(1-r^2_2)}\right].
\ee
Both results vanish when the two states are equal ($r_1=r_2$ and $\theta_{12}=0$). However, contrary to the $\mathrm{SU}(2)$ example, both of the excesses are always non-positive. It is natural to expect that this is due to the absence of the state with maximal  entanglement entropy. We will test this observation further for our link complement states. 
%%%%%%%%%%%%%%%%%%%%%%%%%%%
%%%%%%%%%%%%%%%%%%%%%%%%%%%
\subsection{Tripartite GHZ and W states}
%%%%%%%%%%%%%%%%%%%%%%%%%%%
%%%%%%%%%%%%%%%%%%%%%%%%%%%
As another class of examples we consider tripartite systems that consist of 3 qubits.  
Specifically, we focus on generalizations of the $3$-qubit GHZ state~\cite{mermin_1990} and the W state~\cite{D_r_2000}, which may be regarded as representatives of two classes of non-separable $3$-qubit states. These two classes differ in separability after one component is traced out: states in the GHZ class are separable, whereas those in the W class are not.
The generalized GHZ and W states that we consider are
\begin{eqnarray}
    |\mathrm{GHZ}(p) \rangle &=& \sqrt{p}|000 \rangle+\sqrt{1-p}|111 \rangle, \label{defn:GHZ} \\
    |\mathrm{W}(p_{1},p_{2}) \rangle &=& \sqrt{p_{1}}|100 \rangle+\sqrt{p_{2}}|010 \rangle+\sqrt{1-p_{1}-p_{2}}|001 \rangle, \label{defn:W} 
\end{eqnarray}
respectively, where the numbers $p,p_{1},p_{2}$, as well as $p_1+p_2$ lie in $[0,1]$.
The canonical GHZ and W states correspond to the special cases $p=\frac{1}{2}$ and $p_{1}=p_{2}=\frac{1}{3}$ respectively.

It is immediate to observe that~$\langle \mathrm{GHZ}(p) |\mathrm{W}(p_{1},p_{2}) \rangle=0$, i.e. any state in the GHZ class is orthogonal to any in the W class.
Hence an inter-class notion of pseudo-entropy cannot be realized.
However, we can consider the pseudo-entropy within each family, and evaluate the entropy excess in both cases.
The SVD entropy between a generalized GHZ and a generalized W state cannot be defined because the transition matrix, when reduced, is nilpotent, and hence its eigenvalues cannot be normalized.
%%%%%%%%%%%%%%%%%%%%%%%%%%%
\subsubsection*{GHZ class}
%%%%%%%%%%%%%%%%%%%%%%%%%%%
To evaluate the pseudo-entropy for two generalized GHZ states $|\mathrm{GHZ}(p) \rangle$ and $|\mathrm{GHZ}(q) \rangle$, the transition matrix,
\begin{multline}
    |\mathrm{GHZ}(p) \rangle \langle \mathrm{GHZ}(q)|=\sqrt{pq}|000\rangle \langle000|+\sqrt{(1-p)(1-q)}|111\rangle \langle111| \\ +\sqrt{p(1-q)}|000\rangle \langle111|+\sqrt{q(1-p)}|111\rangle \langle000| ,\label{transitionmatrix:GHZ}
\end{multline}
may be reduced over either one or two qubits. 
Without loss of generality because of the symmetry of the state~\eqref{defn:GHZ}, we may consider the respective cases of reduction in the first qubit, and the last two qubits.
Straightforward calculations show that the resultant pseudo-entropy, identical in both cases, is
\begin{equation}
S_{\mathrm{P}}^{\mathrm{GHZ}}(p,q)=-\lambda_{1} \log \lambda_{1}-\lambda_{2} \log \lambda_{2}, \label{eq:psent_GHZ}
\end{equation}
where
\begin{equation}
\lambda_{1}= \frac{\sqrt{pq}}{\sqrt{pq}+\sqrt{(1-p)(1-q)}},\qquad\lambda_{2}= \frac{\sqrt{(1-p)(1-q)}}{\sqrt{pq}+\sqrt{(1-p)(1-q)}}.
\end{equation}
It is a standard two-dimensional entropy, positive real and is symmetric in $p$ and $q$.
The condition for the inner product vanishing is $p=0$ and $q=1$, or $p=1$ and $q=0$, hence~\eqref{eq:psent_GHZ} can be defined on $[0,1]\times [0,1]\setminus \{(0,1),(1,0)\}$. 
The special case of $p=q$ corresponds to the entanglement entropy 
\begin{equation}
S_{\mathrm{E}}^{\mathrm{GHZ}}(p)=S_{\mathrm{P}}^{\mathrm{GHZ}}(p,p)=-p \log p -(1-p)\log(1-p).
\end{equation}
For the GHZ state, i.e. $p=\frac{1}{2}$, it attains the maximal value $S_{\mathrm{E}}^{\mathrm{GHZ}}\left(\frac{1}{2}\right)=\log 2$. 

\begin{figure}[h!]
\centering
    \includegraphics[width=7cm]{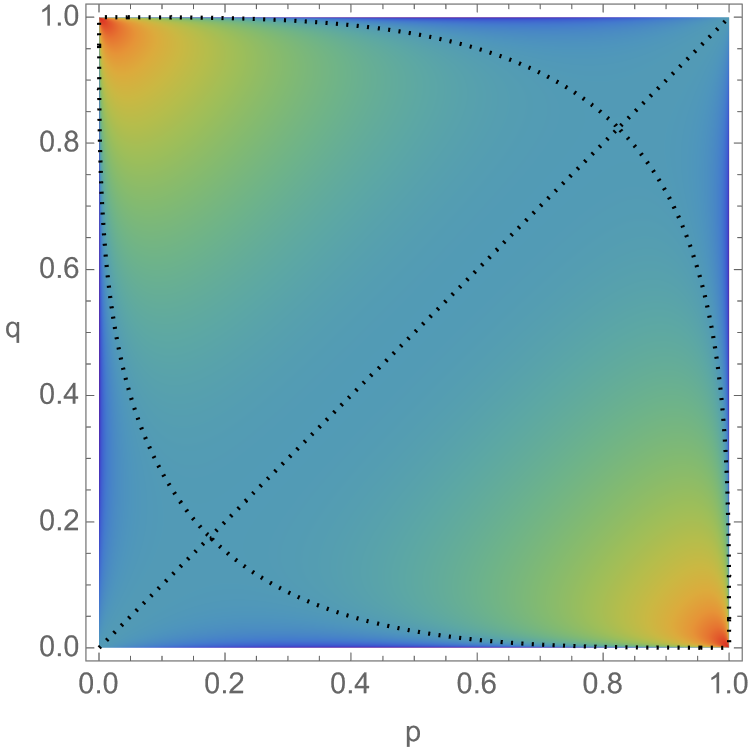}
    \hspace{0.8 cm}
    \includegraphics[width=1cm]{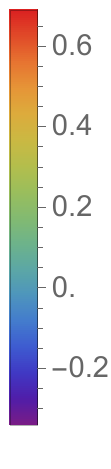}
    \caption{The entropy excess ~$\Delta S^{\mathrm{GHZ}}_{\mathrm{P}}(p,q)$ plotted on $[0,1]\times [0,1]\setminus \{(0,1),(1,0)\}$.}
    \label{fig:ent_entropy_excess_GHZ}
\end{figure}

In Fig. \ref{fig:ent_entropy_excess_GHZ} we present a heat plot of the pseudo-entropy excess $\Delta S^{\mathrm{GHZ}}_{\mathrm{P}}(p,q)$, as defined in~\eqref{defn:entropy_excess}, plotted in the region $[0,1]\times [0,1]\setminus \{(0,1),(1,0)\}$. 
The dotted black lines indicate where the excess vanishes, dividing the domain into six regions; the excess is positive in the two lobe-like regions, and negative in the four other regions. It is non-differentiable at $(1,0)$ and $(0,1)$, hence different angles of approach to these points give different limiting values. 
The highest such limiting value is $\log 2$ approaching along the line $p+q=1$. 
The lowest value of $-\frac{1}{2} \log 2$ is attained at $(\frac{1}{2},0)$ and $(0,\frac{1}{2})$.

It can be shown from~\eqref{transitionmatrix:GHZ} that the SVD entropy for these two states is identical to the pseudo-entropy in both cases, i.e.
\begin{equation}
S_{\mathrm{SVD}}^{\mathrm{GHZ}}(p,q)=S^{\mathrm{GHZ}}_{\mathrm{P}}(p,q), \label{eq:SVD_GHZ}
\end{equation}
and has the same domain of definition, $[0,1]\times [0,1] \setminus \{(0,1),(1,0) \}$.

\begin{figure}[H]
    \centering
    \includegraphics[width=7cm]{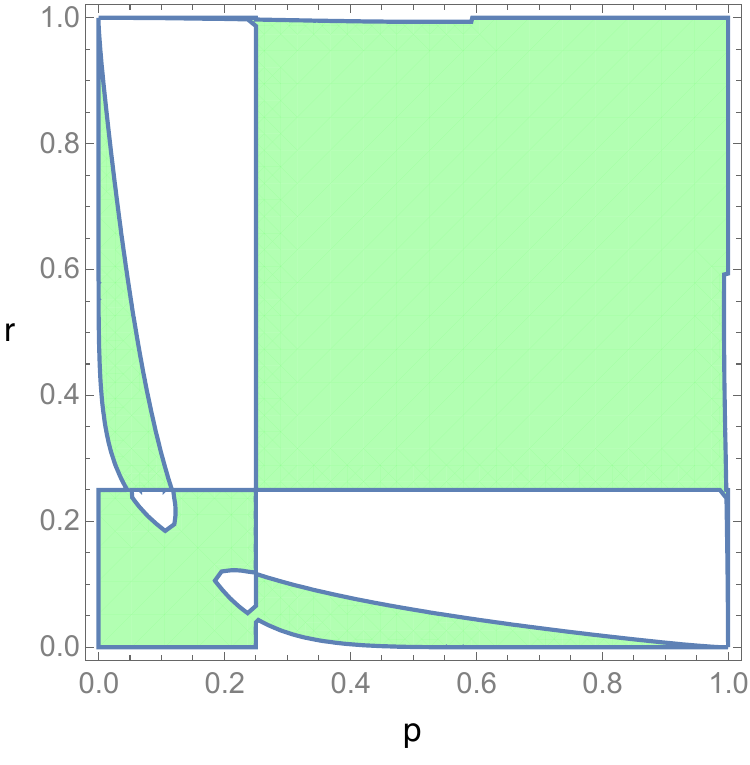}
    \caption{The range (in green) of $p,r \in (0,1)$ for which the triangle inequality~\eqref{eq:SVD_GHZ-triangle} is satisfied for generalized GHZ states, with $q=0.25$.}
    \label{fig:ghztri_main}
\end{figure}

We have also verified when the absolute value of the entropy excess $|\Delta S_{\mathrm{SVD}}^{\mathrm{GHZ}}(p,q)|$ has a metric interpretation. It is clearly non-negative and symmetric with respect to the exchange of $p$ and $q$. The triangle inequality 
\begin{equation}
|\Delta S_{\mathrm{SVD}}^{\mathrm{GHZ}}(p,q)| + |\Delta S_{\mathrm{SVD}}^{\mathrm{GHZ}}(q,r)| \geq |\Delta S_{\mathrm{SVD}}^{\mathrm{GHZ}}(p,r)| ,\label{eq:SVD_GHZ-triangle}
\end{equation}
is however satisfied only for some specific values of parameters $p,q,r\in (0,1)$ determining the three states involved, as shown in figure \ref{fig:ghztri_main} for $q=0.25$.

%%%%%%%%%%%%%%%%%%%%%%%%%%%
\subsubsection*{W class}
%%%%%%%%%%%%%%%%%%%%%%%%%%%
To evaluate the pseudo-entropy for two generalized W states $\ket{W(p_{1},p_{2})}$ and $\ket{W(q_{1},q_{2})}$, as in the previous case for the GHZ states, the transition matrix may be reduced in either one or two qubits.
Let us consider the respective cases of reduction in the first qubit, and the last two qubits.
Straightforward calculations show that the pseudo-entropy is identical in both cases and we obtain
\begin{equation}
    S^{\mathrm{W}}_{\mathrm{P}}(p_{1},p_{2};q_{1},q_{2})|_{(23|1)}=S^{\mathrm{W}}_{\mathrm{P}}(p_{1},p_{2};q_{1},q_{2})|_{(1|23)}=-\lambda_{1} \log \lambda_{1}-\lambda_{2} \log \lambda_{2}, \label{eq:psent_W}
\end{equation}
where 
\bea
\lambda_{1}&=&\frac{\sqrt{p_{1}q_{1}}}{\sqrt{p_{1}q_{1}}+\sqrt{p_{2}q_{2}}+\sqrt{(1-p_{1}-p_{2})(1-q_{1}-q_{2})}},\nonumber \\ \lambda_{2}&=&\frac{\sqrt{p_{2}q_{2}}+\sqrt{(1-p_{1}-p_{2})(1-q_{1}-q_{2})}}{\sqrt{p_{1}q_{1}}+\sqrt{p_{2}q_{2}}+\sqrt{(1-p_{1}-p_{2})(1-q_{1}-q_{2})}}, \label{eq:eigenvalues_PE_W}
\eea
and the notation $(a|b)$ is understood to mean that the qubits corresponding to the variables $b$ are first reduced.
The pseudo-entropy~\eqref{eq:psent_W} is a standard two-dimensional entropy, positve real, but unlike~\eqref{eq:psent_GHZ}, is not invariant under changes to the qubit(s) reduced first (i.e. $ab|c \leftrightarrow bc|a \leftrightarrow ca|b$ or $a|bc \leftrightarrow b|ca \leftrightarrow c|ab$).
Operationally, taking different qubits to reduce first amounts to permuting the entries $\sqrt{p_{1}q_{1}}$, $\sqrt{p_{2}q_{2}}$ and $\sqrt{(1-p_{1}-p_{2})(1-q_{1}-q_{2})}$ in~\eqref{eq:eigenvalues_PE_W}. 

The inner product vanishes when $\sqrt{p_{1}q_{1}}+\sqrt{p_{2}q_{2}}+\sqrt{(1-p_{1}-p_{2})(1-q_{1}-q_{2})}=0$, which describes a $3$-dimensional surface $S$ in $[0,1]^{4}$, hence the domain of definition is $[0,1]^{4} \setminus S$.
From the form of the generalized W state, we see that this domain is restricted to the region of $[0,1]^{4}$ delineated by $0<p_{1}+p_{2}<1$, $0<q_{1}+q_{2}<1$.

\begin{figure}[t!]
\centering
    \includegraphics[width=7cm]{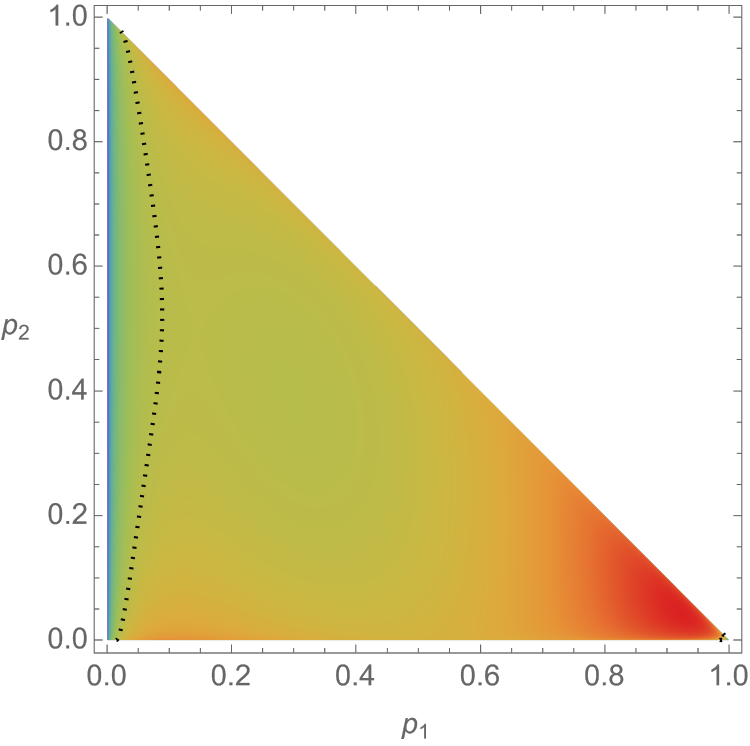}
    \hspace{0.8 cm}
    \includegraphics[width=1cm]{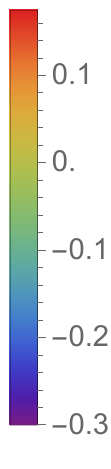}
    \caption{The entropy excess $\Delta S^{\mathrm{W}}_{\mathrm{P}}(p_{1},p_{2},q_{1},q_{2})$ at $q_{1}=0.3,q_{2}=0.4$.}
    \label{fig:ent_entropy_excess_W}
\end{figure}

As an example, Fig. \ref{fig:ent_entropy_excess_W} shows a heat plot of the entropy excess $\Delta S^{\mathrm{W}}_{\mathrm{P}}(p_{1},p_{2};q_{1},q_{2})$, as defined in~\eqref{defn:entropy_excess}, plotted for $q_{1}=0.3$ and $q_{2}=0.4$ in the region $0< p_{1}+p_{2}<1$. The dotted black lines indicate where the entropy excess vanishes.
It is straightforward to show that, when two qubits are first reduced, the SVD entropy between these two states is identical to the corresponding pseudo-entropy, for example,
\begin{equation}
S^{\mathrm{W}}_{\mathrm{SVD}}(p_{1},p_{2};q_{1},q_{2})|_{(1|23)}=S^{\mathrm{W}}_{\mathrm{P}}(p_{1},p_{2};q_{1},q_{2})|_{(1|23)},
\end{equation}
but they differ when one qubit is first reduced, for example we have
\begin{equation}
S^{\mathrm{W}}_{\mathrm{SVD}}(p_{1},p_{2};q_{1},q_{2})|_{(23|1)}= -\mu_{1} \log \mu_{1}-\mu_{2} \log \mu_{2},
\end{equation}
where 
\bea
\mu_{1}&=& \frac{\sqrt{p_{1}q_{1}}}{\sqrt{p_{1}q_{1}}+\sqrt{(1-p_{1})(1-q_{1})}},\nonumber \\ 
\mu_{2}&=& \frac{\sqrt{(1-p_{1})(1-q_{1})}}{\sqrt{p_{1}q_{1}}+\sqrt{(1-p_{1})(1-q_{1})}}. \label{eq:eigenvalues_PE_W_SVD}
\eea
Hence we observe that the SVD entropy between two W states can change when the ordering of reduction is reversed, for example
\begin{equation}
S^{\mathrm{W}}_{\mathrm{SVD}}(p_{1},p_{2};q_{1},q_{2})|_{(1|23)} \ne S^{\mathrm{W}}_{\mathrm{SVD}}(p_{1},p_{2};q_{1},q_{2})|_{(23|1)}.
\end{equation}
It is worth noting that the eigenvalues~\eqref{eq:eigenvalues_PE_W_SVD} do not contain the parameters $p_{2}$ and $q_{2}$.
This is a consequence of the qubit corresponding to the variable $1$, which attains value $1$ in the form~\eqref{defn:W} for the first ket, being traced out first.
Unlike for the pseudo-entropy, in this case taking different qubits to reduce first amounts to permuting the pairs of entries  $(\sqrt{p_{1}},\sqrt{q_{1}})$, $(\sqrt{p_{2}},\sqrt{q_{2}})$ and $(\sqrt{1-p_{1}-p_{2}},\sqrt{1-q_{1}-q_{2}})$ in the functional form~\eqref{eq:eigenvalues_PE_W_SVD}.

Let us look at the special case of $p_{1}=q_{1}$, $p_{2}=q_{2}$ in detail.
The pseudo-entropy reduces to the entanglement entropy, and from~\eqref{eq:psent_W} we see that this is
\begin{equation}
S^{\mathrm{W}}_{\mathrm{E}}(p_{1},p_{2})=S^{\mathrm{W}}_{\mathrm{P}}(p_{1},p_{2};p_{1},p_{2})=-\beta \log \beta-(1-\beta)\log(1-\beta),
\end{equation}
where $\beta=p_{1},p_{2} \textrm{ or } p_{1}+p_{2}$, depending on the order of qubits traced out.
The entanglement entropy for the W state, for which $p_{1}=p_{2}=\frac{1}{3}$, does not attain the maximal value of $\log 2$: it instead is $\log 3- \frac{2}{3} \log 2$ in all cases.
Turning to the SVD entropy, the analysis is identical to the above when two qubits are first reduced, and so it is when one qubit is first reduced, since the eigenvalues~\eqref{eq:eigenvalues_PE_W_SVD} become equal to their counterparts~\eqref{eq:eigenvalues_PE_W}.

We also analyzed a metric interpretation of the absolute value of the SVD entropy excess
for generalized W states. In both cases of one or two variables first traced out, the absolute excess is symmetric over the whole four-dimensional parameter space under interchange of points.
However, it satisfies the triangle inequality only in some specific regions of the parameter space. 

\bigskip

In conclusion, we see that for both  examples of generalized GHZ and W states with tripartite entanglement the pseudo-entropy is real. 
It is typically sub-maximal, i.e. less than or equal to $\log 2$ in both cases. 
Where these two classes differ is the behaviour under the interchange of subregion, both in the order of tracing out (i.e. $a|bc \leftrightarrow bc|a$) and in the particular subregion(s) traced out (i.e. $ab|c \leftrightarrow bc|a \leftrightarrow ca|b$ or $a|bc \leftrightarrow b|ca \leftrightarrow c|ab$).
In the GHZ class, the pseudo-entropy and SVD entropy are invariant under interchanges of both types, and are equal to each other. 
In the W class, the pseudo-entropy is invariant under interchanges of the first kind, but not of the second.
The SVD entropy is not in general invariant under both types of interchange -- only in the case of two qubits first reduced is it equal to the pseudo-entropy.  
This clearly shows that differences between these two generalized entropies are interestingly correlated with the amount of entanglement as well as entanglement patterns in the pre- and post-selected states used to define them. Finally, the absolute value of the entropy excess for GHZ and W states can be interpreted as a metric on the space of states in some specific subregions of the whole parameter space.

%%%%%%%%%%%%%%%%%%%%%%%%%%%%%%%%%%%%%
%%%%%%%%%%%%%%%%%%%%%%%%%%%%%%%%%%%%%
\section{Entropy measures for link complement states} \label{sec-links}
%%%%%%%%%%%%%%%%%%%%%%%%%%%%%%%%%%%%%
%%%%%%%%%%%%%%%%%%%%%%%%%%%%%%%%%%%%%
In this section we study pseudo-entropy, SVD entropy and their excess for link complement states (\ref{link-state}) for various families of links, and for U(1) or $\mathrm{SU}(2)$ gauge group. We also provide metric interpretation of such results. For U(1) gauge group we consider arbitrary links, while for $\mathrm{SU}(2)$ we focus on specific infinite families: composite links $\mathcal{K}\#{2^{2}_{1}}$ and certain torus links, and other examples including Borromean links. 

%%%%%%%%%%%%%%%%%%%%%%%%%%%%%%%%%%%%%%%%%%%%%%
\subsection{Two-component links in U(1) Chern--Simons theory}    \label{ssec-U1}
%%%%%%%%%%%%%%%%%%%%%%%%%%%%%%%%%%%%%%%%%%%%%%
To start with, we consider arbitrary links $\mathcal{L}$ with two components in Chern--Simons theory with U(1) gauge group and level $k$, which in this case is equal to the dimension of the Hilbert space, $d=k$. In this case the link states are determined  entirely by the linking number $l$ between the two link components that we label as A and B (the dependence on the self-linking factors out for our purposes) and the level $k$ \cite{Balasubramanian_2017}
\begin{equation}
|\mathcal{L}\rangle = \frac{1}{k} \sum_{q_A, q_B=0}^{k-1} e^{2\pi i \left( \frac{q_A q_B}{k} l  \right)} |q_A\rangle \otimes |q_B\rangle. 
\end{equation}
The entanglement entropy for such a state is given by \cite{Balasubramanian_2017}
\begin{equation}
S^{\cL}_{\mathrm{E}}(k) = \log \Big( \frac{k}{\gcd(k, l)} \Big).   \label{SE-U1}
\end{equation}
The plot of the dependence of $S_{\mathrm{E}}$ on $k$ is thus logarithmic with the number of sub-sequences given by the divisor function $\sigma_0(l)$. Since topological properties of the constituent knots are irrelevant, the entropy measures are independent of which subsystem is traced out. Also, due to the factor of $\frac{2 \pi i}{k}$, we simply take linking numbers modulo $k$. 

In what follows we analyze pseudo-entropy and SVD entropy for the pre-selected (reference) and post-selected (target) link states associated to links $\mathcal{L}_{1} (l_{1})$ and $\mathcal{L}_{2} (l_{2})$, with linking numbers $l_1$ and $l_2$ respectively. In this case 
\begin{align}
\begin{split}
    \langle \mathcal{L}_1 (l_{1})| \mathcal{L}_2 (l_{2}) \rangle  & = \frac{1}{k^{2}}\sum_{q_A,q_B}e^{\frac{2 
    \pi i}{k}q_Aq_B(l_1-l_2)} , \\
|\mathcal{L}_1(l_{1})\rangle\langle\mathcal{L}_2(l_{2})| & = \frac{1}{k^2}\sum_{q_A, q_B, q'_A, q'_B} e^{\frac{2 \pi i}{k} (l_{1}q_Aq_B - l_{2}q'_Aq'_B)}|q_A q_B\rangle\langle q'_A q'_B|.
\end{split}
\end{align} 
We first consider SVD entropy and its excess for arbitrary $k$. Our main statement is that in this case the SVD entropy excess provides a pseudo-metric on the space of two-component link states. A pseudo-metric means that while other axioms of the metric are satisfied, a distance between two different links may be zero even if they are not identical; this is indeed so, as the entropy measures in U(1) case depend only on the linking numbers and not on other topological details. Nonetheless, as we argue below, in this case other axioms of the metric are satisfied, in particular the triangle inequality. To see that, we find numerically that the SVD entropy for two (pre-selected and post-selected) two-component link states takes the form
\begin{equation}
 S_{\mathrm{SVD}} = \log \Big( \frac{k}{\gcd(k, l_1 l_2)} \Big),    \label{SSVD-U1}
\end{equation}
whenever $\gcd(k,l_1l_2)\neq np^2$ for $n,p \in \mathbb{N}$ (thus this holds for all $k \neq np^2$ irrespective of linking numbers). Let us discuss basic features of this expression. First, it follows from (\ref{SSVD-U1}) that $S_{\textrm{SVD}}$ and its derivatives do not depend on ordering of states in U(1) case, which is  not true in general. Second, note that when $l_1=l_2\equiv l$ and $\gcd(k, l^2)\neq np^2$, then $\gcd(k,l^2)=\gcd(k,l)$, see (\ref{gcd-k-l}), so that (\ref{SSVD-U1}) reduces then to (\ref{SE-U1}) as expected. If $\gcd(k,l_1l_2) = np^2$ for some $n$ and $p$ then the expression for SVD entropy is more involved and we cannot write its analytic form, however we verified various statements that follow for a large range of parameters $k,l_1,l_2$. Furthermore, it follows from (\ref{SSVD-U1}) that
\begin{equation} 
\Delta S_{\mathrm{SVD}} = \frac{1}{2}\log \left( \frac{{\gcd(k, l_1) \cdot \gcd(k, l_2)}}{(\gcd(k, l_1 l_2))^2} \right). \label{excess_u1}
\end{equation}
Since $\gcd(k,l_{1}),\gcd(k,l_{2}) \le \gcd(k,l_{1}l_{2})$, it follows from~\eqref{excess_u1} that whenever $\gcd(k,l_1l_2) \ne np^2$,  $\Delta S_{\mathrm{SVD}} \le 0$, and
\begin{equation}
  | \Delta S_{\mathrm{SVD}} | = \frac{1}{2} \log \left( \frac{(\gcd(k, l_1 l_2))^2}{\gcd(k, l_1) \cdot \gcd(k, l_2)} \right). % \quad  \textrm{for} \ \gcd(k,l_1l_2)\neq np^2  . 
\label{SSVD-U1-abs}
\end{equation} 
As mentioned above, this expression provides a pseudo-metric on the space of two-component links. Indeed, it is obviously non-negative and symmetric under the exchange of $l_1$ and $l_2$, and as we show in~\cref{sec-app-U1}, $\Delta S_{\mathrm{SVD}}=0$ for $l_1=l_2$ (which is consistent with general properties of the SVD excess) and it satisfies the triangle inequality. We verified for a broad range of parameters $k,l_1,l_2$ that these properties also hold when $\gcd(k,l_1l_2) = np^2$. We thus claim that \eqref{SSVD-U1-abs} indeed provides a pseudo-metric. This is a prototype example that motivates us to analyze metric axioms also for other classes of links and other Chern--Simons gauge groups.

\bigskip

Let us now consider SVD entropy and also pseudo-entropy for more specific examples. First, consider the level $k=2$, for which we get 2-qubit systems. A link state (with linking number $l$) in this case takes the form
\begin{equation}
|\mathcal{L}\rangle = \begin{cases}
|00\rangle+|01\rangle+|10\rangle+ |11\rangle & \ \mathrm{for}\;  l \; \mathrm{even}\,, \\
|00\rangle+|01\rangle+|10\rangle - |11\rangle &\ \mathrm{for}\;  l \; \mathrm{odd}\,.
\end{cases}  
\end{equation}
The states arising from even linking numbers $l$ are unentangled ($S_\mathrm{E}=0$), and for odd $l$ they are maximally entangled ($S_\mathrm{E}=\log2$). For the reference and target states within the same class $S_{\mathrm{P}}=S_{\mathrm{SVD}}=S_{\mathrm{E}}$, i.e. 0 and $\log 2$ for the first and second classes respectively,  and otherwise $S_{\mathrm{P}}=S_{\mathrm{SVD}}=0$. The entropy excesses $\Delta S_\mathrm{P} = \Delta S_\mathrm{SVD} = 0$  when both states belong to the same class, while $\Delta S_\mathrm{P} = \Delta S_\mathrm{SVD} = -\frac{\log 2}{2}$ when they belong to different classes. In this case both $|\Delta S_\mathrm{P}|$ and $|\Delta S_\mathrm{SVD}|$ provide a  pseudo-metric for this class of states. 

Furthermore, upon fixing the level $k=3$, we get unnormalised 2-qutrit states \begin{align}
\begin{split}
& \ket{\mathcal{L}} = \\
& =\begin{cases}
|00\rangle + |01\rangle + |10\rangle + |11\rangle + |02\rangle + |20\rangle + |12\rangle + |21\rangle + |22\rangle, & l \equiv 0 \text{ mod } 3, \\
|00\rangle + |01\rangle + |10\rangle + \omega|11\rangle + |02\rangle + |20\rangle + \overline{\omega}|12\rangle + \overline{\omega}|21\rangle + \omega|22\rangle, & l \equiv 1 \text{ mod } 3, \\
|00\rangle + |01\rangle + |10\rangle + \omega^2|11\rangle + |02\rangle + |20\rangle + \overline{\omega}^2|12\rangle + \overline{\omega}^2|21\rangle + \omega^2|22\rangle, & l \equiv 2 \text{ mod } 3,
\end{cases}
\end{split}
\end{align}
where $\omega=\exp \left(\frac{2\pi i}{3}\right)$.
The  states from the first class are unentangled, and those from the second and third classes are maximally entangled ($S_\mathrm{E} = \log 3$). 
As previously, for the choice of the reference and target states within the same class $S_{\mathrm{P}}=S_{\mathrm{SVD}}=S_{\mathrm{E}}$, i.e. 0, $\log 3$ and $\log 3$ respectively. For the choice of states from two different classes, if the first class is involved then $S_\mathrm{P} =  S_\mathrm{SVD} = 0$, and otherwise $\text{Re}(S_{\mathrm{P}})=0$ and $ S_\mathrm{SVD} = \log 3$. Therefore once again the entropy excesses $\Delta S_\mathrm{P} = \Delta S_\mathrm{SVD} = 0$ when both the states belong to the same class. For two states from two different classes $\Delta S_\mathrm{P} = \Delta S_\mathrm{SVD} = -\frac{\log 3}{2}$ when one of the states is from the first class, and otherwise $\Delta S_\mathrm{P} = -\log 3$ and $\Delta S_\mathrm{SVD} = 0$. In this case we checked (also numerically up to $k=20$) that  $|\Delta S_\mathrm{SVD}|$ satisfies axioms of the pseudo-metric. On the other hand, $|\Delta S_\mathrm{P}|$ can be interpreted as a pseudo-metric not for all $k$, as for higher values of $k$ some violations of the triangle inequality appear.

\begin{figure}[H]
\centering
\begin{subfigure}[b]{0.5\textwidth}
\includegraphics[width=7 cm]{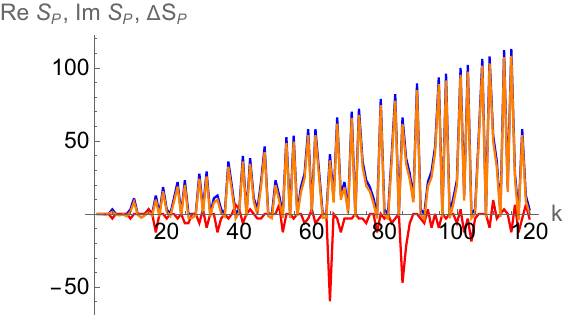}   
\caption{Pseudo-entropy}
\label{fig:U1pe24_28}
\end{subfigure}%
\begin{subfigure}[b]{0.5\textwidth}
\includegraphics[width=7 cm]{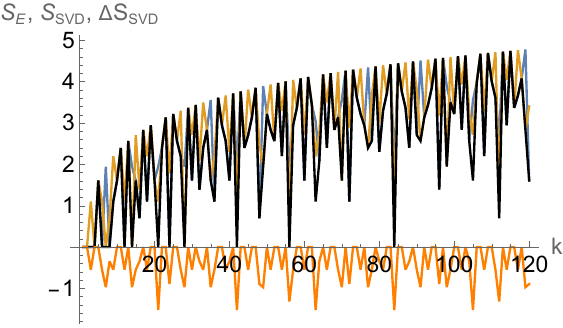}    
\caption{SVD entropy}
\label{fig:U1svd24_28}
\end{subfigure}
\caption{(a) The real (blue) and imaginary (red) parts of the pseudo-entropy (b) SVD entropy (black) between torus links $\mathrm{T}(6,16)$  ($S_\mathrm{E}$ in light blue) and $\mathrm{T}(8,14)$ ($S_\mathrm{E}$ in yellow). The respective entropy excesses are given in orange.}
\label{u1_example}
\end{figure}

Finally, let us illustrate the dependence of pseudo-entropy, SVD entropy and their excess on the level $k$, for specific examples of two-component links with higher linking numbers. An interesting source of such examples are torus links, whose link states in U(1) Chern--Simons theory were analyzed also in \cite{Balasubramanian:2018por}. Recall that $\mathrm{T}(p,q)$ torus link has gcd$(p,q)$ components and the linking number between each two of them is $\frac{pq}{\textrm{gcd}(p,q)^2}$. The link states depend only on this linking number (and no other topological features of links under consideration) and can be written as
\begin{equation}
\left|\mathrm{T}(p,q)\right\rangle = \frac{1}{k} \sum_{n=0}^{k-1} 
c_n
|n, n, \ldots, n\rangle,\qquad
c_n = \sum_{j=0}^{k-1} \exp \Big(\frac{\pi i (j+1)}{k}\Big(2 (n+1) + \frac{ pq (j+1)}{  \gcd(p,q)^2}\Big)\Big).
\end{equation}
In particular, for two-component links this expression is of the form (\ref{psi-2-component}), so to find pseudo-entropy and SVD entropy the formulae from section \ref{sec:general 2 component} can be immediately used. For definiteness, we present on \cref{u1_example} an example with two-component $(6,16)$ and $(8,14)$ torus links, with linking numbers $24$ and $28$ respectively. 

%%%%%%%%%%%%%%%%%%%%%%%%%%%%%%%%%%%%%%%%%%%%%%
\subsection{$\mathcal{K}\#2^2_1$ links in non-abelian Chern--Simons theory}
%%%%%%%%%%%%%%%%%%%%%%%%%%%%%%%%%%%%%%%%%%%%%%
In turn, we consider links of the form $\mathcal{K}\#2^2_1$, i.e. connected sums of a knot $\mathcal{K}$ and the Hopf-link $2^2_1$ (so effectively one unknot component of the Hopf-link is replaced by $\mathcal{K}$). By adjusting the value of the level $k$ we consider qubit, qutrit, and more involved states. 
Apart from providing some general statements we focus on connected sums $\mathcal{K}_p\#2^2_1$
involving twist knots $\mathcal{K}_p$, compute various entropy measures, and in particular show that at low levels $k$ the excess measures satisfy the metric axioms for this class of links. 

To start with, we consider $\mathrm{SU}(2)$ Chern--Simons theory at level $k=1$, so that the dimension of the one-component Hilbert space is $d=k+1=2$ and the resulting link states are (tensor products of) qubits. In this case the parameter \eqref{def:q} takes value $q=\exp\left(\frac{2\pi i}{3}\right)\equiv \omega$. The Jones polynomial in trivial representation is normalized to 1, $\tilde{C}^\mathcal{K}_0 = 1$. Interestingly, for any knot, for the fundamental representation and the argument $q=\omega$ it also evaluates to 1, $\tilde{C}^\mathcal{K}_1|_{q=\omega}= V^\mathcal{K}(\omega)=1$ \cite{murakami2016periodicity}. It follows that for any link of the form $\mathcal{K}\#{2^{2}_{1}}$, its 2-qubit quantum state 
\begin{equation}
    | \mathcal{K}\#{2^{2}_{1}} \rangle = |0,0\rangle + |1,1\rangle,
\end{equation}
is maximally entangled. In the above equation we made the transformation (\ref{eqn:connected_sum_firstinstance}).

Further, for $k=2$ the parameter \eqref{def:q} is $q=\exp\left(\frac{2\pi i}{4}\right) = i$ and the one-component Hilbert space is of dimension $d=3$, so that the link states are (tensor products of) qutrits. With the Jones polynomial normalized to 1 for the trivial representation, its value for the fundamental representation and $q=i$ turns out to depend only on the so-called Arf invariant of a knot $\mathcal{K}$ in question   \cite{Jones:1985dw}
\begin{equation}
\tilde{C}^{\mathcal{K}}_1|_{q=i} = V^\mathcal{K}(i)= \begin{cases}
1, & \text { if } \operatorname{Arf}(\mathcal{K})=0, \\
-1, & \text { if } \operatorname{Arf}(\mathcal{K})=1.
\end{cases}  
\end{equation} 

It follows that link states in this qutrit case, after the diagonalization transformation  \eqref{eqn:connected_sum_firstinstance}, take the form
\begin{equation}
|\mathcal{K} \#{2^{2}_{1}}\rangle = |0,0\rangle \pm |1,1\rangle + l |2,2\rangle, \label{K_hash_2_state}
\end{equation}
where $\pm$ is chosen for $\mathrm{Arf}(\mathcal{K}) = 0,1$  respectively, and $l = \tilde{C}^{\mathcal{K}}_2(i)$. Recall that the 3-dimensional representation of $\mathrm{SU}(2)$ is the same as the fundamental representation of $\mathrm{SO}(3)$, and $\mathrm{SO}(N)$ link invariants are given by the Kauffman polynomial $F^\mathcal{K}(a,z)$ \cite{kauffman1990invariant}, so that a change of variable \cite{ramadevi1993chirality} $a=iq^2$ and $z=-i(q-q^{-1})$ yields $l \equiv \tilde{C}^{\mathcal{K}}_2(i) = F^{\mathcal{K}}\left(-i, 2\right)$.

For the state~\eqref{K_hash_2_state} we find the entanglement entropy
\begin{equation}
S_{\mathrm{E}}= \log(|l|^2 + 2) - {|l|^2 \log(|l|^2)}{\left(|l|^2 + 2\right)^{-1}}, 
\end{equation}
and for two such states $\ket{\mathcal{K}^{1}\#2^{2}_{1}}$ and $\ket{\mathcal{K}^{2}\#2^{2}_{1}}$ the SVD entropy reads
\begin{equation}
S_{\mathrm{SVD}}^{1|2}= \log(|l_1 \bar{l}_2| + 2) - {|l_1 \bar{l}_2| \log(|l_1 \bar{l}_2|)}{\left(|l_1 \bar{l}_2| + 2\right)^{-1}},
\end{equation}
where $l_{j}=\tilde{C}^{\mathcal{K}^{j}}_{2}(i)$, $j=1,2$.
Note that both the entanglement and SVD entropies are independent of the $\mathrm{Arf}$ invariants of the knots under consideration. 
This is not the case for the pseudo-entropy $S_{\mathrm{P}}$ between $\ket{\mathcal{K}^{1}\#2^{2}_{1}}$ and $\ket{\mathcal{K}^{2}\#2^{2}_{1}}$ in general, since for $\mathrm{Arf}(\mathcal{K}^1) = \mathrm{Arf}(\mathcal{K}^2)$ we get
\begin{equation}
    S_{\mathrm{P}}^{1|2} = \log(l_1 \bar{l}_2 + 2) - {l_1 \bar{l}_2 \log(l_1 \bar{l}_2)}{\left(l_1 \bar{l}_2 + 2\right)^{-1}},
\end{equation}
while for $\mathrm{Arf(\mathcal{K}^1)} \neq \mathrm{Arf}(\mathcal{K}^2)$ we get
\begin{equation}
    S_{\mathrm{P}}^{1|2} = {i\pi}{\left(l_1 \bar{l}_2\right)^{-1}}.
\end{equation}

We examined the above result for the case of connected sums involving twist knots $\mathcal{K}=\mathcal{K}_p$. For all hyperbolic twist knots upto 10 crossings (see Table 11 in \cite{nawata2012super}) we find that in \eqref{K_hash_2_state} the coefficient $l=1$  for all $p=-1,2,-2,3,-3,4,-4$, and we expect it holds in general. Therefore the states (\ref{K_hash_2_state}) and their entropy properties depend only on the Arf invariant. In tables \ref{tab:first_table} and \ref{tab:second_table} we present the excess pseudo-entropy \eqref{defn:entropy_excess} and excess SVD entropy  \eqref{defn:SVDentropy_excess} for the pairs of (reference and target) links $\mathcal{K}^1_p\#2^2_1$ and $\mathcal{K}^2_p\#2^2_1$, with $p=-1,2,-2,3$ (so that $\mathcal{K}^1_p$ and $\mathcal{K}^2_p$ are $4_1,5_2,6_1,7_2$ knots, for which Arf-invariants are equal to 1,0,0,1 respectively). The states are maximally entangled.

\begin{table}[t!]
    \centering
    \begin{minipage}{0.45\textwidth}
        \centering
        \begin{tabular}{|c|c|c|c|c|}
            \hline
            $S_\mathrm{P}$ & $4_1$ & $5_2$ & $6_1$ & $7_2$ \\ \hline
            $4_1$ & \textcolor{blue}{$\log 3$} & $\pi i$ & $\pi i$ & {$\log 3$} \\ \hline
            $5_2$ & $\pi i$ & \textcolor{blue}{$\log 3$} & {$\log 3$} & $\pi i$ \\ \hline
            $6_1$ & $\pi i$ & {$\log 3$} & \textcolor{blue}{$\log 3$} & $\pi i$ \\ \hline
            $7_2$ & {$\log 3$} & $\pi i$ & $\pi i$ & \textcolor{blue}{$\log 3$} \\ \hline
        \end{tabular}
        \caption{Values of $S_\mathrm{P}$. $S_\mathrm{SVD} = \log 3$ for all cases. $S_\mathrm{E}$ given by blue diagonal.} \label{tab:first_table}
    \end{minipage}
    \hfill
    \begin{minipage}{0.45\textwidth}
        \centering
        \begin{tabular}{|c|c|c|c|c|}
            \hline
            $\Delta S_\mathrm{P}$ & $4_1$ & $5_2$ & $6_1$ & $7_2$ \\ \hline
            $4_1$ & 0 & -$\log 3$ & -$\log 3$ & 0 \\ \hline
            $5_2$ & -$\log 3$ & 0 & 0 & -$\log 3$ \\ \hline
            $6_1$ & -$\log 3$ & 0 & 0 & -$\log 3$ \\ \hline
            $7_2$ & 0 & -$\log 3$ & -$\log 3$ & 0 \\ \hline
        \end{tabular}
        \caption{Analogous values of $\Delta S_\mathrm{P}$.  $\Delta S_\mathrm{SVD} = 0$ for all cases.}  \label{tab:second_table}
    \end{minipage}
\end{table}

We find that the pseudo-entropy is purely imaginary $\pi i$ when the reference state and the target state have different Arf invariants, and purely real $\log 3$ when they are same. These tables extend in the same way up to $p=-3,4,-4$, i.e., the $8_1,9_2,10_1$ knots. To conclude, we note that the pseudo-entropy excess takes only two values that satisfy the triangle inequality, so that this excess can be reinterpreted as providing the discrete pseudo-metric on the space of $\mathcal{K}_p\#2^2_1$ links (this is a pseudo-metric, as e.g. the distance between links $4_1\#2^2_1$ and $7_2\#2^2_1$ vanishes). On the other hand, the SVD excess vanishes for all pairs of links from the class under consideration, so its metric interpretation is trivial.

We present below an example of entropy measures using the (torus) $3_{1}$ and (hyperbolic) $4_{1}$ knots (i.e. $\mathcal{K}_p$ twist knots with $p=1,-1$ respectively), at arbitrary levels $k$. 

\begin{figure}[H]
\centering
\begin{subfigure}[b]{.5\textwidth}
\centering
\includegraphics[width=7cm]{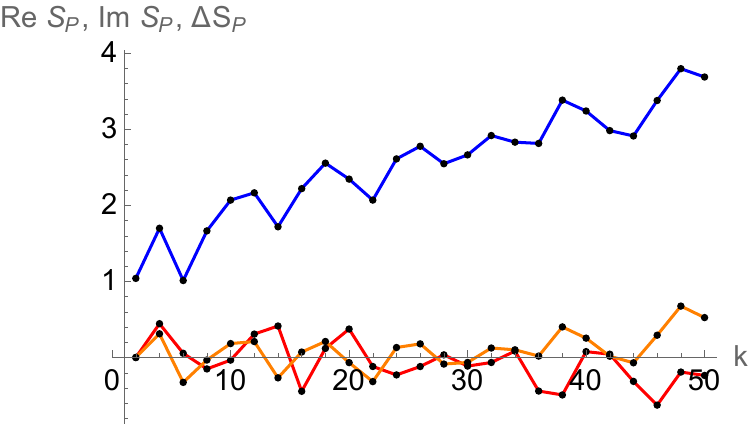}
\label{fig_connsum_pe}
\caption{Pseudo-entropy}
\end{subfigure}%
\begin{subfigure}[b]{.5\textwidth}
\centering
\includegraphics[width=7cm]{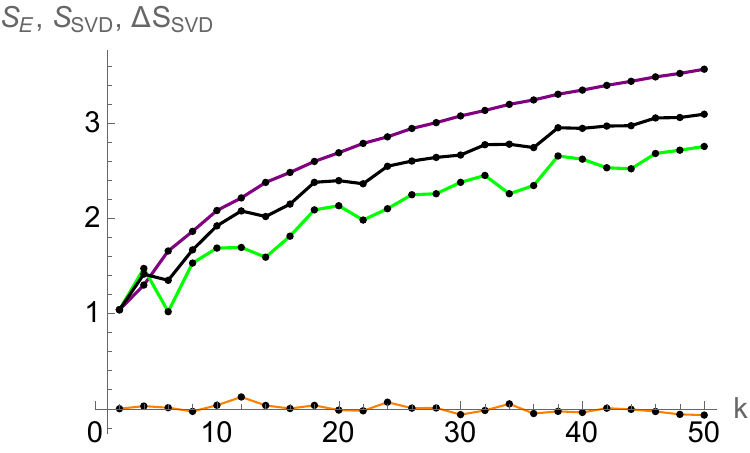}
\caption{Entanglement and SVD entropies}
\end{subfigure}
\label{fig_PEtrefoilfig8_SU2}
\caption{(a) The real (blue) and imaginary (red) parts of the pseudo-entropy (b) The SVD entropy (black) between $3_{1}\#2^2_1$ ($S_\mathrm{E}$ in purple) and $4_{1}\#2^2_1$ ($S_\mathrm{E}$ in green). The respective entropy excesses are presented in orange.}
\end{figure}
We can generalize the above results to other gauge groups. After the diagonalization the quantum state associated to a connected sum $\mathcal{K}\# 2^2_1$ takes the form 
(\ref{eqn:connected_sum_firstinstance}), which is a special case of (\ref{psi-2-component}) with $c^{(i)}_m$ identified with reduced polynomials $\tilde{C}^{\mathcal{K}}_m$ of a knot $\mathcal{K}$. In particular, for $\mathrm{SU}(N)$ the reduced knot polynomials are given by substituting $a=q^N$ in \cref{eqn: twist knots master formula}. It follows that pseudo-entropy and SVD entropy respectively take the form (\ref{PEGen2Q}) and (\ref{SSVD-2-component}), with the same identification of $c^{(i)}_m$.
However, 2-qubit/2-qutrit systems are not possible for $N>3$ due to the dimension of the Hilbert space. On the other hand, when $N=2$ (as done in this section), we can also obtain the corresponding link states and entropy measures for SO(3) Chern--Simons theory for $k \in 2\mathbb{N}$, which take the form \cite{SO3}
\begin{equation}
    \left|\mathcal{K} \#{2^{2}_{1}}\right\rangle_{\textrm{SO}(3)}=\sum_{m=0}^{k} V^\mathcal{K}_{2m}\big(q=e^{\frac{4 \pi i}{k+1}}\big)\left|m, m\right\rangle,
\end{equation}
where $V^\mathcal{K}_{2m}(q)$ is an ordinary colored Jones polynomial of $\mathcal{K}$. In particular the 2-qutrit examples in this section can be analogously worked out for the SO(3) case. 

%%%%%%%%%%%%%%%%%%%
%%%%%%%%%%%%%%%%%%%
\subsection{$(p,pn)$ torus links in non-abelian Chern--Simons theory}
\label{subsec:toruslinks}
%%%%%%%%%%%%%%%%%%%
%%%%%%%%%%%%%%%%%%%

In this section we focus on a specific class of torus links in non-abelian Chern--Simons theory (as we discussed in section \ref{ssec-U1}, entropy measures in U(1) theory depend only on the linking number between various link components and not on other topological details of links under consideration). Consider pre-selected and post-selected link states corresponding respectively to $(p_1,q_1)$ and $(p_2,q_2)$ torus links with the same number of components $d=\gcd(p_{1},q_{1})=\gcd(p_{2},q_{2})$. Recalling that link states for $(p,q)$ torus links can be written in the form (\ref{eq:general_torus_link_component_state_firstinstance}), it follows that pseudo-entropy and SVD entropy of our interest can be written respectively in the form (\ref{PEGen2Q}) and (\ref{SSVD-2-component}) with the summand involving 
\begin{equation}
c^{(1)}_m \bar{c}^{(2)}_m 
=\left(\cS X\left(\frac{p_{1}}{d}\right) \cT^{\frac{q_{1}}{p_{1}}} \cS \right)_{m0} \left(\cS X\left(\frac{p_{2}}{d}\right) \cT^{-\frac{q_{2}}{p_{2}}} \cS \right)_{m0} \frac{1}{\cS_{0m}^{2d-2}},\qquad 0\le m \le k.
\label{def:theta_eigenvalue}
\end{equation} 

\begin{figure}[b!]
    \centering
    \begin{subfigure}[b]{0.48\textwidth}
        \centering
        \includegraphics[width=\linewidth, height=4cm]{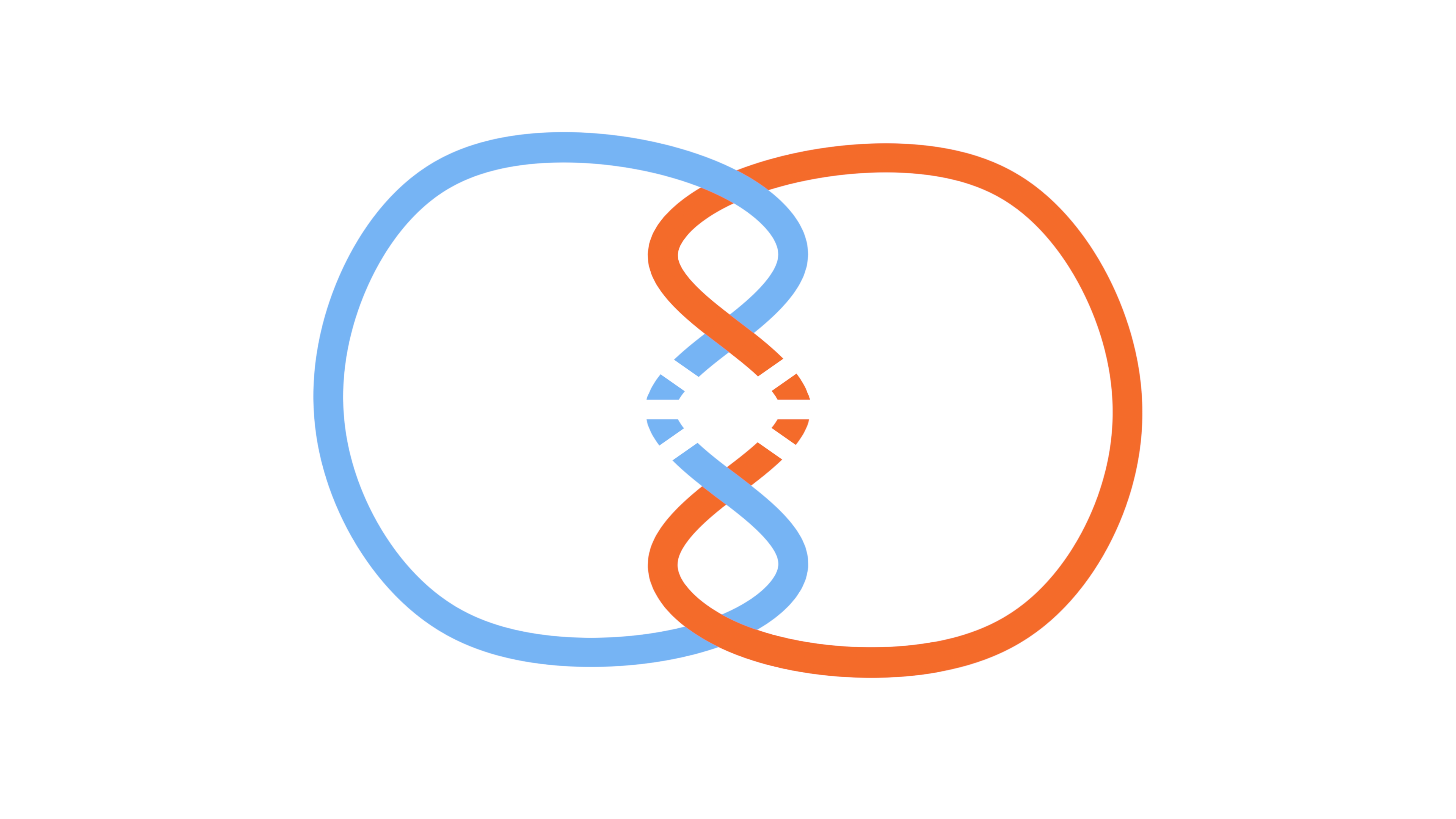}
    \end{subfigure}
    \caption{ $\mathrm{T}(2,2n)$ link is a generalization of a Hopf-link to $2n$ twists.}   \label{T22n}
\end{figure}

In what follows we focus on $(p,pn)$ torus links with $p\ge 2$ components, which were also studied in \cite{Dwivedi2018,Dwivedi_2020}. The linking number between any two components of such links is $n$; in particular, the $\mathrm{T}(2,2n)$ links can be thought of as generalizing the two-component Hopf link $\mathrm{T}(2,2)$ to $2n$ twists, see~\cref{T22n}. In fact, $n$ may be generalized to negative integer values to encode information about the orientation of the link; mirroring the link is equivalent to the transformation $n \rightarrow -n$. For $(p_1,q_1)=(p,pn_1)$ and $(p_2,q_2)=(p,pn_2)$ torus links, the factor (\ref{def:theta_eigenvalue}) reduces to 
\begin{equation}
\Gamma^{n_{1}|n_{2}}_{m}(p;k) \equiv  %\Theta_{m}^{(p,n_{1}p)|(p,n_{2}p)}(k)
c^{(1)}_m \bar{c}^{(2)}_m =
\left(\cS \cT^{n_{1}} \cS \right)_{0m}\left(\cS \cT^{-n_{2}} \cS \right)_{0m} \frac{1}{\cS_{0m}^{2p-2}}=\alpha_{m}^{n_{1}}(k)\alpha_{m}^{-n_{2}}(k), \label{eq:2N|2M_gamma}
\end{equation}
where we define
\begin{equation}
    \alpha_{m}^{n}(p;k)=\frac{\left(\cS \cT^{n}\cS\right)_{0m}}{\cS_{0m}^{p-1}}. \label{def:2N_alpha}
\end{equation}
We denote pseudo-entropy and SVD entropy for this class of links by $S_{\mathrm{P}}^{n_{1}|n_{2}}(p;k)$ and $S_{\mathrm{SVD}}^{n_{1}|n_{2}}(p;k)$. Again, they are given by formulae (\ref{PEGen2Q}) and (\ref{SSVD-2-component}) with $\Gamma^{n_{1}|n_{2}}_{m}(p;k)\equiv c^{(1)}_m \bar{c}^{(2)}_m$ given by (\ref{eq:2N|2M_gamma}), and the factors (\ref{f-overlap}) and (\ref{f-tilde-overlap}) respectively taking form
\begin{equation}
f^{n_{1}|n_{2}}(p;k)= \sum_{m=0}^{k} \Gamma^{n_{1}|n_{2}}_{m}(p;k),\qquad \tilde{f}^{n_{1}|n_{2}}(p;k)= \sum_{m=0}^{k} |\Gamma^{n_{1}|n_{2}}_{m}(p;k)|.
 \label{eq:2N|2M_innerproduct} 
\end{equation}
Note that the SVD entropy $S_{\mathrm{SVD}}^{n_{1}|n_{2}}(p;k)$ depends only on $|n_{1}|$ and $|n_{2}|$, so it is insensitive to mirroring. However, the pseudo-entropy $S_{\mathrm{P}}^{n_{1}|n_{2}}(p;k)$ is sensitive to mirroring and in general $S_{\mathrm{P}}^{n_{1}|n_{2}}(p,k) \ne S_{\mathrm{P}}^{n_{1}|-n_{2}}(p;k)$. 

As usual, for $n_{1}=n_{2}=n$ we get the entanglement entropy that in the current case we denote by $S_{\mathrm{E}}^{n}(p;k)\equiv S_{\mathrm{P}}^{n|n}(p;k)$. For the Hopf-link $\mathrm{T}(2,2)$ it is maximally entangled (and equal to the logarithm of the dimension of the Hilbert space): $S_{\mathrm{E}}^{1}(2;k)=\log(k+1)$, 
as follows immediately from \eqref{eq:SandTidentities} in\eqref{eq:2N|2M_gamma} (see also 
\cite{Balasubramanian_2017}). The entanglement entropy of $\mathrm{T}(p,p)$ links for $p \ge 3$ has been conjectured in \cite{Dwivedi_2020} to take the form
\begin{equation}
S_{\mathrm{E}}^{n}(p;k)=\log(k+1)+\log(k+3)+\log P_{p-2}(k^{2}+4k)-(p-2) \frac{P'_{p-2}(k^{2}+4k)}{P_{p-2}(k^{2}+4k)}, \label{eq:ent_entropy_T(p,p)}
\end{equation}
where $P_{m}(x)$ are certain polynomials. In general, both the pseudo-entropy and SVD entropy vanish if one of the links is the unlink, i.e. $S^{0|n}_{\mathrm{P}}(p;k)=S^{0|n}_{\mathrm{SVD}}(p;k)=0$ for all $p,k$, and for all $n \in \mathbb{Z}$. 

\bigskip

Let us analyze the growth of the entanglement and SVD entropy for various positive values of $n_{1}$ and $n_{2}$ for $p \ge 2$, and for $k$ up to $120$. First, we consider how the SVD entropy $S_{\mathrm{SVD}}^{n_{1}|n_{2}}(p;k)$ mixes information between two states by comparing it to individual entanglement entropies $S_{\mathrm{E}}^{n_{1}}(p;k)$ and $S_{\mathrm{E}}^{n_{2}}(p;k)$.
Fixing one of the links to be of the form $\mathrm{T}(p,p)$, for $p=2$ and for some values $p\ge 3$, for $n>1$ and up to $n=13$, and for sufficiently large $k$, we observe the pattern
\be
S_{\mathrm{E}}^{n}(p;k) < S_{\mathrm{SVD}}^{1|n}(p;k) < S_{\mathrm{E}}^{1}(p;k),
\ee
i.e. the SVD entropy between two links appears to interpolate the individual entanglement entropies. 
In section \ref{subsec:largek} we also discuss limiting values  $\lim_{k\to\infty} S_{\mathrm{E}}^{n}(p;k)$ generalizing the discussion in \cite{Dwivedi_2020}.

\begin{figure}[H]
\centering
\begin{subfigure}[b]{.5\textwidth}
\centering
\includegraphics[width=7cm]{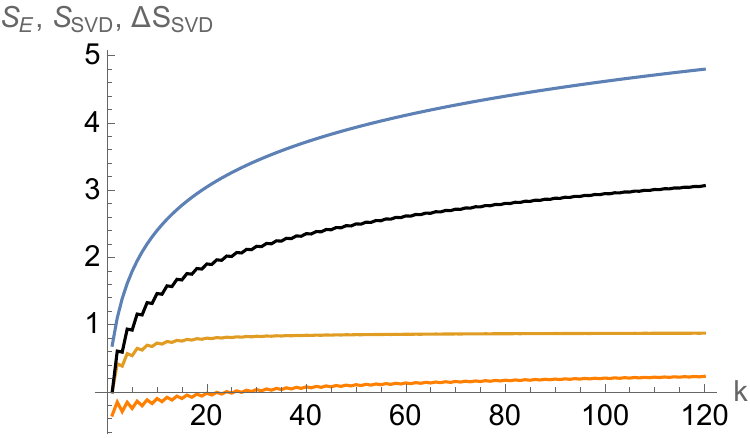}
\caption{$p=2$: $n_{1}=1$ vs $n_{2}=2$}
\label{fig_N=1_N=2}
\end{subfigure}%
\begin{subfigure}[b]{.5\textwidth}
\centering
\includegraphics[width=7cm]{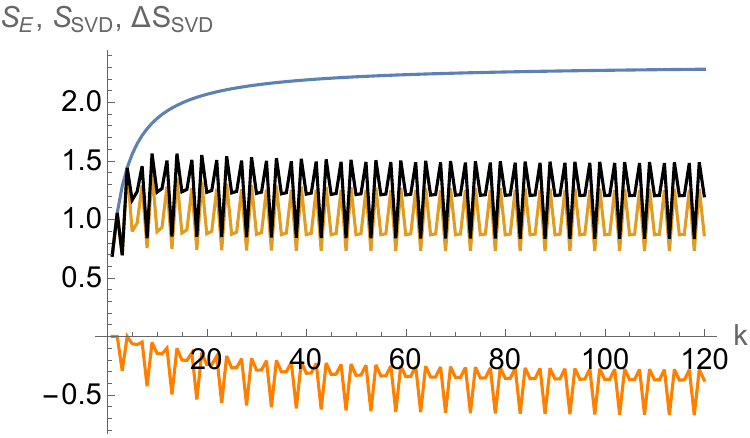}
\caption{$p=3$: $n_{1}=1$ vs $n_{2}=5$}
\label{fig_N=1_N=5}
\end{subfigure}%
\caption{The SVD entropy of (a) $\mathrm{T}(2,2)$ vs $\mathrm{T}(2,4)$, and (b) $\mathrm{T}(3,3)$ vs $\mathrm{T}(3,15)$, interpolates between the respective entanglement entropies. Also plotted in each case is the entropy excess. For simplicity just the interpolating curves are shown.}
\label{fig_ent_svd_1}
\end{figure}

As an example, in \cref{fig_N=1_N=2} the SVD entropy (black curve) of the $\mathrm{T}(2,2)$ link with respect to the $\mathrm{T}(2,4)$ link is plotted against the respective (blue and olive curves) entanglement entropies. Analogous results are shown in \cref{fig_N=1_N=5} for
 $\mathrm{T}(3,3)$ (blue curve) and the $\mathrm{T}(3,15)$ (olive curve) entanglement entropies.
An interesting difference in the two cases is the entropy excess (orange curves). In~\cref{fig_N=1_N=2} the excess is initially negative and eventually appears to stay positive as $k$ increases. In \cref{fig_N=1_N=5} the excess is briefly positive or close to zero at low $k$, and thereafter appears to stay negative (from large $k$ arguments we discuss in~\cref{subsec:largek}, the excess should saturate to one or more limit points).

\begin{figure}[H]
\centering
\begin{subfigure}[b]{0.5\textwidth}
\includegraphics[width=7cm]{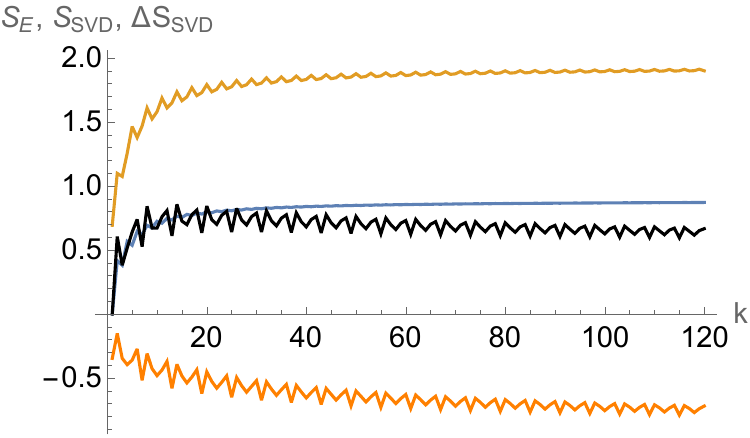}
\caption{$p=2$: $n_{1}=2$ vs $n_{2}=3$.}
\label{fig_p=2_n=2_3}
\end{subfigure}%
\begin{subfigure}[b]{0.5\textwidth}
\includegraphics[width=7cm]{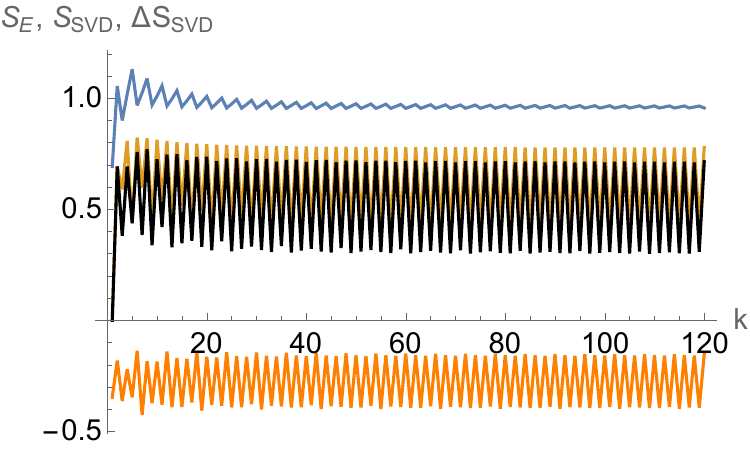}   
\caption{$p=3$: $n_{1}=3$ vs $n_{2}=4$.}
\label{fig_p=3_n=3_4}
\end{subfigure}
\caption{The SVD entropy of (a) $\mathrm{T}(2,4)$ vs $\mathrm{T}(2,6)$, and (b) $\mathrm{T}(3,9)$ vs $\mathrm{T}(3,12)$, plotted against the individual entanglement entropies. Also plotted in each case is the entropy excess. For simplicity just the interpolating curves are shown.}
\label{fig_ent_svd_2}
\end{figure}

The observed interpolation fails to hold if links other than $\mathrm{T}(p,p)$ are chosen; there appear to be instances where the SVD entropy periodically or eventually grows slower than either entanglement entropy.
Presented below are two such examples in which the SVD entropy (black curves) appears to eventually grow slower than the respective entanglement entropies.
In~\cref{fig_p=2_n=2_3} the SVD entropy between the $\mathrm{T}(2,4)$ and $\mathrm{T}(2,6)$ links is plotted against the respective (blue and olive curves) entanglement entropies.
In~\cref{fig_p=3_n=3_4} analogous results are shown for $\mathrm{T}(3,9)$ (blue curve) and the $\mathrm{T}(3,12)$ (olive curve) entanglement entropies.
In both these cases, the entropy excess (orange curves) appears to always be negative (in large $k$ limit discussed in~\cref{subsec:largek} we should observe saturation to one or more limit points).

In all these observations we notice that the growth curves, apart from  those for the entanglement entropies of the $\mathrm{T}(p,p)$ links with $p \ge 2$, all display a characteristic zig-zag pattern, whose complexity appears to increase with the twist number $n$.
In addition, the SVD entropy growth curves for sufficiently large $n_{1}$ or $n_{2}$ appear to have more than one limit point, as has been observed for the entanglement entropy in~\cite{Dwivedi_2020}.
We further discuss this in~\cref{subsec:largek}.

\bigskip

\begin{figure}[H]
\centering
\begin{subfigure}[b]{0.5\textwidth}
\centering
\includegraphics[width=7cm]{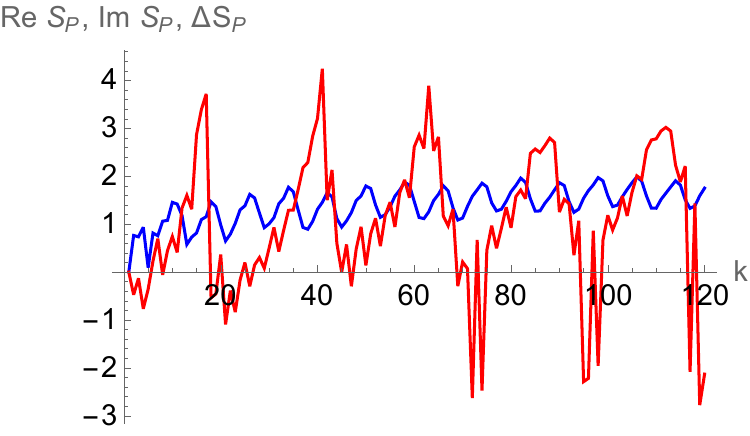}
\caption{$N=1$, $M=2$ pseudo-entropy}
\label{fig_N=1_M=2}
\end{subfigure}%
\begin{subfigure}[b]{0.5\textwidth}
\centering   
\includegraphics[width=7cm]{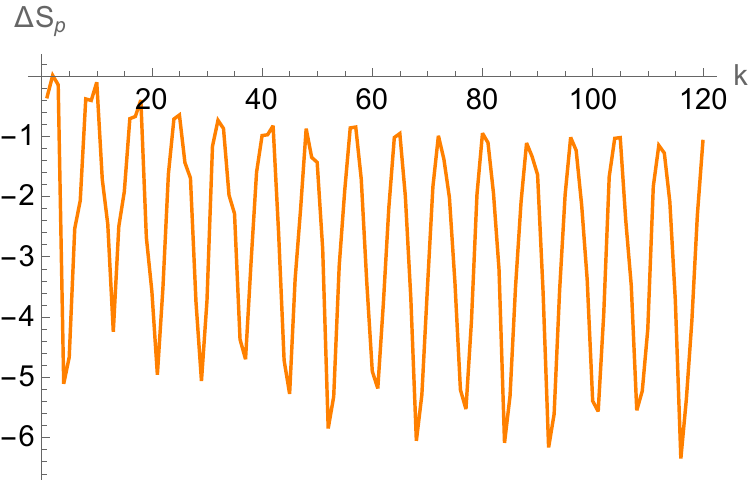}
\caption{$N=1$, $M=2$ entropy excess}
\label{fig_excess1_1}
\end{subfigure}
\caption{The (a) real (blue curve) and imaginary (red curve) parts of the pseudo-entropy and (b) entropy excess, for $\mathrm{T}(2,2)$ vs $\mathrm{T}(2,4)$. For simplicity just the interpolating curves are shown.}
\label{fig_psent_1}
\end{figure}

Let us also discuss the growth of the pseudo-entropy for $\mathrm{T}(p,pn)$ links, for $k$ up to 120. 
For the pseudo-entropy to be defined, the inner product $f^{n_{1}|n_{2}}(p;k)$ in (\ref{eq:2N|2M_innerproduct}) must be non-vanishing. However, we observe that for certain choices of parameters this product vanishes; e.g. for $n_{1},n_{2} \in \mathbb{Z}$, $|n_{1}-n_{2}| \equiv 2 \mod{4}$, and for all $p \ge 2$, we have $f^{n_{1}|n_{2}}(p;2k-1)=0$. 
In other cases the pseudo-entropy appears to grow sub-logarithmically with some spikes, see e.g. ~\cref{fig_N=1_M=2}. The complexity of the spiking pattern increases with an increase in $N$ and $M$.

We now note some observations on the entropy excess~\eqref{defn:entropy_excess}, which is invariant under the interchange of states $N \leftrightarrow M$.
In various examples we have checked -- for example the case in~\cref{fig_excess1_1} -- the entropy excess appears to eventually settle to negative values as $k$ grows. 
In several cases, it appears to be vanishing or a small positive value at a few small values of $k$ -- notably $k=2$ -- but appears to be negative thereafter.
There are a few cases where the excess is weakly positive for larger $k$ -- however it is unclear if this is a persistent trend.
Presented below is an example for the pseudo-entropy and excess between $\mathrm{T}(2,2)$ and $\mathrm{T}(2,4)$.
The above discussions center around examples where $n_{1},n_{2}>0$, i.e. the two links have the same chirality.
We observe similar patterns in the pseudo-entropy and entropy excess growth when the two links are of opposite chirality.
In this case however we find examples of $n_{1},n_{2}$ for which for several values of $k$, the excess attains a positive value.

Finally, we analyzed the metric interpretation of the absolute value of the SVD entropy excess for $(p,pn)$ torus links. The absolute value of the excess is clearly non-negative and symmetric.
The triangle inequality holds true for all sufficiently large $k$ for $p=2$, and one of the links being the Hopf link, for the triplets $(n_{1},n_{2},n_{3})=(1,2,3)$ and $(1,3,4)$.
For $p=2$, and $(n_{1},n_{2},n_{3})=(1,2,5),(1,6,7)$, the triangle inequality is satisfied periodically for sufficiently large $k$.
For $p=3$, the triangle inequality is not satisfied for all sufficiently large $k$ for $(n_{1},n_{2},n_{3})=(1,2,3)$, and is periodically satisfied for $(1,3,4)$.

\bigskip

In conclusion, we see that the behaviour of the growth of the SVD entropy for the $\mathrm{T}(p,pn)$ links is similar to that of the entanglement entropy -- the growth curves have individual sub-sequences of growth, and appear to have more than one limit point for sufficiently large $n$ (see also ~\cref{subsec:largek}). 
If one of the links is taken to be of the form $\mathrm{T}(p,p)$, the SVD entropy appears to interpolate between the two entanglement entropies.
The SVD entropy excess in such a case is negative at low $k$ and changes sign to be eventually positive at sufficiently large $k$.
From the viewpoint of quantum statistical mechanics, this may possibly indicate that the two link complement states belong to different quantum phases.
The interpolation manifestly fails when we do not consider any $\mathrm{T}(p,p)$ link complement state; the SVD entropy eventually appears to decrease below either entanglement entropy, and the excess appears to be always non-positive.
Finally, the pseudo-entropy in general for dissimilar links is more difficult to interpret, but it appears to be oscillatory and growing sub-logarithmically. The pseudo-entropy excess appears to eventually become negative for several cases.
%%%%%%%%%%%%%%%%%%%
%%%%%%%%%%%%%%%%%%%
\subsection{Example with Borromean links}
\label{subsec:borromean}

\begin{figure}[H]
    \centering
        \includegraphics[width=0.5\linewidth,height=4cm]{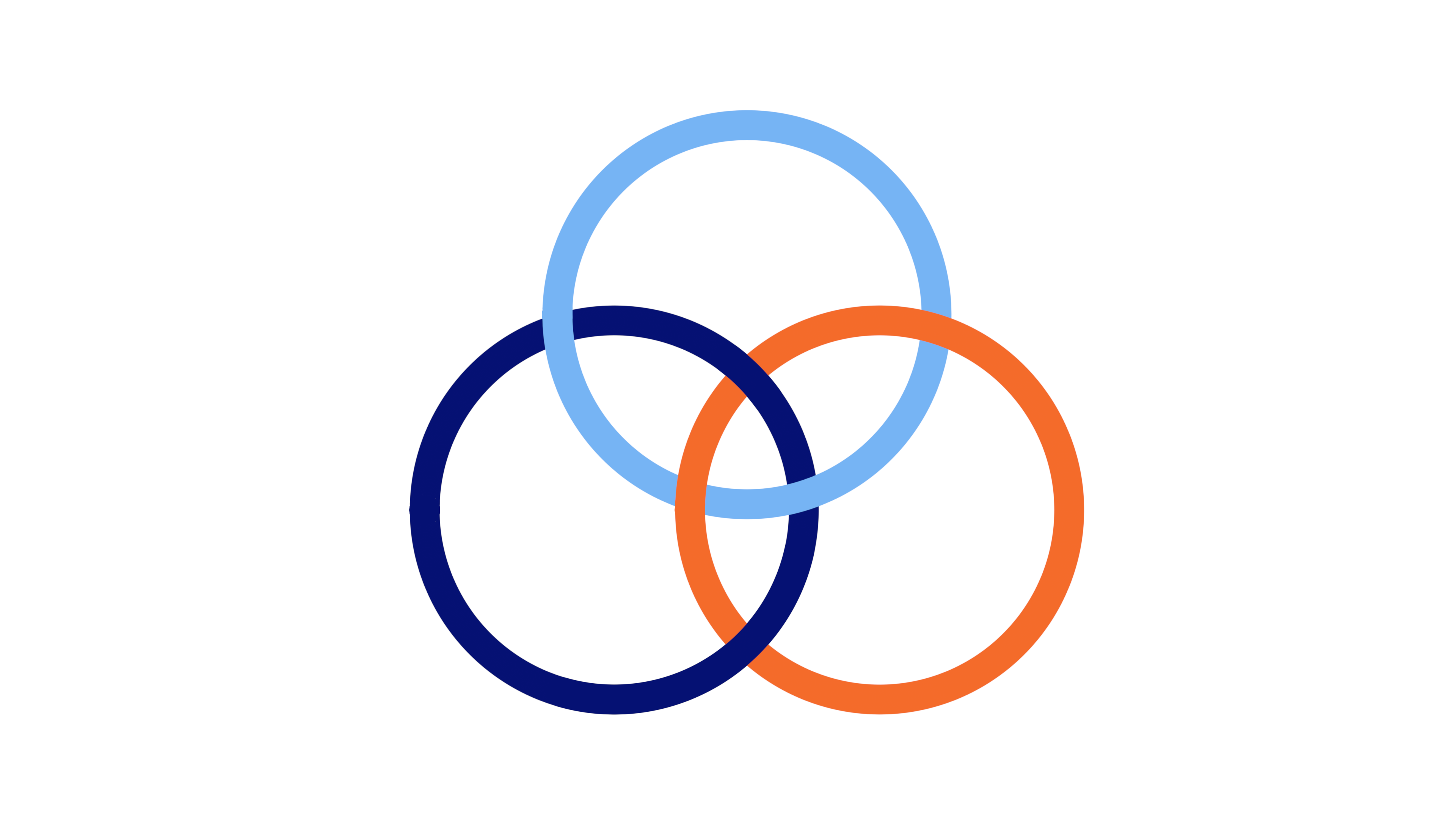}
        \caption{Borromean links $6^3_2$}
        \label{borromean link}
\end{figure}

We consider one other more complicated example involving two three-component links: a connected sum of Hopf-links ${2^2_1}\#{2^2_1}$ and Borromean links $6^3_2$, see \cref{borromean link}.

The coefficients of the link complement state coming from the connected sum of Hopf-links (in vertical framing) is given by palindromic (i.e. invariant under $q\mapsto q^{-1}$) polynomials  \cite{Balasubramanian_2017}
\begin{equation}
    |{2^2_1}\#{2^2_1} \rangle = \sum_{l,m,n} \frac{S_{lm} S_{nm}}{S_{0m}} |l, m, n\rangle .
\end{equation}
The coefficients of the (amphichiral) Borromean link complement state is given by \cite{Habiro} 
\begin{equation}
\begin{aligned}
    |6^3_2\rangle = \sum_{l,m,n} \sum_{i=0}^{\min \left(l, m, n \right)}(-1)^i\left(q^{1 / 2}-q^{-1 / 2}\right)^{4 i} \frac{\left[l+i+1\right] !\left[m+i+1\right] !\left[n+i+1\right] !([i] !)^2}{\left[l-i\right] !\left[m-i\right] !\left[n-i\right] !([2 i+1] !)^2} |l, m, n\rangle.
\end{aligned}
\label{borro eqn}
\end{equation}
\begin{figure}[H]
\centering
\begin{subfigure}[b]{0.5\textwidth}
\centering
\includegraphics[width=7cm]{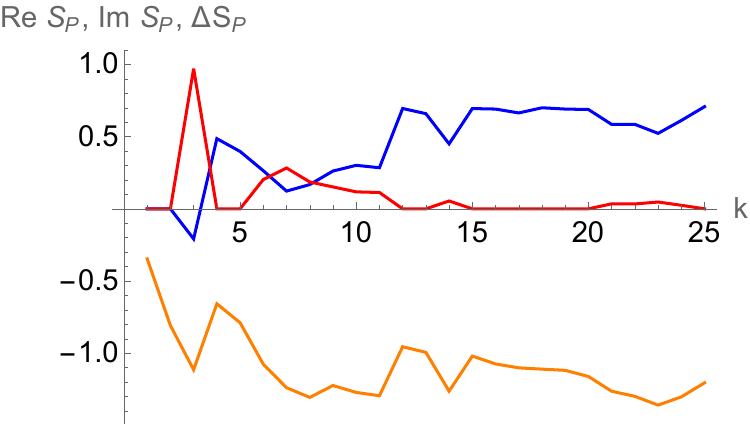}
\caption{Pseudo-entropy}
\label{fig:borr_pe}
\end{subfigure}%
\begin{subfigure}[b]{0.5\textwidth}
\includegraphics[width=7cm]{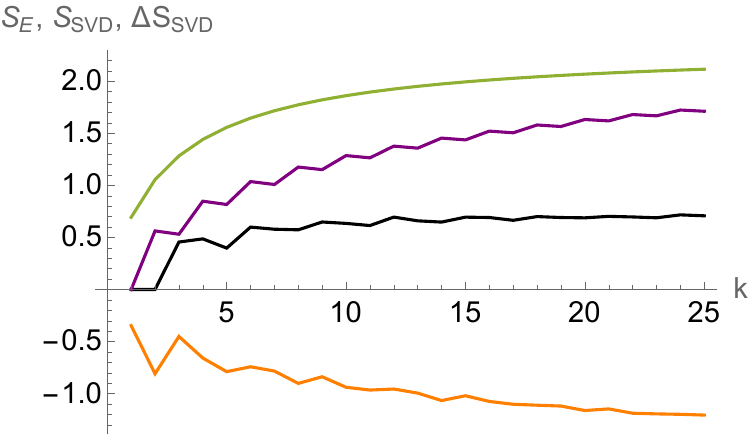}
\caption{Entanglement and SVD entropy}
\label{fig:borr_svd}
\end{subfigure}
\caption{Real (blue) and imaginary (red) parts of pseudo-entropy and SVD entropy (black) between the connected sum of two Hopf links ($S_\mathrm{E}$ in green), and the Borromean rings ($S_\mathrm{E}$ in purple). The respective entropy excesses are given in orange.}
\label{lab}
\end{figure}

We present various entropy measures for these two states (taken as the reference and target states) in \cref{lab}. In particular, both the excess SVD and pseudo-entropy are entirely negative. The imaginary part in \cref{fig:borr_pe} (in red) is entirely non-negative and equal to $r\pi$ where $r =-\sum_i \lambda_i ,\forall \lambda_i \in \mathbb{R} \text{ s.t. } \lambda_i<0$. This is because in general the amphichiral and palindromic link states have all real coefficients (see \cref{sec_chirality}) and thus the transition matrix has its complex  eigenvalues occurring in conjugate pairs. Therefore the contributions from the imaginary parts in $S_\textrm{P}$ cancel out except for $r\pi$.
%%%%%%%%%%%%%%%%%%%%%%%%%%
%%%%%%%%%%%%%%%%%%%%%%%%%%
\section{Large \texorpdfstring{$k$}{k} asymptotics for \texorpdfstring{$\mathrm{T}(p,pn)$}{T(p,pn)} links}
\label{subsec:largek}
%%%%%%%%%%%%%%%%%%%%%%%%%%
%%%%%%%%%%%%%%%%%%%%%%%%%%
In this section we study the large $k$  or semi-classical limit of entropy measures for $\mathrm{T}(p,pn)$ links. One motivation to study the large $k$ limit is its role in the volume conjecture \cite{kashaev1996,murakami1999,SO3}.

The semi-classical limits of the entanglement entropy $S_{\mathrm{E}}^{n}(p;k)$ for $\mathrm{T}(p,pn)$ links have been discussed in \cite{Dwivedi2018,Dwivedi_2020}. It was found that 
for $n=1$ and $p \ge 3$  
\begin{equation}
S_{\mathrm{E}}^{1}(p;k) \xrightarrow[]{k \rightarrow \infty} \log \zeta(2p-4)-(2p-4) \frac{\zeta'(2p-4)}{\zeta(2p-4)}+\log 2,
\end{equation}
whereas for $n \ge 2$ there appear to be either a limiting form or, for $n \ge 4$, a set of limiting forms, each for sub-sequences composed of $k$ modulo specific integers, 
\begin{equation}
\begin{split}
S_{\mathrm{E}}^{n \ge 2}(p;k) &\xrightarrow[k \equiv l\operatorname{mod} r(p,n)]{k \rightarrow \infty} \log \zeta(2p-2)-(2p-2) \frac{\zeta'(2p-2)}{\zeta(2p-2)} \\
&+ \left[ \log a^{(l)}_{m}(p,n)-\frac{d}{dm} \log a^{(l)}_{m}(p,n) \right]_{m=1},
\end{split}
\end{equation}
where $\zeta(s)=\sum_{n=1}^{\infty} \frac{1}{n^{s}}$ is the Riemann zeta function, the $a^{(l)}_{m}(p,n)$, for real $m \ge 1$ are real functions of $m$, $p$ and $n$~\footnote{The $m$ arises from the consideration of the R\'enyi entropy at order $m$; see~\cite{Dwivedi_2020} for details.}, and $l \in \{0,\ldots,r(p,n)-1\}$ for some integer $r \ge 2$ dependent on $p$ and $n$ (for the case of one unique limit, we may simply choose $r=2$ and set $a^{(0)}_{m}(p,n)=a^{(1)}_{m}(p,n)$).
Notably, the limiting entanglement entropy in all cases appears to consist of a universal term composed of zeta functions, which is independent of $n$ and the sub-sequence parameter $l$, and a second additive term which is dependent on $n$ and $l$ for $n \ge 2$. 
For $n \ge 4$, the number of limit points increases with an increase in $n$ (for example see~\cref{fig_N=1_N=5} and~\cref{fig_p=3_n=3_4}). 
It was also observed that the partition functions appear to be polynomials in $k$ up to $n=4$ (see~\cref{appendix_evnum}), but this pattern is broken from $n=5$ onwards.

Based on computational observations (detailed in~\cref{appendix_evnum}), we obtain similar conjectures for the SVD entropy between $\mathrm{T}(p,pn)$ links with $2 \le n \le 4$. 
We have been unable to find limiting forms for the SVD entropy when one of the links is of the $\mathrm{T}(p,p)$ form, i.e. when $n=1$, and we do not consider cases of $n>4$ because an analysis of the large $k$ growth of these information-theoretic parameters for $n=5$, $6$ and beyond has so far only proven partially successful; there is no known closed form for the large $k$ entanglement entropy for $n=5$, and a closed form is known only for $n=6$ only for $p=2$~\cite{Dwivedi_2020}.
Our conjectures are as follows:
\begin{align}
S_{\mathrm{SVD}}^{2|3}(p;k)  & \xrightarrow[]{k \rightarrow \infty}  \log \zeta(2p-2)-(2p-2) \frac{\zeta'(2p-2)}{\zeta(2p-2)}+\log \left(1-\frac{1}{2^{2p-2}}-\frac{1}{3^{2p-2}}+\frac{1}{6^{2p-2}} \right) \nonumber \\ 
& \qquad \quad - \frac{(2p-2)\left( \frac{1}{2^{2p-2}} \log 2+\frac{1}{3^{2p-2}} \log 3-\frac{1}{6^{2p-2}} \log 6 \right)}{1-\frac{1}{2^{2p-2}}-\frac{1}{3^{2p-2}}+\frac{1}{6^{2p-2}}},  \label{eq:SVD_large_23}
\end{align}
\begin{align}
S_{\mathrm{SVD}}^{2|4}(p;k)  \xrightarrow[k \equiv 0 \operatorname{mod}2]{k \rightarrow \infty} && & \log \zeta(2p-2)-(2p-2) \frac{\zeta'(2p-2)}{\zeta(2p-2)}+\log \left(1-\frac{1}{2^{2p-2}} \right) \nonumber \\ && & -\frac{(2p-2) \log 2}{2^{2p-2}-1}, 
\label{eq:SVD_large_24_even}
\end{align}
\begin{align}
S_{\mathrm{SVD}}^{3|4}(p;k)  \xrightarrow[k \equiv 0 \operatorname{mod}2]{k \rightarrow \infty} && & \lim_{k \rightarrow \infty} S_{\mathrm{SVD}}^{2|3}(p;k)+ \zeta(2p-2)\left(1-\frac{1}{2^{2p-2}}-\frac{1}{3^{2p-2}}+\frac{1}{6^{2p-2}} \right) \log 2, \label{eq:SVD_large_34_even}
\end{align}
\begin{align}
S_{\mathrm{SVD}}^{3|4}(p;k)  \xrightarrow[k \equiv 1 \operatorname{mod}2]{k \rightarrow \infty} && & \lim_{k \rightarrow \infty} S_{\mathrm{SVD}}^{2|3}(p;k)+ \zeta(2p-2)\left(1-\frac{1}{2^{2p-2}}-\frac{1}{3^{2p-2}}+\frac{1}{6^{2p-2}} \right) \nonumber \\ && & \times \left( \frac{2^{2p-\frac{5}{2}}}{1+2^{2p-\frac{5}{2}}} \log \left(1+2^{-2p+\frac{5}{2}}\right)+ \frac{1}{1+2^{2p-\frac{5}{2}}} \log \left(1+2^{2p-\frac{5}{2}} \right) \right). \label{eq:SVD_large_34_odd}
\end{align}
In fact,~\eqref{eq:SVD_large_24_even} is equal to the conjectured large $k$ limit of the $\mathrm{T}(p,2p)$ entanglement entropy~\cite{Dwivedi_2020} 
\begin{equation}
\lim_{k \rightarrow \infty} S_{\mathrm{SVD}}^{2|4}(p;2k) = \lim_{k \rightarrow \infty} S_{\mathrm{E}}^{2}(p;k), 
\end{equation}
even though at finite $k$ we appear to have a strict inequality
\begin{equation}
S_{\mathrm{SVD}}^{2|4}(p;2k) > S_{\mathrm{E}}^{2}(p;2k).
\end{equation}
\begin{figure}[H]
\centering
\begin{subfigure}[b]{.5\textwidth}
\centering
\includegraphics[width=7cm]{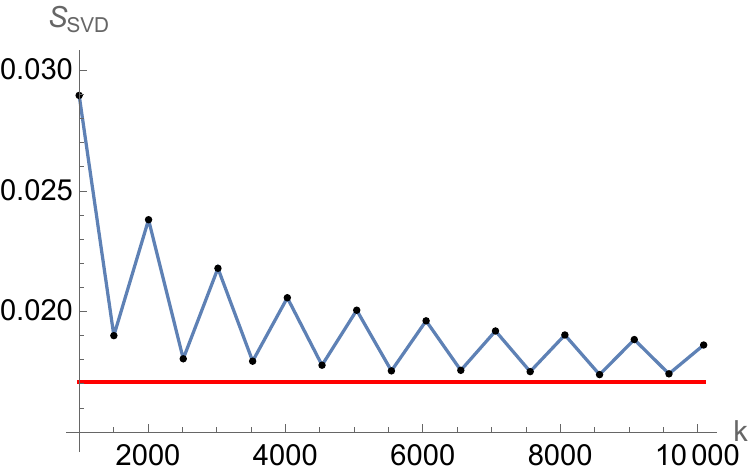}
\caption{$S_{\mathrm{SVD}}^{2|3}(3;k)$}
\label{fig:SVD_2|3_p=3}
\end{subfigure}%
\begin{subfigure}[b]{.5\textwidth}
\centering
\includegraphics[width=7cm]{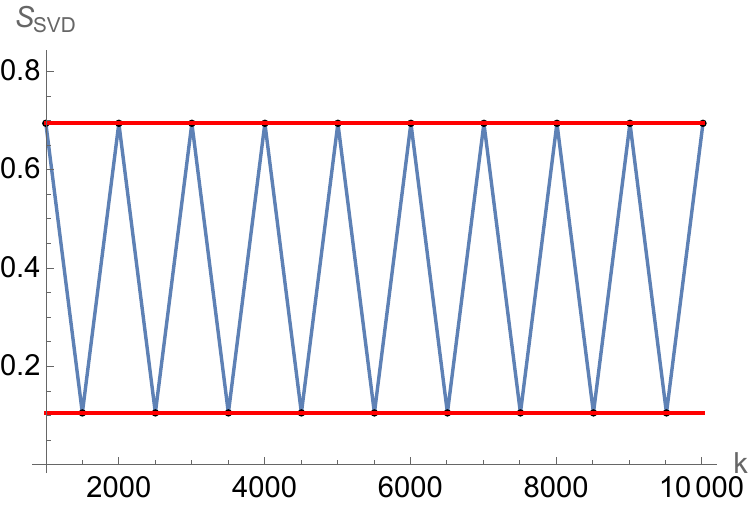}
\caption{$S_{\mathrm{SVD}}^{3|4}(4;k)$}
\label{fig:SVD_3|4_p=4}
\end{subfigure}%
\caption{The large $k$ behaviour of (a) $S_{\mathrm{SVD}}^{2|3}(3;k)$ and (b) $S_{\mathrm{SVD}}^{3|4}(4;k)$, converging to the conjectured limits~\eqref{eq:SVD_large_23} for $p=3$, and~\eqref{eq:SVD_large_34_even} and~\eqref{eq:SVD_large_34_odd} for $p=4$ respectively (red lines). }
\label{fig_ent_svd_largek}
\end{figure}
In~\cref{fig:SVD_2|3_p=3} we show $k$ at intervals of $505$, to account for all six possible sub-sequences, as convergence appears to be relatively slow; in~\cref{fig:SVD_3|4_p=4} we show $k$ at intervals of $501$ as convergence to the two limit points is much quicker.

The expressions~\eqref{eq:SVD_large_23} through~\eqref{eq:SVD_large_34_odd} are obtained by adapting the method used to obtain the large $k$ limit of the $\mathrm{T}(2,4)$ entanglement entropy in~\cite{Dwivedi2018} (and further generalized to $\mathrm{T}(p,np)$ links in~\cite{Dwivedi_2020}),
\begin{equation}
S^{2}_{\mathrm{E}}(k) \xrightarrow{k \rightarrow \infty} 24 \log A-2\gamma-\frac{17}{3}\log 2 \approx 0.887842096...\,, \label{eq_conjecture:SEE_N=2_largek_1orig_conj}
\end{equation}
where $A$ is the Glaisher constant and $\gamma$ the Euler--Mascheroni constant (\eqref{eq_conjecture:SEE_N=2_largek_1orig_conj} is identical to~\eqref{eq:SVD_large_24_even}~\cite{Dwivedi_2020}). 
We obtain asymptotic forms of the eigenvalues $|\Gamma^{n_{1}|n_{2}}(p;k)|$ at large $k$, for $2 \le n_{1},n_{2} \le 4$, and are able to write the SVD entropy as one or more infinite sums.
We detail this procedure in~\cref{appendix_evnum}.
We are unable to obtain any odd $k$ asymptotic value for $S^{2|4}_{\mathrm{SVD}}(p;k)$ to complement~\eqref{eq:SVD_large_24_even} because, for odd $k$, the eigenvalues $|\Gamma_{m}^{2|4}(p;k)|$ appear to vanish for all $m$.

Our conjectures, in conjunction with those in~\cite{Dwivedi_2020}, permit us to obtain large $k$ limits for the SVD entropy excess for the links under consideration. Notably the universal part of the expressions vanishes.
For example, 
\begin{align}
\Delta S_{\mathrm{SVD}}^{2|3}(p;k) \xrightarrow[]{k \rightarrow \infty}&& & \frac{1}{2}\log \left(1-\frac{1}{2^{2p-2}}-\frac{1}{3^{2p-2}}+\frac{1}{6^{2p-2}} \right)-\frac{1}{2} \log 2 \nonumber \\ && & 
-\frac{\left(3^{2p-2}+1\right)\log 2+\left(2^{2p-2}+1\right)\log 3-2 \log 6}{2\left(2^{2p-2}-1\right)\left(3^{2p-2}-1\right)}\,, \label{excess:2|3_generalp}
\end{align}
and similarly for the other cases.
The limiting excess~\eqref{excess:2|3_generalp} appears to be always negative, and viewed as an analytic function of a positive real $p>1$, is strictly increasing and appears to limit to $-\frac{1}{2} \log 2$ as $p \rightarrow \infty$, and to $-\infty$ as $p\rightarrow 1$.

Though our conjectures so far are based on guessing asymptotic functional forms from numerics, and are further well-corroborated by numeric computation of the transition matrix eigenvalues involved up to $k$ of order $10^{4}$ using Fortran, we have also ventured into an analytic survey of these entropies. 
In particular we have looked at the entanglement entropy of the $\mathrm{T}(2,4)$ link, whose transition matrix eigenvalues in un-normalized form appear to be given by~\cite{Dwivedi2018}\footnote{See~\eqref{eq_conjecture:alpha^2_m} for our conjecture on the $\alpha^{2}_{m}(2;k)$.}
\begin{equation}
    \Gamma_{m}^{2|2}(2;k)=\frac{1+(-1)^{m-k}}{2+q^{\frac{m+1}{2}}+q^{-\frac{m+1}{2}}}=\frac{1+(-1)^{m-k}}{4\cos^{2}\left( \frac{\pi}{2} \frac{m+1}{k+2} \right)}. ~\label{eq_conjecture:gamma^2_m}
\end{equation}
In~\cref{appendix_intform} we prove that they are resummed into (\ref{f-overlap})
\begin{equation}
   f^{2|2}(2;k)= \left\lfloor \frac{(k+2)^2}{4} \right\rfloor= \begin{cases} \frac{(k+2)^2}{4} & k \, \mathrm{even}, \\ \frac{(k+1)(k+3)}{4} & k \, \mathrm{odd}. \end{cases} \label{eq_conjecture:partitionfunction_N=2}
\end{equation}
Further, we develop an integral representation of $S^{2}_{\mathrm{E}}(2;k)$ for finite $k$, given by~\eqref{eq_SEE_N=2_rearranged_integrals_even} and~\eqref{eq_SEE_N=2_rearranged_integrals_odd} respectively for even and odd $k$, which appear to have a unique limit,~\eqref{eq_conjecture:SEE_N=2_largek_1_integral_form}, at $k \rightarrow \infty$.
This limit is exactly equal to the conjectured values in~\cite{Dwivedi2018} and~\cite{Dwivedi_2020}, and sheds light on the zig-zag behaviour of the $S_{\mathrm{E}}^{2}(2;k)$ interpolation curve.
We hope that the technique involved in developing this representation can be modified, improved upon or generalized to evaluate various entropies for other torus links in particular, and be used develop a large $k$ expansion of these entropies.

In conclusion, in this subsection we conjecture the limiting forms of the SVD entropy between two $\mathrm{T}(p,pn)$ links, for $2 \le n \le 4$. 
These forms appear similar to those observed for the entanglement entropy in~\cite{Dwivedi_2020} -- notably the presence of a universal part composed of zeta functions, but with some zeta function terms  now also appearing in the non-universal part in some cases (see e.g.~\eqref{eq:SVD_large_24_even} and~\eqref{eq:SVD_large_34_odd}).
Finally, to get an analytic handle on these numerical observations, we propose (and detail in~\cref{appendix_intform}) an integral form for the $\mathrm{T}(2,4)$ entropy, which we hope may be extended to other links in the future. 
%%%%%%%%%%%%%%%%%%%%%%%%%%%%%%%%
%%%%%%%%%%%%%%%%%%%%%%%%%%%%%%%%
\section{Chirality and imaginary part}%\texorpdfstring{$S_\mathrm{P}$}{SP}
\label{sec_chirality}
%%%%%%%%%%%%%%%%%%%%%%%%%%%%%%%%
%%%%%%%%%%%%%%%%%%%%%%%%%%%%%%%%
It has been noted that $S_{\mathrm{P}}$ can distinguish two different quantum phases \cite{Mollabashi:2020yie,Mollabashi:2021xsd}. Since it is generally complex-valued due to the transition matrix being non-hermitian, its imaginary part is an interesting quantity. To generalize these observations, let us consider how various entropy measures behave upon taking the chiral version of a quantum state, defined as a state with conjugate coefficients
\begin{equation}
\sum_i a_i |\psi_i\rangle
\xrightarrow[]{\text{chiral}}
\sum_i \overline{a}_i |\psi_i\rangle.
\end{equation}
One can ask whether both a quantum state and its chiral version co-exist in the scope of a given theory. For example, for generalized $\mathrm{SU}(2)$ and $\mathrm{SU}(1,1)$ coherent states given by a  point on the unit sphere  and the hyperbolic disc  respectively,  parametrized by $z_i$ according to~\eqref{zi-tan} or~\eqref{zi-tanh}, the above transformation produces a state given by the opposite point parametrized by $\bar{z}_i$. In case of link states in $\mathrm{U}(1)$ Chern--Simons theory the answer is also yes, as the link invariants are simply the writhes of the links, and taking the mirror image of a link switches the linking number $l_{ab} \mapsto -l_{ab}$, thus the coefficients of link states are complex conjugated. 
In knot theory, a knot or link is called chiral if it is not ambient isotopic to its mirror image, and amphichiral otherwise. Therefore flipping the chirality of the link gives us the chiral version of the link state in U(1) theory. Moreover, for two-component amphichiral links the above statement suggests that $l_{ab}=0$, which then implies that the $\mathrm{U}(1)$ states take the form $\frac{1}{k}\sum_{m,n}|m,n\rangle$, i.e. they are unentangled. Chiral states also exists in $\mathrm{SU}(2)$ Chern--Simons theory, which follows from properties of colored Jones polynomials, as we discuss below.

\begin{figure}[H]
    \centering
\includegraphics[width=0.7\linewidth,height=4cm]{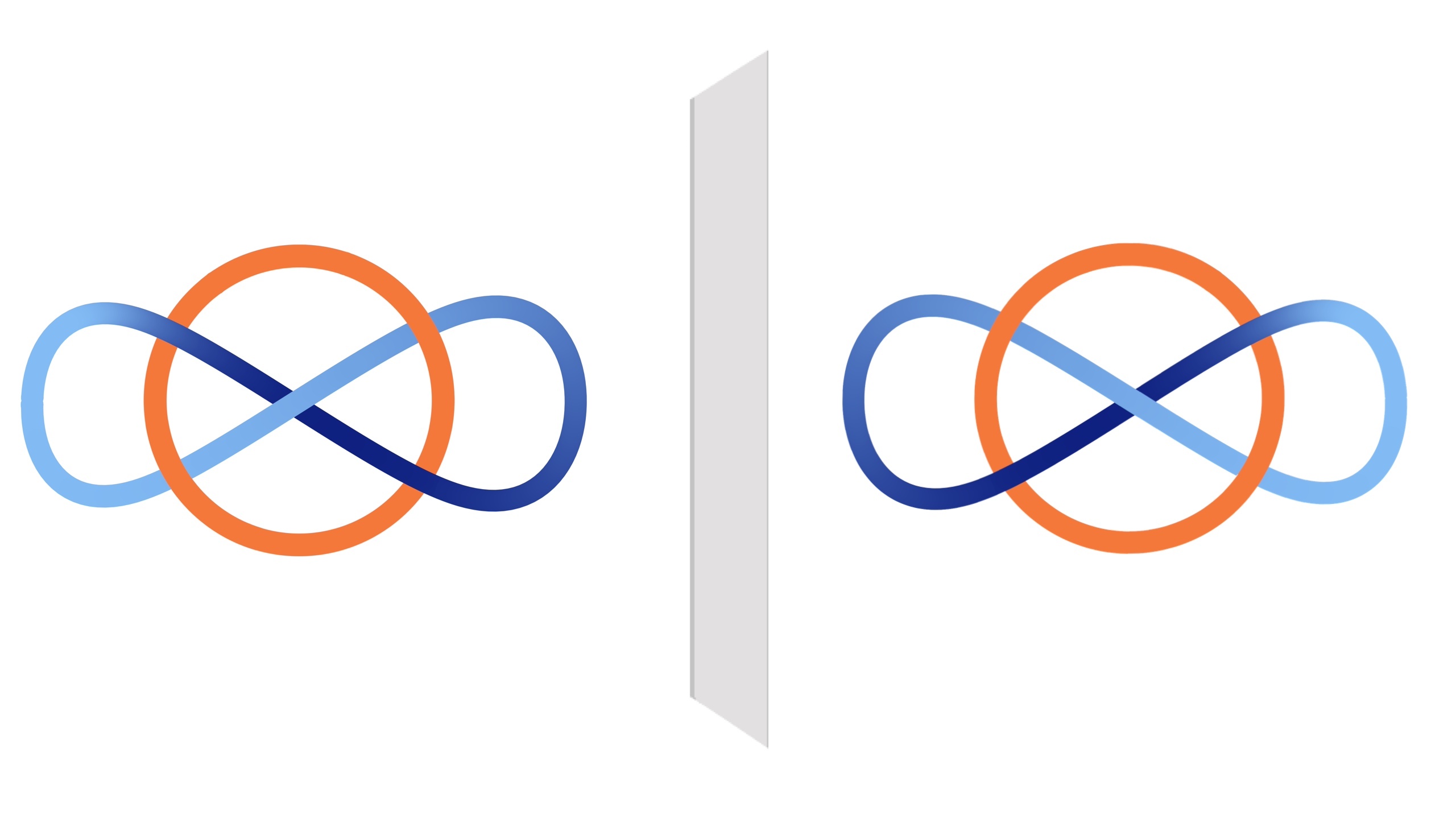}
    \caption{The left-handed Whitehead link $\mathcal{W}$ (left figure) and its mirror image, the right-handed Whitehead link $\mathcal{W}^\star$ (right figure).}
    \label{whiteheadpics}
\end{figure}

Note that the entanglement entropy does not detect chirality as it only depends on the magnitude of the transition matrix entries. However, $S_{\mathrm{P}}$ and $S_{\mathrm{SVD}}$ depend on two distinct choices of reference and target states $\phi$ and $\psi$ in the transition matrix $\tau^{\phi|\psi}$, and thus in principle might detect chirality. In fact, if chirality of both the reference and target states is flipped, the $S_{\mathrm{SVD}}$ will also be unable to detect the chirality change. This is because for the reduced transition matrix $\tau$ we can obtain its singular values $\sigma_i$ by computing the eigenvalues $\sqrt{\lambda_i}$ of $\sqrt{\tau \tau^{\dag}}$, where
$ \tau \rightarrow \overline{\tau} \implies   (\tau \tau^{\dag})^{\frac{1}{2}} \rightarrow  (\overline{\tau \tau^{\dag}})^{\frac{1}{2}}$. Once again these are Hermitian matrices where ${\lambda_i}^\frac{1}{2} ={\overline{\lambda_i}}^\frac{1}{2}$ are real and non-negative. Thus the singular values do not change, and so $S_{\mathrm{SVD}}$ remains the same. It will however detect chirality change when only either the reference or target state is flipped. On the other hand $S_{\mathrm{P}}$ detects chirality always, i.e., when the chirality of either one or both states are flipped.

\begin{figure}[t!]
\centering
\begin{subfigure}[b]{0.5\textwidth}
\centering
\includegraphics[width=7cm]{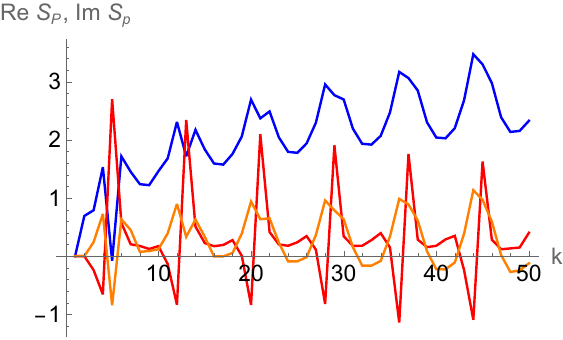}
\caption{$S_{\mathrm{P}}^{\mathcal{W}|\mathcal{W}^{\star}}$}
\label{fig_pe_w|-w}
\end{subfigure}%
\begin{subfigure}[b]{0.5\textwidth}
\centering   
\includegraphics[width=7cm]{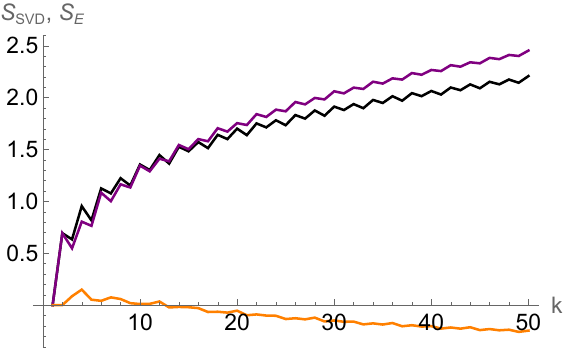}
\caption{$S_{\mathrm{SVD}}^{\mathcal{W}|\mathcal{W}^{\star}}$ and $ S_{\mathrm{E}}^\mathcal{W}=S_{\mathrm{E}}^{\mathcal{W}^\star}$}
\label{fig_svd_w|-w}
\end{subfigure}
\caption{(a) Real (red) and imaginary (blue) parts of pseudo-entropy, and (b) The SVD entropy (black), between the Whitehead link and its mirror, both of which have the same entanglement entropy (violet). Also plotted are the respective entropy excesses (orange).}
\label{fig_whitehead_1}
\end{figure}

Let us focus on link states $|\cL\rangle$ in $\mathrm{SU}(2)$ Chern--Simons theory. The coefficients of these states involve colored Jones  polynomials $V^{\cL}_{mn\ldots}(q)$ of a link $\cL$, colored by symmetric representations $m,n,\ldots$. It is known that colored Jones polynomials of the mirror image $\cL^\star$ are given by $V^{\cL}_{mn\ldots}(q^{-1})$, i.e. with the parameter $q = \exp{\frac{2 \pi i}{k+2}}$ inverted. Furthermore, coefficients of colored Jones polynomials are real, so transforming the link $\cL$ to its mirror $\cL^\star$ changes the colored Jones polynomials as follows
\begin{equation}
%    |\cL\rangle \longrightarrow |\cL^{\star}\rangle \implies    
    V^\cL_{mn\ldots}(q) \longrightarrow V^{\cL^\star}_{mn\ldots}(q) =  V^\cL_{mn\ldots}(q^{-1}) = V^\cL_{mn\ldots}(\overline{q}) = \overline{V^\cL_{mn\ldots}(q)}.
\end{equation}
It follows that when we take a chiral version of both the pre-selected and post-selected link state, the reduced transition matrix (up to normalisation) and ultimately the pseudo-entropy change as follows\footnote{Principal branch of complex logarithm is discontinuous at the negative real axis. As per the relation \cite{Nishioka:2021cxe} for the pseudo R\'enyi entropy $S^{(n)}(\tau_A^{\varphi|\psi}) = S^{(n)}(\tau_A^{\psi|\varphi})^*$, complex conjugation of logarithm can be handled by fixing the branch cut appropriately. Or more simply one can consider systems where $\lambda_i \in \mathbb{C}\symbol{92}\mathbb{R}^{-}$ \cite{Nakata:2020luh}.}
\begin{equation}
\text{Tr}_\mathrm{B} \left(|\phi\rangle \langle \psi|\right) \longrightarrow \text{Tr}_\mathrm{B} \left(|\phi^\star\rangle \langle \psi^\star|\right) = \overline{\text{Tr}_\mathrm{B} \left(|\phi\rangle \langle \psi|\right)} \implies  \lambda \longrightarrow \overline{\lambda} \implies S_{\mathrm{P}} \longrightarrow \overline{S_{\mathrm{P}}} . 
\end{equation}
Thus flipping the chirality of both links does not affect the real part of $S_{\mathrm{P}}$, while the imaginary part also flips. Moreover, 
since we know that $S_\mathrm{P}^{\phi|\psi} = \overline{S_\mathrm{P}^{\psi|\phi}}$, a physical interpretation of the above statement is that flipping the chirality of the links is equivalent to swapping the order of the states in $S_\mathrm{P}$.

\begin{figure}[H]
\centering
\begin{subfigure}[b]{.5\textwidth}
\centering
\includegraphics[width=7cm]{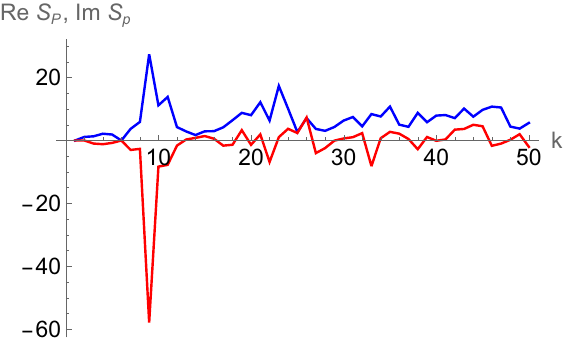}
\caption{$S_{\mathrm{P}}^{1|\mathcal{W}}$}
\label{fig_pe_2_2|w}
\end{subfigure}%
\begin{subfigure}[b]{.5\textwidth}
\centering
\includegraphics[width=7cm]{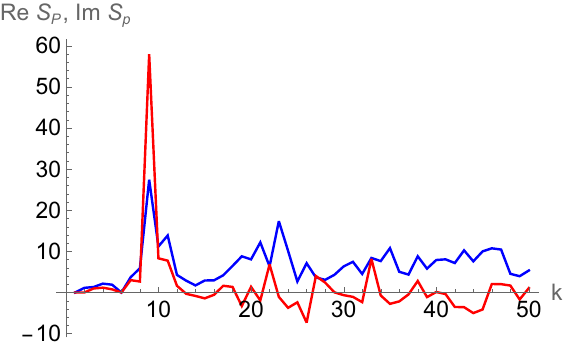}
\caption{$S_{\mathrm{P}}^{-1|\mathcal{W}^{\star}}$}
\label{fig_pe_2_2|-w}
\end{subfigure}
\caption{The real (blue) and imaginary (red) parts of the pseudo-entropy of (a) $\mathrm{T}(2,2)$ vs Whitehead (b) mirrored $\mathrm{T}(2,2)$ vs mirrored Whitehead.}
\label{fig_whitehead_2}
\end{figure}

We illustrate the above statements in examples involving the Whitehead link $\mathcal{W}$ and its mirror, see \cref{whiteheadpics}. Colored Jones polynomial of a Whitehead link takes the form \cite{Habiro}
\begin{equation}
\begin{aligned}  \label{Whitehead-Jones}
 V_{mn}^{\mathcal{W}} &= \sum_{i=0}^{\min(m,n)} (-1)^i q^{\frac{i}{4}(i+3)} (q^{\frac{1}{2}} - q^{-\frac{1}{2}})^{3i}  \frac{[m + i + 1]![n + i + 1]![i]!}{[m - i]! [n - i]! [2i + 1]!}. 
\end{aligned}
\end{equation}

In \cref{fig_whitehead_1} we show how the pseudo and SVD entropies grow for the Whitehead link and its mirror. The pseudo-entropy excess oscillates around the horizontal axis, whereas the SVD entropy excess is negative at large $k$. On the other hand, in \cref{fig_whitehead_2} we show how the imaginary part (red curves) of the pseudo-entropy between the Hopf and Whitehead links changes sign if both links are mirrored. The real part (blue curves) remain unaffected. Thus we see that $S_\mathrm{P}^{1|\mathcal{W}} \rightarrow {S_\mathrm{P}^{-1|\mathcal{W}^\star}} = \overline{S_\mathrm{P}^{1|\mathcal{W}}} = S_\mathrm{P}^{\mathcal{W}|1}$.

This is automatically extendable to all $n$-component links with arbitrary colours, and also to $\mathrm{SU}(N)$ where the link invariants are given by the two variable HOMFLY-PT polynomial $P^\mathcal{K}(a,q)$. In that case, for the mirror image of the knot $\mathcal{K}^\star$, we take $q\rightarrow q^{-1}$ and $a \rightarrow q^{-N}$. In principle the result extends to any general (compact) gauge group.

%%%%%%%%%%%%%%%%%%%%%%%%%%%
%%%%%%%%%%%%%%%%%%%%%%%%%%%
 \section{Conclusions and future directions}\label{Conclusions}
%%%%%%%%%%%%%%%%%%%%%%%%%%%
%%%%%%%%%%%%%%%%%%%%%%%%%%%

In this work we explored two recent generalizations of entanglement entropy, i.e. pseudo-entropy and SVD entropy, as well as their excess, as potential tools for quantifying differences between the two quantum states used in their definitions. Building intuitions in quantum mechanical examples of states with different patterns of entanglement, such as qubits and generalised coherent states, we performed numerical studies of these quantities using link complement states in Chern--Simons theory.   

Interestingly, we found that for certain gauge groups and families of states, pseudo and SVD entropy excess plays the role of a metric and serves as a new tool to distinguish and classify Chern--Simons states from a new, quantum-informational, perspective. We note that for both the $\mathrm{SU}(1,1)$ coherent states and the U(1) link complement states, the SVD entropy excess is shown to be always non-positive. We have mostly focused on several examples of composite, torus and hyperbolic links where expansion coefficients are known analytically (e.g. in terms of the modular $\cS$ and $\cT$ matrices).  It will be very interesting to follow our program more systematically and test for which states the triangle inequality is satisfied or violated and how useful these (pseudo)-metrics are in classifying complexity of the knot states.

Even though the path integral definition of pseudo-entropy is very natural and allowed for its interesting applications in QFTs and holography, the physical meaning of its imaginary part remains mysterious. Here, we managed to shed a new light on its sensitivity to complex phases that also distinguish the information about chirality of knots. This information is rather non-trivial and decoding it using e.g. topological invariants such as usual knot polynomials is is not always possible. However, Chern-Simons invariants such as the coloured Jones polynomials used in this paper are known to be sensitive to chirality \cite{ramadevi1993chirality,Kaul:1998ye}. We hope that our finding will help to develop a more systematic path to understanding the imaginary part of pseudo entropy and analyzing its correlations with different choices of framing will be an important future step in this direction.

We note that, though we have focused on the SVD entropy in these works, it should be straightforward to generalize our computations to the so-called SVD R\'enyi entropy~\cite{Parzygnat:2023avh}. 
In particular, the large $k$ limiting values of the SVD R\'enyi entropy for the systems we considered may be analyzed similar to how the entanglement version has been in~\cite{Dwivedi2018}.
A physical interpretation of the behaviour of these entropies at large $k$ -- notably, that several of them saturate to finite limiting values -- is also something interesting to look at.

Last but not the least, the quantities that we studied involve two, pre- and post-selected or reference and target quantum states and, as we showed, may serve as metrics to quantify distances between them. It is then very natural to wonder if they could also play a role of complexity measures for quantum states\footnote{We thank Jackson Fliss for stressing this connection.}. Indeed, different choices of cost functions in Nielsen's geometric approach \cite{Nielsen:2005mkt} generically give rise to Finsler geometries \cite{antonelli2003handbook} and it would be interesting to make a connection with our analysis more precise. On a similar note, the SVD decomposition has already been applied to quantifying complexity \cite{strydom2021svd} and it will be interesting to revisit it in the recent context of complexity in quantum field theories. 
%%%%%%%%%%%%%%%%%%%%%%%%%%%%%
%%%%%%%%%%%%%%%%%%%%%%%%%%%%%
\acknowledgments
%%%%%%%%%%%%%%%%%%%%%%%%%%%%%
%%%%%%%%%%%%%%%%%%%%%%%%%%%%%
We would like to thank Onkar Parrikar and Tadashi Takayanagi for discussions and comments, Siddharth Dwivedi and Jackson Fliss for discussions and sharing Mathematica codes, and Anuvab Kayaal for the knot/link graphics. We especially thank Pichai Ramadevi for several helpful discussions and comments on the draft. SP thanks the Galileo Galilei Institute of Theoretical Physics and the \'Ecole de Physique des Houches for hospitality, and the INFN Italy and the Universit\'e Grenoble-Alpes for partial support during the completion of this work. 
This work is supported by the OPUS grant no. 2022/47/B/ST2/03313 ``Quantum geometry and BPS states'' and NCN Sonata Bis 9 grant no. 2019/34/E/ST2/00123 funded by the National Science Centre, Poland, and NAWA ``Polish Returns 2019'' grant no. PPN/PPO/2019/1/00010/U/0001. This research was supported in part by grant NSF PHY-2309135 to the Kavli Institute for Theoretical Physics (KITP). The work of PS is also supported  by the U.S. Department of Energy, Office of Science, Office of High Energy Physics, under Award Number DE-SC0011632. 

%%%%%%%%%%%%%%%%%%%%%%%%%%%%%
%%%%%%%%%%%%%%%%%%%%%%%%%%%%
\appendix
%%%%%%%%%%%%%%%%%%%%%%%%%%%%%
%%%%%%%%%%%%%%%%%%%%%%%%%%%%

%%%%%%%%%%%%%%%%%%%%%%%%%%%
%%%%%%%%%%%%%%%%%%%%%%%%%%%
\section{Two-qubit states}\label{App:2-qubits}
%%%%%%%%%%%%%%%%%%%%%%%%%%%
%%%%%%%%%%%%%%%%%%%%%%%%%%%
Here we provide analytic formulas for eigenvalues and singular values of the transition matrix for a general case of two-qubit systems \cite{2-qubit}, which belong to the Hilbert space  $\mathcal{H}_A \otimes \mathcal{H}_B$, where dim$(\mathcal{H}_i) = 2$. The various entanglement entropy measures are determined by the Schmidt coefficients of the chosen reference and target states. 
Two general unnormalized states in this setup can be written with complex coefficients as
\begin{equation}
\begin{aligned}
& |\phi\rangle=a_1|00\rangle+a_2|01\rangle+a_3|10\rangle+a_4 |11\rangle, \\
& |\psi\rangle=b_1|00\rangle+b_2|01\rangle+b_3|10\rangle+b_4 |11\rangle.
\end{aligned}  
\label{eqn: generic 2qubits}
\end{equation}
The pseudo-entropy $S_\mathrm{P}$ and SVD entropy $S_{\mathrm{SVD}}$ are obtained  from the eigenvalues $\lambda_\pm$ and normalised singular values $\hat{\lambda}_\pm$ respectively of the reduced transition matrix $\tau^{\phi|\psi}_A = \Tr_B (\tau^{\phi|\psi})$ by tracing out the second qubit. 
They evaluate to be
\begin{eqnarray}
\lambda_{\pm} &=&   \frac{1}{2} \pm \frac{ \left((l_1 + l_4)^2 - 4 (l_1 l_4 - l_2 l_3)\right)^\frac{1}{2}}{2 (l_1 + l_4)},\qquad   
  \hat{\lambda}_{\pm} = \frac{\Delta_{\pm}}{\Delta_{+}+\Delta_{-}},
  \label{eqn: eigen and singular values 2qubits}
\end{eqnarray} 
where
\begin{equation}
l_1 = \sum_{i=1,2} a_i \overline{b}_i,  \; \  l_2 = (a_1 \overline{b}_3 + a_2 \overline{b}_4), \; \  l_3 = (a_3 \overline{b}_1 + a_4 \overline{b}_2), \; \  l_4 = \sum_{i=3,4} a_i \overline{b}_i,   
\label{eqn: l values}
\end{equation}
and 
\begin{equation}
\Delta_{\pm}= \left(  \sum_{1}^4 |l_i|^2  \pm \left((\sum_{1}^4 |l_i|^2)^2 - 4 \left((|l_1|^2+|l_3|^2)(|l_2|^2+|l_4|^2) - |l_1 \overline{l}_2 + l_3 \overline{l}_4|^2\right)\right)^{\frac{1}{2}} \right)^\frac{1}{2}.
\label{eqn: delta values}
\end{equation}

Similarly, we can instead trace out the first qubit to get $\tau^{\phi|\psi}_B = \Tr_A (\tau^{\phi|\psi})$. The eigenvalues and singular values are given by the same expressions \eqref{eqn: eigen and singular values 2qubits} but by replacing $l_i \rightarrow l_i^\prime$, where
\begin{equation}
l^\prime_1 = \sum_{i=1,3} a_i \overline{b}_i,  \; \  l^\prime_2 = (a_1 \overline{b}_2 + a_3 \overline{b}_4), \; \  l^\prime_3 = (a_2 \overline{b}_1 + a_4 \overline{b}_3), \; \  l^\prime_4 = \sum_{i=2,4} a_i \overline{b}_i,   
\end{equation}
that is, by simply interchanging the 2nd and 3rd coefficients in the original setup ($a_2 \leftrightarrow a_3$ and $b_2 \leftrightarrow b_3$ in \eqref{eqn: generic 2qubits} through \eqref{eqn: delta values}).

%%%%%%%%%%
%%%%%%%%%%
\section{Pseudo-metric for two-component U(1) link states} \label{sec-app-U1}
%%%%%%%%%%
%%%%%%%%%%
In this appendix we consider the SVD entropy for two-component link states in $\mathrm{U}(1)$ Chern--Simons theory, given by \eqref{SSVD-U1}, which leads to the expression~\eqref{SSVD-U1-abs} for absolute value of the SVD entropy excess, \eqref{SSVD-U1-abs}, for $n,p\in \mathbb{N}$.
This excess satisfies axioms of a pseudo-metric.
It is both non-negative (recall \eqref{excess_u1} was shown to be non-positive) and symmetric under interchanging $l_1$ and $l_2$. 
Furthermore, $|\Delta S_{\mathrm{SVD}}| = 0$ for $l_1=l_2\equiv l$, since in this case \eqref{SSVD-U1-abs} reduces to 
\begin{equation}
    |\Delta S_{\mathrm{SVD}}|=\frac{1}{2} \log   \frac{(\gcd(k, l^2))^2}{ (\gcd(k, l))^2 }, \label{excessSVDU1_equal}
\end{equation} 
and so for $k,n,p,p_i,a_i \in \mathbb{N}$, and any factorisation of $l=p_1^{a_1}p_2^{a_2}\cdots p_n^{a_n}$, we have 
\begin{equation}
\gcd(k, l^2=p_1^{2a_1}p_2^{2a_2}\cdots p_n^{2a_n})\neq np^2\implies \gcd(k,l^2)=\gcd(k,l)    \label{gcd-k-l}
\end{equation}
as there are no common (multiples of) square factors. Thus from~\eqref{excessSVDU1_equal} we get 
$|\Delta S_{\mathrm{SVD}}|=0$.

Finally, let us prove the triangle inequality
\begin{equation}
 \log \left( \frac{(\gcd(k, l_1 l_2))^2}{\gcd(k, l_1)  \gcd(k, l_2)} \right) + \log \left( \frac{(\gcd(k, l_2 l_3))^2}{\gcd(k, l_2)  \gcd(k, l_3)} \right) \geq \log \left( \frac{(\gcd(k, l_1 l_3))^2}{\gcd(k, l_1)  \gcd(k, l_3)} \right),
\end{equation}
or equivalently,
\begin{equation}
\gcd(k,l_1l_2)\gcd(k,l_2l_3) \geq \gcd(k, l_1 l_3)\gcd(k, l_2). \label{triangle-check}
\end{equation}

To prove~\eqref{triangle-check}, we use the associative property of the greatest common divisor function 
\begin{equation}
\gcd(a_{1},\ldots,a_{n},\gcd(p,q))=\gcd(a_{1},\ldots,a_{n},p,q),
\end{equation}
for positive integers $a_{1},a_{2},\ldots,a_{n},p,q$.
Rewriting the quantities on both sides of~\eqref{triangle-check} using this property, we get
\begin{eqnarray}
\gcd(k,l_{1}l_{2})\gcd(k,l_{2}l_{3}) &=& \gcd(k^{2},kl_{1}l_{2},kl_{2}l_{3},l_{1}l_{2}^{2}l_{3}) \nonumber
\\ &=& \gcd(k^{2},l_{1}l_{2}^{2}l_{3},kl_{2} \gcd(l_{1},l_{3})),
\end{eqnarray}
and
\begin{eqnarray}
\gcd(k,l_{1}l_{3})\gcd(k,l_{2}) &=& \gcd(k^{2},kl_{2},kl_{1}l_{3},l_{1}l_{2}l_{3}) \nonumber
\\ &=& \gcd(k^{2},l_{1}l_{2}l_{3},k \gcd(l_{2},l_{1}l_{3})),
\end{eqnarray}
respectively.
\eqref{triangle-check} immediately follows from observing that
\begin{equation}
l_{1}l_{2}^{2}l_{3} \ge l_{1}l_{2}l_{3}, \; l_{2} \ge \gcd(l_{2},l_{1}l_{3}),
\end{equation}
the second entry in each of these inequalities also dividing the first.

%%%%%%%%%%%%%%%%%%%%%%%%%%%%%%%%%%%%%%%%%%%%%%
\section{Details on large \texorpdfstring{$k$}{k} calculations}
\label{appendix_largek}
%%%%%%%%%%%%%%%%%%%%%%%%%%%%%%%%%%%%%%%%%%%%%%
%%%%%%%%%%%%%%%%%%%%%%%%%%%%%%%%%%%%%%%%%%%%%%
\subsection{Asymptotic eigenvalue numerics}
\label{appendix_evnum}
%%%%%%%%%%%%%%%%%%%%%%%%%%%%%%%%%%%%%%%%%%%%%%

We now discuss how we obtain the asymptotic values~\eqref{eq:SVD_large_23} through~\eqref{eq:SVD_large_34_odd} employing the method used to obtain~\eqref{eq_conjecture:SEE_N=2_largek_1orig_conj} in~\cite{Dwivedi2018}.
This method may also be used to obtain limiting forms of the entanglement entropy discussed in~\cite{Dwivedi_2020}.
In each case, we observe regularities in the eigenvalue distribution and guess a point-wise limiting form for each eigenvalue at large $k$.
Then we normalize the eigenvalues using partition functions of the form $f^{n_{1}|n_{2}}(p;k)$. For $n_{1}=n_{2}=n$ they are given by~\eqref{eq_conjecture:partitionfunction_N=2} for $p=2$ and $n=2$, and by
\begin{subequations}
\begin{eqnarray}
f^{3|3}(2;k) &=& \frac{1}{2} \left\lfloor \frac{2(k+2)}{3} \right\rfloor \left\lceil \frac{2(k+2)}{3} \right\rceil. \label{eq_conjecture:Z_N=3_largek_1} \\
f^{4|4}(2;k) &=& \begin{cases}
\frac{(k+2)^{2}}{4} & k \, \mathrm{even}, \\
\frac{(k+1)(3k+5)}{16} & k \equiv 1 \mod 4, \\
\frac{(k+1)(3k+7)}{16} & k \equiv 3 \mod 4. \label{eq_conjecture:Z_N=4_largek_1}
\end{cases}
\end{eqnarray}
\end{subequations}
for $n=3,4$ respectively (these were obtained with finite sums in~\cite{Dwivedi_2020}\footnote{One can obtain exact rational values for these partition functions by using roots of unity in--built into SageMath or Mathematica.}.) For $n_{1} \ne n_{2}$ and $p>2$, we are unable to conjecture exact polynomial forms for the partition functions in $k$, and instead rely on a large $k$ approximation at leading order obtained by summing over the $|\Gamma^{n_{1}|n_{2}}_{m}(p;k)|$.
Finally, based on how the eigenvalues taper off at the starting or ending index, we are able to write the limiting entanglement or SVD entropy as one or more infinite sums.
Relative to the indexing convention~\eqref{eq:2N|2M_gamma}, these sums may indexed from the conventional `forward' end, i.e. $0,1\ldots$, or from the `backward' end, i.e. $k, k-1,\ldots$.
For example, let us detail how we obtain~\eqref{eq:SVD_large_23},~\eqref{eq:SVD_large_34_even} and~\eqref{eq:SVD_large_34_odd}.

For the first case, we found that the eigenvalue distribution peaks only at the backward end, and is given by
\begin{equation}
|\Gamma^{2|3}_{m}(p;k)| \xrightarrow[]{k \rightarrow \infty} \frac{2^{\frac{3}{2}-p}\sqrt{3}k^{3p-4}}{\pi^{2p-2}m^{2p-2}}, m \equiv \pm 1 \mod{6}, \label{eq:gamma_2|3_asymp_backwd}
\end{equation}
where (and subsequently throughout this Appendix), we begin indexing from $1$ for convenience, the scaling limit is assumed at fixed $m$, and we adopt the convention that only those eigenvalues satisfying the indicated congruence relation are non-zero.
Then we normalize the eigenvalues with their sum, and get
\begin{equation}
|\Gamma^{2|3}_{m}(p;k)|_{\mathrm{nz.}} \xrightarrow[]{k \rightarrow \infty} \frac{1}{\left(1-\frac{1}{2^{2p-2}}-\frac{1}{3^{2p-2}}+\frac{1}{6^{2p-2}}\right)}\frac{1}{m^{2p-2}}, m \equiv \pm 1 \mod{6},
\label{eq:gamma_2|3_asymp_backwd_norm}
\end{equation}
only from the backward end, the subscript $\mathrm{nz}.$ denoting normalization.
In this calculation we use the result
\begin{equation}
\sum_{\substack{m \ge 1, \\ m \equiv \pm 1 \operatorname{mod} 6}} \frac{1}{m^{\beta}}=\left(1-\frac{1}{2^{\beta}}-\frac{1}{3^{\beta}}+\frac{1}{6^{\beta}} \right) \zeta(\beta), \label{zeta_identity_1}
\end{equation}
for $\beta> 1$, which may be derived from the definition of the zeta function.
Finally we evaluate the SVD entropy as a sum of the eigenvalues~\eqref{eq:gamma_2|3_asymp_backwd_norm}.
This computation is straightforward and outputs~\eqref{eq:SVD_large_23}, using the result
\begin{equation}
\sum_{\substack{m \ge 1, \\ m \equiv \pm 1 \operatorname{mod} 6}} \frac{1}{m^{\beta}} \log \frac{1}{m^{\beta}}=\left(\frac{\log 2}{2^{\beta}}+\frac{\log 3}{3^{\beta}}-\frac{\log 6}{6^{\beta}} \right) \zeta(\beta)+\left(1-\frac{1}{2^{\beta}}-\frac{1}{3^{\beta}}+\frac{1}{6^{\beta}} \right) \zeta'(\beta), \label{zeta_identity_2}
\end{equation}
for $\beta>1$, which may be obtained by taking derivatives with respect to $\beta$ in~\eqref{zeta_identity_1}.

Similarly, for the subsequent two cases, we found that the eigenvalue distribution peaks at both ends for even and odd $k$ but has different forms in both cases.
For even $k$, the non-zero eigenvalues are given by
\begin{equation}
|\Gamma^{3|4}_{m}(p;k)| \xrightarrow[]{k \rightarrow \infty} \frac{2^{1-p}\sqrt{3}k^{3p-4}}{\pi^{2p-2}m^{2p-2}}, m \equiv \pm 1 \mod{6},
\end{equation}
from both ends, whereas for odd $k$ they are given by
\begin{equation}
|\Gamma^{3|4}_{m}(p;k)| \xrightarrow[]{k \rightarrow \infty}  
\begin{cases}
\frac{2^{1-p}\sqrt{3}k^{3p-4}}{\pi^{2p-2}m^{2p-2}}, & m \equiv \pm 1 \mod{6}\,\textrm{(forward indexing)},\\
\frac{2^{1-p}\sqrt{6}k^{3p-4}}{\pi^{2p-2}m^{2p-2}}, & m \equiv \pm 2 \mod{12}\,\textrm{(backward indexing)},
\end{cases}
\end{equation}
respectively.
Upon normalizing the eigenvalues with their sum in each case, we get
\begin{equation}
|\Gamma^{3|4}_{m}(p;k)|_{\mathrm{nz.}} \xrightarrow[]{k \rightarrow \infty} \frac{1}{2}|\Gamma^{2|3}_{m}(p;k)|_{\mathrm{nz.}}, m \equiv \pm 1 \mod{6},
\end{equation}
from both ends for even $k$, and
\begin{equation}
|\Gamma^{3|4}_{m}(p;k)|_{\mathrm{nz.}} \xrightarrow[]{k \rightarrow \infty}
\begin{cases}
 \frac{2^{2p-\frac{5}{2}}}{\left(1+2^{2p-\frac{5}{2}}\right)} \Gamma^{2|3}_{m}(p;k)|_{\mathrm{nz.}}, & m \equiv \pm 1 \mod{6}, \\  
  \frac{2^{2p-2}}{\left(1+2^{2p-\frac{5}{2}}\right)} \Gamma^{2|3}_{m}(p;k)|_{\mathrm{nz.}},& m \equiv \pm 2 \mod{12},
\end{cases}
\end{equation}
from the forward and backward ends respectively.
In both cases, these eigenvalues can be conveniently expressed in terms of the functional form~\eqref{eq:gamma_2|3_asymp_backwd_norm} but interpreted as non-zero as indicated.
Finally, we evaluate the SVD entropy and obtain the respective results~\eqref{eq:SVD_large_34_even} and~\eqref{eq:SVD_large_34_odd}.
We proceed similarly for obtaining~\eqref{eq:SVD_large_24_even}, and are further able to corroborate the results in~\cite{Dwivedi2018,Dwivedi_2020} for the entanglement entropies $S_{\mathrm{E}}^{n}(p;k)$ for $n=2,3,4$. 
We note down a few asymptotic functional forms for the $\Gamma_{m}^{n|n}(p;k)$ which we do not directly use in our work.
First, the $\Gamma^{2|2}(p;k)$, which are known to admit the finite $k$ functional form~\eqref{eq_conjecture:gamma^2_m} for $p=2$, appear to have the forms 
\begin{equation}
\Gamma^{2|2}_{m}(p;k) \xrightarrow[]{k \rightarrow \infty} \frac{2^{3-p}\sqrt{3}k^{3p-4}}{\pi^{2p-2}m^{2p-2}}, m \equiv 1 \mod 2, \label{eq:gamma_2_asymptotic}
\end{equation}
for general $p$, from the backward end. 
For the $\Gamma^{3|3}(p;k)$, we have
\begin{equation}
\Gamma^{3|3}_{m}(p;k) \xrightarrow[]{k \rightarrow \infty} \frac{2^{-p}3k^{3p-4}}{\pi^{2p-2}m^{2p-2}},  m \equiv 1,2 \mod 3,
\end{equation}
from both ends. 
For the $\Gamma^{4|4}_{m}(p;k)$, for even $k$ we have
\begin{equation}
\Gamma^{4|4}_{m}(p;k) \xrightarrow[]{k \rightarrow \infty} \frac{2^{2-p}k^{3p-4}}{\pi^{2p-2}m^{2p-2}}, m \equiv 1 \mod{2},
\end{equation}
from both ends, whereas for odd $k$ we have
\begin{equation}
\Gamma^{4|4}_{m}(p;k) \xrightarrow[]{k \rightarrow \infty}  
\begin{cases}
\frac{2^{2-p}k^{3p-4}}{\pi^{2p-2}m^{2p-2}}, & m \equiv 1 \mod{2}\,\textrm{(forward indexing)},\\
\frac{2^{3-p}k^{3p-4}}{\pi^{2p-2}m^{2p-2}}, & m \equiv 2 \mod{4}\,\textrm{(backward indexing)}.
\end{cases}
\end{equation}
One can check the consistency of all these obtained asymptotic forms $\Gamma^{n_{1}|n_{2}}_{m}(p;k)$, $2 \le n_{1},n_{2} \le 4$, with the identity $|\Gamma^{n_{1}|n_{2}}_{m}(p;k)|=\sqrt{\Gamma^{n_{1}|n_{1}}_{m}(p;k)\Gamma^{n_{2}|n_{2}}_{m}(p;k)}$.
We also note that we have numerically obtained a functional form for the $\alpha_{m}^{2}(2;k)$,
\begin{equation}
    \alpha^{2}_{m}(k)= i \frac{e^{\frac{\pi i}{4}}}{\sqrt{2}}e^{-\frac{\pi i}{4}\frac{m^{2}+2m+6}{k+2}} \frac{1+(-1)^{m-k}}{q^{\frac{m+1}{4}}+q^{-\frac{m+1}{4}}}. \label{eq_conjecture:alpha^2_m}
\end{equation}
This is consistent with the form~\eqref{eq_conjecture:gamma^2_m} numerically observed in~\cite{Dwivedi2018} for the $\Gamma^{2|2}_{m}(2;k)$.
Finally, we remark on how one may possibly derive these asymptotic forms for general $n_{1},n_{2}$. 
Each $\Gamma_{m}^{n_{1}|n_{2}}(p;k)$, being constructed out of $\cS$ and $\cT$ matrix element entries, is a sum of products of some expressions involving roots of unity.
These sums are of the particular kind known as generalized quadratic Gauss sums in the number theory literature. 
In fact, we report obtaining the exact form~\eqref{eq_conjecture:gamma^2_m} (and not just the asymptotic form~\eqref{eq:gamma_2_asymptotic} for $p=2$) using the techniques developed in~\cite{paris2014asymptotic} for a particular such class of quadratic Gauss sums. 
It may be possible to extend these arguments to other values of $n_{1}$ and $n_{2}$. 
\subsection{\texorpdfstring{$\mathrm{T}(2,4)$}{T(2,4)} entanglement entropy integral form}
\label{appendix_intform}
First we prove~\eqref{eq_conjecture:partitionfunction_N=2} assuming~\eqref{eq_conjecture:gamma^2_m}.
This is equivalent to proving
\begin{equation}
\sum_{\substack{m\ge 0 \\ k+1-2m >0}} \frac{1}{2+q^{\frac{k+1}{2}-m}+q^{-\frac{k+1}{2}+m}}= \begin{cases} \frac{(k+2)^2}{8} & k \, \mathrm{even}, \\ \frac{(k+1)(k+3)}{8} & k \, \mathrm{odd}. \end{cases} \label{eq_result:sumofalternaterootsbeginningfromlast_upperhalf}
\end{equation}
In order to prove~\eqref{eq_result:sumofalternaterootsbeginningfromlast_upperhalf}, we use a resolvent~\footnote{This appears to be a standard technique often employed in evaluating such sums; see e.g.~\cite{2653689} for a similar derivation which motivates our use of this technique.}
\begin{equation}
g(z)= \prod_{\substack{m\ge 0 \\ k+1-2m >0}} \left(z+z^{-1}-q^{\frac{k+1}{2}-m}-q^{-\frac{k+1}{2}+m}\right), \label{def:resolvent_g_N=2}    
\end{equation}
which is analytic for $z \in \mathbb{C}\setminus \{0\}$.
Using properties of the roots of unity, it is straightforward to establish that
\begin{equation}
g(z)= \begin{cases} z^{\frac{k+2}{2}}+z^{-\frac{k+2}{2}} & k \, \mathrm{even}, \\ \frac{1}{z-1} \left[ z^{\frac{k+3}{2}}-z^{-\frac{k+1}{2}} \right] & k \, \mathrm{odd}, \end{cases} \label{eq_result:resolvent_g_N=2_simplified form}
\end{equation}
with the singularites at $z=\pm 1$ understood to be removable.
\eqref{eq_result:sumofalternaterootsbeginningfromlast_upperhalf} then follows by evaluating $\lim_{z \rightarrow -1} \frac{z^2}{1-z^{2}} \frac{d}{dz} \log g(z)$.

We proceed on to provide mathematical arguments for the following integral form
\begin{equation}
S^{2}_{\mathrm{E}}(2;k) \xrightarrow{k \rightarrow \infty} 2 \int_{1}^{\infty} \dd \phi \,\frac{\tanh \phi}{\phi^{2}}-2\int_{0}^{1}  \frac{\dd \phi}{\phi} \left(1-\frac{\tanh \phi}{\phi}\right) -\log 2, \label{eq_conjecture:SEE_N=2_largek_1_integral_form}
\end{equation}
of the large $k$ limit of the $\mathrm{T}(2,4)$ entanglement entropy.
This form is numerically equal to~\eqref{eq_conjecture:SEE_N=2_largek_1orig_conj} and to~\eqref{eq:SVD_large_24_even} for $p=2$; we provide a brief proof at the end of this Appendix. 

Looking at the decomposition~\eqref{PEGen2Q}, it remains to evaluate, upto a multiplicative factor, the quantity
\begin{equation}
\sum_{\substack{m\ge 0 \\ k+1-2m >0}} \frac{2}{2+q^{\frac{k+1}{2}-m}+q^{-\frac{k+1}{2}+m}} \log \left( \frac{2}{2+q^{\frac{k+1}{2}-m}+q^{-\frac{k+1}{2}+m}} \right). \label{eq:2|2_avgenergy}
\end{equation}
To work with~\eqref{eq:2|2_avgenergy} we utilize a different resolvent,
\begin{equation}
h(w)=-\sum_{\substack{m\ge 0 \\ k+1-2m >0}} \frac{1}{1-\frac{q^{\frac{k+1}{2}-m}+q^{-\frac{k+1}{2}+m}}{w}} \log \left(1-\frac{q^{\frac{k+1}{2}-m}+q^{-\frac{k+1}{2}+m}}{w}\right), \label{eq:ln(x-roots)/(x-roots)_resolvent}
\end{equation}
which has the following expansion in $w^{-1}$ valid in the domain $w \in \mathbb{C}, \, |w| \ge 2$ for any $k$,
\begin{equation}
h(w)= \sum_{n \ge 1} H_{n}\underbrace{ \left[ \sum_{\substack{m\ge 0 \\ k+1-2m >0}} \left( q^{\frac{k+1}{2}-m}+q^{-\frac{k+1}{2}+m}  \right)^{n} \right]}_{=C_{n}} w^{-n}, \label{eq:resolvent_2|2_formalseries}
\end{equation}
where $H_{n}=\sum_{l=1}^{n}\frac{1}{l}$ is the $n$-th harmonic number, and outputs~\eqref{eq:2|2_avgenergy} at $w=-2$.
Employing a standard integral representation $H_{n}=\int_{0}^{1} \dd t \, \frac{1-t^{n}}{1-t}$ of the harmonic numbers, the resolvent~\eqref{eq:resolvent_2|2_formalseries} may be written as
\begin{equation}
h(w)=\int_{0}^{1} \frac{\dd t}{1-t} \left[\hat{h}(w)-\hat{h}\left(\frac{w}{t}\right) \right], \label{eq:resolvent_2_2_integraltransform}
\end{equation}
where $\hat{h}(w)= \sum_{n \ge 1}C_{n}w^{-n} \label{def:hat_resolvent_2|2}$ is related to~\eqref{eq_result:resolvent_g_N=2_simplified form} by
\begin{equation}
\hat{h}(w)=w \left[ \frac{z^{2}}{z^{2}-1} \frac{g'(z)}{g(z)} \right]_{z=\frac{1}{2}\left(\omega-\sqrt{\omega^{2}-4}\right)} - \left\lfloor \frac{k+2}{2} \right\rfloor, \label{eq:hath_intermsof_g(z)}
\end{equation}
and we have selected a particular branch for $\omega$ such that $z$ decreases in $(-\infty,-1]$ as $\omega$ decreases in $(-\infty,-2]$.
Evaluating~\eqref{eq:hath_intermsof_g(z)} at $w=-2$ and $w=-\frac{2}{t}$, and making a substitution of $\frac{1}{t}=\cosh \theta$, i.e. $z=-e^{\theta}$, we obtain a hyperbolic representation\footnote{This representation also appears related to the Chebyshev polynomials of the first and second kind.} of the average energy sector,
\begin{equation}
h(-2)=\begin{cases}
\int_{0}^{\infty} \frac{\dd \theta \, (\cosh \theta+1)}{\sinh \theta \cosh \theta} \left( \frac{(k+2)^{2}}{4}-\frac{k+2}{2} \frac{\tanh \left( \frac{k+2}{2} \theta \right)}{\tanh \theta}  \right) & k \, \mathrm{even},
 \\
\int_{0}^{\infty} \frac{\dd \theta \, (\cosh \theta+1)}{\sinh \theta \cosh \theta} \left( \frac{(k+1)(k+3)}{4}-\frac{k+2}{2} \frac{\tanh \left( \frac{k+2}{2} \theta \right)}{\tanh \theta} + \frac{\cosh \theta}{2(1+\cosh \theta)}\right) & k \, \mathrm{odd}. \label{eq:N=2_h(-2)_explicitform_hypertrig}
\end{cases}
\end{equation}
We observe that both~\eqref{eq_conjecture:partitionfunction_N=2} and~\eqref{eq:N=2_h(-2)_explicitform_hypertrig} have slightly different forms depending on the parity of $k$.
This appears to explain the two-subsequence zig-zag pattern observed for $S_{\mathrm{E}}^{2}(2;k)$ (e.g. see~\cref{fig_N=1_N=2}).

We now carefully split the integrals in~\eqref{eq:N=2_h(-2)_explicitform_hypertrig}, and incorporate~\eqref{eq_conjecture:partitionfunction_N=2} in integral form to obtain an integral form of $S_{\mathrm{E}}^{2}(2;k)$ via~\eqref{PEGen2Q}\footnote{The eventual goal is to get an expression for $S^{2}_{\mathrm{E}}(2;k)$ in terms of integrals without problematic singularities in the integrand.}.
First let us show this for the even $k$ case.
We obtain
\begin{multline}
S_{\mathrm{E}}^{2}(2;k)|_{k \, \mathrm{even}}= \int_{1}^{\frac{k+2}{2}} \dd \phi \, \left(\frac{2}{\phi} -\frac{2}{k+2} \frac{1+\cosh \left( \frac{2 \phi}{k+2} \right)}{\sinh \left( \frac{2 \phi}{k+2} \right) \cosh \left( \frac{2 \phi}{k+2} \right)} \right) 
\\ + \int_{1}^{\frac{k+2}{2}} \dd \phi \, \frac{4}{(k+2)^{2}} \frac{1+\cosh \left( \frac{2 \phi}{k+2} \right)}{\sinh^{2} \left( \frac{2 \phi}{k+2} \right)\cosh^{2} \left( \frac{2 \phi}{k+2} \right)} \tanh \phi
\\
-\int_{0}^{1} \dd \phi \, \frac{2}{k+2} \frac{1+\cosh \left( \frac{2 \phi}{k+2} \right)}{\sinh \left( \frac{2 \phi}{k+2} \right) \cosh \left( \frac{2 \phi}{k+2} \right)} \left(1- \frac{2}{k+2} \frac{\tanh \phi}{\tanh \left( \frac{2 \phi}{k+2} \right)} \right)
\\ -\int_{1}^{\infty} \dd \theta \, \frac{1+\cosh \theta}{\sinh \theta \cosh \theta} + \frac{2}{k+2}\int_{1}^{\infty} \frac{(1+\cosh{ \theta})\tanh( \frac{k+2}{2}\theta)}{\sinh^{2} \theta}.
\label{eq_SEE_N=2_rearranged_integrals_even}
\end{multline}
We propose a large $k$ approximation of these integrals as follows\footnote{What follows is an exact value of the leading order expression at large $k$. We leave the  analysis of subleading terms to future work.}.
For the first integral, we make the substitution $\theta = \frac{2}{k+2} \phi$, and computer numerics indicates that we may approximate the new lower limit by $0$.
For the second and third integrals, we make the usual small parameter substitutions for the hyperbolic functions $\sinh x, \tanh x \sim x; \, \cosh x \sim 1$ for $x \ll 1$, and approximate the upper limit of the second integral by $\infty$; computer numerics indicates that this is admissible even though the range of integration in the second integral incorporates a region where the dummy variable $\phi$ is not small.
The fourth integral is exactly computable in terms of elementary functions, and the fifth is indicated by numerics to be sub-leading, of $\mathcal{O}\left( \frac{1}{k}\right)$.
Hence we obtain the expression 
\begin{multline}
S_{\mathrm{E}}^{2}(2;k)|_{k \, \mathrm{even}} \approx \int_{0}^{1} \dd \theta \, \left( \frac{2}{\theta}- \frac{1+\cosh \theta}{\sinh \theta \cosh \theta} \right)+\int_{1}^{\infty} \dd \phi \, \frac{2}{\phi^{2}} \tanh \phi \\
-\int_{0}^{1} \dd \phi \, \frac{2}{\phi} \left(1- \frac{\tanh \phi}{\phi} \right) 
- \log \left(\frac{1+e^{2}}{(1-e)^{2}}\right). \label{eq_conjecture:S2_largek_even}
\end{multline}
which, after further simplification -- the first integral in~\eqref{eq_conjecture:S2_largek_even} is once again expressible in terms of elementary functions -- yields~\eqref{eq_conjecture:SEE_N=2_largek_1_integral_form}.

The odd $k$ expression may be obtained similarly, and is nearly identical to~\eqref{eq_SEE_N=2_rearranged_integrals_even} apart for some residual terms,
\begin{multline}
S_{\mathrm{E}}^{2}(2;k)|_{k \, \mathrm{odd}}= S_{\mathrm{E}}^{2}(2;k)|_{k \, \mathrm{even}}
\\+ \frac{2}{(k+2)^{3}} \int_{0}^{\frac{k+2}{2}} \dd \phi \, \frac{1+\cosh \left( \frac{2 \phi}{k+2} \right)}{\sinh \left( \frac{2 \phi}{k+2} \right) \cosh \left( \frac{2 \phi}{k+2} \right)} \left(1- \frac{2 \cosh \left( \frac{2\phi}{k+2} \right)}{1+\cosh \left( \frac{2\phi}{k+2} \right)} \right)
\\+\log \left(1-\frac{1}{(k+2)^{2}} \right)+\mathcal{R}(k).
\label{eq_SEE_N=2_rearranged_integrals_odd}
\end{multline}
Computer numerics indicates that the integral in~\eqref{eq_SEE_N=2_rearranged_integrals_odd} appears to vanish under a small parameter approximation.
The two residual terms are subleading; they arise from the consideration of $\frac{(k+1)(k+3)}{4}$ instead of $\frac{(k+2)^{2}}{4}$ as partition function ($\mathcal{R}$ collects contributions from\eqref{eq:N=2_h(-2)_explicitform_hypertrig}), and appear to go as $\mathcal{O}\left( \frac{1}{k^{2}}\right)$ and $\mathcal{O}\left( \frac{ \log k}{k^{2}}\right)$ respectively.
Hence the conjectured large $k$ limit of the entanglement entropy in the odd $k$ case is also~\eqref{eq_conjecture:SEE_N=2_largek_1_integral_form}.

The equivalence of~\eqref{eq_conjecture:SEE_N=2_largek_1_integral_form} and~\eqref{eq_conjecture:SEE_N=2_largek_1orig_conj} may be established by showing that
\begin{equation}
I=\int_{1}^{\infty} \dd \phi \,\frac{\tanh \phi}{\phi^{2}}-\int_{0}^{1}  \frac{\dd \phi}{\phi} \left(1-\frac{\tanh \phi}{\phi}\right)=12 \log A-\gamma-\frac{7}{3} \log 2. \label{eq:to_show_int_tanh}
\end{equation}
We carefully manipulate the integrals, using the known results\footnote{The first is a well-known integral representation of $\gamma$; for the second, see e.g.~\cite{2467077}.}
\begin{equation}
    \int_{0}^{1} \dd \phi \, \left(\frac{1}{\phi}+\frac{1}{\log(1-\phi)} \right)=\gamma, \; \int_{0}^{\infty} \dd \phi \, \left( \frac{\tanh \phi}{\phi^{2}}-\frac{1}{\phi e^{2\phi}} \right)=12 \log A-\frac{4}{3} \log 2.
\end{equation}
to extract out the Euler--Mascheroni and Glaisher constants respectively.
Eventually we get 
\begin{equation}
I=12 \log A -\gamma-\frac{4}{3} \log 2 + \int_{1}^{\infty} \frac{\dd \phi}{\phi e^{2\phi}}+\int_{0}^{1} \dd \phi \left( \frac{1}{\log(1-\phi)}+\frac{1}{\phi e^{2 \phi}} \right),
\end{equation}
and further evaluation of the remnant integrals outputs~\eqref{eq:to_show_int_tanh}.

The equivalence of~\eqref{eq_conjecture:SEE_N=2_largek_1orig_conj} and~\eqref{eq:SVD_large_24_even} for $p=2$ may be seen from the identity\footnote{See e.g.~\cite{1577823}.} 
\begin{equation}
\frac{\zeta'(2)}{\zeta(2)}=\gamma-12\log A+\log(2\pi).
\end{equation}

%%%%%%%%%%%%%%%%%%%%%%%%%%%%%
%%%%%%%%%%%%%%%%%%%%%%%%%%%%
\bibliography{Refs}

\providecommand{\href}[2]{#2}\begingroup\raggedright\begin{thebibliography}{10}

\bibitem{Horodecki:2009zz}
R.~Horodecki, P.~Horodecki, M.~Horodecki and K.~Horodecki, \emph{{Quantum
  entanglement}}, \href{http://dx.doi.org/10.1103/RevModPhys.81.865}{\emph{Rev.
  Mod. Phys.} {\bf 81} (2009) 865--942},
  [\href{http://arxiv.org/abs/quant-ph/0702225}{{\tt quant-ph/0702225}}].

\bibitem{Ryu:2006bv}
S.~Ryu and T.~Takayanagi, \emph{{Holographic derivation of entanglement entropy
  from AdS/CFT}},
  \href{http://dx.doi.org/10.1103/PhysRevLett.96.181602}{\emph{Phys. Rev.
  Lett.} {\bf 96} (2006) 181602},
  [\href{http://arxiv.org/abs/hep-th/0603001}{{\tt hep-th/0603001}}].

\bibitem{Takayanagi:2017knl}
T.~Takayanagi and K.~Umemoto, \emph{{Entanglement of purification through
  holographic duality}},
  \href{http://dx.doi.org/10.1038/s41567-018-0075-2}{\emph{Nature Phys.} {\bf
  14} (2018) 573--577}, [\href{http://arxiv.org/abs/1708.09393}{{\tt
  1708.09393}}].

\bibitem{Dutta:2019gen}
S.~Dutta and T.~Faulkner, \emph{{A canonical purification for the entanglement
  wedge cross-section}},
  \href{http://dx.doi.org/10.1007/JHEP03(2021)178}{\emph{JHEP} {\bf 03} (2021)
  178}, [\href{http://arxiv.org/abs/1905.00577}{{\tt 1905.00577}}].

\bibitem{Dong:2016fnf}
X.~Dong, \emph{{The Gravity Dual of Renyi Entropy}},
  \href{http://dx.doi.org/10.1038/ncomms12472}{\emph{Nature Commun.} {\bf 7}
  (2016) 12472}, [\href{http://arxiv.org/abs/1601.06788}{{\tt 1601.06788}}].

\bibitem{Kitaev:2005dm}
A.~Kitaev and J.~Preskill, \emph{{Topological entanglement entropy}},
  \href{http://dx.doi.org/10.1103/PhysRevLett.96.110404}{\emph{Phys. Rev.
  Lett.} {\bf 96} (2006) 110404},
  [\href{http://arxiv.org/abs/hep-th/0510092}{{\tt hep-th/0510092}}].

\bibitem{levin2006detecting}
M.~Levin and X.-G. Wen, \emph{{Detecting topological order in a ground state
  wave function}},
  \href{http://dx.doi.org/10.1103/PhysRevLett.96.110405}{\emph{Phys. Rev.
  Lett.} {\bf 96} (2006) 110405}.

\bibitem{aharonov1988result}
Y.~Aharonov, D.~Z. Albert and L.~Vaidman, \emph{How the result of a measurement
  of a component of the spin of a spin-1/2 particle can turn out to be 100},
  \href{http://dx.doi.org/10.1103/PhysRevLett.60.1351}{\emph{Physical review
  letters} {\bf 60} (1988) 1351}.

\bibitem{Horowitz:2003he}
G.~T. Horowitz and J.~M. Maldacena, \emph{{The Black hole final state}},
  \href{http://dx.doi.org/10.1088/1126-6708/2004/02/008}{\emph{JHEP} {\bf 02}
  (2004) 008}, [\href{http://arxiv.org/abs/hep-th/0310281}{{\tt
  hep-th/0310281}}].

\bibitem{PasqualeCalabrese_2009}
P.~Calabrese, J.~Cardy and B.~Doyon, \emph{Entanglement entropy in extended
  quantum systems},
  \href{http://dx.doi.org/10.1088/1751-8121/42/50/500301}{\emph{Journal of
  Physics A: Mathematical and Theoretical} {\bf 42} (dec, 2009) 500301}.

\bibitem{Nakata:2020luh}
Y.~Nakata, T.~Takayanagi, Y.~Taki, K.~Tamaoka and Z.~Wei, \emph{{New
  holographic generalization of entanglement entropy}},
  \href{http://dx.doi.org/10.1103/PhysRevD.103.026005}{\emph{Phys. Rev. D} {\bf
  103} (2021) 026005}, [\href{http://arxiv.org/abs/2005.13801}{{\tt
  2005.13801}}].

\bibitem{Parzygnat:2023avh}
A.~J. Parzygnat, T.~Takayanagi, Y.~Taki and Z.~Wei, \emph{{SVD entanglement
  entropy}}, \href{http://dx.doi.org/10.1007/JHEP12(2023)123}{\emph{JHEP} {\bf
  12} (2023) 123}, [\href{http://arxiv.org/abs/2307.06531}{{\tt 2307.06531}}].

\bibitem{Balasubramanian_2017}
V.~Balasubramanian, J.~R. Fliss, R.~G. Leigh and O.~Parrikar,
  \emph{Multi-boundary entanglement in {C}hern-{S}imons theory and link
  invariants}, \href{http://dx.doi.org/10.1007/jhep04(2017)061}{\emph{Journal
  of High Energy Physics} {\bf 2017} (apr, 2017) },
  [\href{http://arxiv.org/abs/1611.05460}{{\tt 1611.05460}}].

\bibitem{Balasubramanian:2018por}
V.~Balasubramanian, M.~DeCross, J.~Fliss, A.~Kar, R.~G. Leigh and O.~Parrikar,
  \emph{{Entanglement Entropy and the Colored Jones Polynomial}},
  \href{http://dx.doi.org/10.1007/JHEP05(2018)038}{\emph{JHEP} {\bf 05} (2018)
  038}, [\href{http://arxiv.org/abs/1801.01131}{{\tt 1801.01131}}].

\bibitem{SO3}
A.~Dwivedi, S.~Dwivedi, B.~P. Mandal, P.~Ramadevi and V.~K. Singh,
  \emph{Topological entanglement and hyperbolic volume},
  \href{http://dx.doi.org/10.1007/jhep10(2021)172}{\emph{Journal of High Energy
  Physics} {\bf 2021} (2021) 1--37},
  [\href{http://arxiv.org/abs/2106.03396}{{\tt 2106.03396}}].

\bibitem{kashaev1996}
R.~M. Kashaev, \emph{The hyperbolic volume of knots from quantum dilogarithm},
  \href{http://arxiv.org/abs/q-alg/9601025}{{\tt q-alg/9601025}}.

\bibitem{murakami1999}
H.~Murakami and J.~Murakami, \emph{The colored jones polynomials and the
  simplicial volume of a knot},  \href{http://arxiv.org/abs/math/9905075}{{\tt
  math/9905075}}.

\bibitem{Shinmyo_2024}
K.~Shinmyo, T.~Takayanagi and K.~Tasuki, \emph{Pseudo entropy under joining
  local quenches},
  \href{http://dx.doi.org/10.1007/jhep02(2024)111}{\emph{Journal of High Energy
  Physics} {\bf 2024} (Feb., 2024) }.

\bibitem{Mollabashi:2020yie}
A.~Mollabashi, N.~Shiba, T.~Takayanagi, K.~Tamaoka and Z.~Wei, \emph{{Pseudo
  Entropy in Free Quantum Field Theories}},
  \href{http://dx.doi.org/10.1103/PhysRevLett.126.081601}{\emph{Phys. Rev.
  Lett.} {\bf 126} (2021) 081601}, [\href{http://arxiv.org/abs/2011.09648}{{\tt
  2011.09648}}].

\bibitem{Mollabashi:2021xsd}
A.~Mollabashi, N.~Shiba, T.~Takayanagi, K.~Tamaoka and Z.~Wei, \emph{{Aspects
  of pseudoentropy in field theories}},
  \href{http://dx.doi.org/10.1103/PhysRevResearch.3.033254}{\emph{Phys. Rev.
  Res.} {\bf 3} (2021) 033254}, [\href{http://arxiv.org/abs/2106.03118}{{\tt
  2106.03118}}].

\bibitem{Nishioka:2021cxe}
T.~Nishioka, T.~Takayanagi and Y.~Taki, \emph{{Topological pseudo entropy}},
  \href{http://dx.doi.org/10.1007/JHEP09(2021)015}{\emph{JHEP} {\bf 09} (2021)
  015}, [\href{http://arxiv.org/abs/2107.01797}{{\tt 2107.01797}}].

\bibitem{Doi:2024nty}
K.~Doi, N.~Ogawa, K.~Shinmyo, Y.-k. Suzuki and T.~Takayanagi, \emph{{Probing de
  Sitter Space Using CFT States}},  \href{http://arxiv.org/abs/2405.14237}{{\tt
  2405.14237}}.

\bibitem{Kawamoto:2023nki}
T.~Kawamoto, S.-M. Ruan, Y.-k. Suzuki and T.~Takayanagi, \emph{{A half de
  Sitter holography}},
  \href{http://dx.doi.org/10.1007/JHEP10(2023)137}{\emph{JHEP} {\bf 10} (2023)
  137}, [\href{http://arxiv.org/abs/2306.07575}{{\tt 2306.07575}}].

\bibitem{He:2024jog}
S.~He, P.~H.~C. Lau and L.~Zhao, \emph{{Detecting quantum chaos via
  pseudo-entropy and negativity}},  \href{http://arxiv.org/abs/2403.05875}{{\tt
  2403.05875}}.

\bibitem{Omidi:2023env}
F.~Omidi, \emph{{Pseudo R\'enyi Entanglement Entropies For an Excited State and
  Its Time Evolution in a 2D CFT}},
  \href{http://arxiv.org/abs/2309.04112}{{\tt 2309.04112}}.

\bibitem{Narayan:2023ebn}
K.~Narayan and H.~K. Saini, \emph{{Notes on time entanglement and
  pseudo-entropy}},
  \href{http://dx.doi.org/10.1140/epjc/s10052-024-12855-x}{\emph{Eur. Phys. J.
  C} {\bf 84} (2024) 499}, [\href{http://arxiv.org/abs/2303.01307}{{\tt
  2303.01307}}].

\bibitem{Chen:2023gnh}
Z.~Chen, \emph{{Complex-valued Holographic Pseudo Entropy via Real-time AdS/CFT
  Correspondence}},  \href{http://arxiv.org/abs/2302.14303}{{\tt 2302.14303}}.

\bibitem{Doi:2023zaf}
K.~Doi, J.~Harper, A.~Mollabashi, T.~Takayanagi and Y.~Taki, \emph{{Timelike
  entanglement entropy}},
  \href{http://dx.doi.org/10.1007/JHEP05(2023)052}{\emph{JHEP} {\bf 05} (2023)
  052}, [\href{http://arxiv.org/abs/2302.11695}{{\tt 2302.11695}}].

\bibitem{Ishiyama:2022odv}
Y.~Ishiyama, R.~Kojima, S.~Matsui and K.~Tamaoka, \emph{{Notes on pseudo
  entropy amplification}},
  \href{http://dx.doi.org/10.1093/ptep/ptac112}{\emph{PTEP} {\bf 2022} (2022)
  093B10}, [\href{http://arxiv.org/abs/2206.14551}{{\tt 2206.14551}}].

\bibitem{Mukherjee:2022jac}
J.~Mukherjee, \emph{{Pseudo Entropy in U(1) gauge theory}},
  \href{http://dx.doi.org/10.1007/JHEP10(2022)016}{\emph{JHEP} {\bf 10} (2022)
  016}, [\href{http://arxiv.org/abs/2205.08179}{{\tt 2205.08179}}].

\bibitem{Miyaji:2021lcq}
M.~Miyaji, \emph{{Island for gravitationally prepared state and pseudo
  entanglement wedge}},
  \href{http://dx.doi.org/10.1007/JHEP12(2021)013}{\emph{JHEP} {\bf 12} (2021)
  013}, [\href{http://arxiv.org/abs/2109.03830}{{\tt 2109.03830}}].

\bibitem{Goto:2021kln}
K.~Goto, M.~Nozaki and K.~Tamaoka, \emph{{Subregion spectrum form factor via
  pseudoentropy}},
  \href{http://dx.doi.org/10.1103/PhysRevD.104.L121902}{\emph{Phys. Rev. D}
  {\bf 104} (2021) L121902}, [\href{http://arxiv.org/abs/2109.00372}{{\tt
  2109.00372}}].

\bibitem{Guo:2024edr}
W.-z. Guo, Y.-z. Jiang and J.~Xu, \emph{{Pseudoentropy sum rule by analytical
  continuation of the superposition parameter}},
  \href{http://arxiv.org/abs/2405.09745}{{\tt 2405.09745}}.

\bibitem{He:2023syy}
S.~He, Y.-X. Zhang, L.~Zhao and Z.-X. Zhao, \emph{{Entanglement and Pseudo
  Entanglement Dynamics versus Fusion in CFT}},
  \href{http://arxiv.org/abs/2312.02679}{{\tt 2312.02679}}.

\bibitem{Kanda:2023jyi}
H.~Kanda, T.~Kawamoto, Y.-k. Suzuki, T.~Takayanagi, K.~Tasuki and Z.~Wei,
  \emph{{Entanglement phase transition in holographic pseudo entropy}},
  \href{http://dx.doi.org/10.1007/JHEP03(2024)060}{\emph{JHEP} {\bf 03} (2024)
  060}, [\href{http://arxiv.org/abs/2311.13201}{{\tt 2311.13201}}].

\bibitem{He:2023wko}
S.~He, J.~Yang, Y.-X. Zhang and Z.-X. Zhao, \emph{{Pseudo entropy of primary
  operators in $ T\overline{T}/J\overline{T} $-deformed CFTs}},
  \href{http://dx.doi.org/10.1007/JHEP09(2023)025}{\emph{JHEP} {\bf 09} (2023)
  025}, [\href{http://arxiv.org/abs/2305.10984}{{\tt 2305.10984}}].

\bibitem{Guo:2023aio}
W.-z. Guo and J.~Zhang, \emph{{Sum rule for the pseudo-R\'enyi entropy}},
  \href{http://dx.doi.org/10.1103/PhysRevD.109.106008}{\emph{Phys. Rev. D} {\bf
  109} (2024) 106008}, [\href{http://arxiv.org/abs/2308.05261}{{\tt
  2308.05261}}].

\bibitem{Singh:2024pbo}
Y.~Singh and R.~Banerjee, \emph{{SVD Entanglement Entropy of Chiral Dirac
  Oscillators}},  \href{http://arxiv.org/abs/2407.10898}{{\tt 2407.10898}}.

\bibitem{witten1989quantum}
E.~Witten, \emph{{Quantum field theory and the Jones polynomial}},
  \href{http://dx.doi.org/10.1007/BF01217730}{\emph{Communications in
  Mathematical Physics} {\bf 121} (1989) 351--399}.

\bibitem{Cabra_Rossini_1997}
D.~C. Cabra and G.~L. Rossini, \emph{Explicit connection between conformal
  field theory and 2+1 chern–simons theory},
  \href{http://dx.doi.org/10.1142/S0217732397001722}{\emph{Modern Physics
  Letters A} {\bf 12} (1997) 1687--1697}.

\bibitem{kohno2002conformal}
T.~Kohno, \emph{{Conformal Field Theory and Topology}}.
\newblock Iwanami Series in Modern Mathematics. American Mathematical Society,
  Providence, RI, 2002.

\bibitem{Kaul:1993hb}
R.~K. Kaul, \emph{{Chern-Simons theory, colored oriented braids and link
  invariants}}, \href{http://dx.doi.org/10.1007/BF02102019}{\emph{Commun. Math.
  Phys.} {\bf 162} (1994) 289--320},
  [\href{http://arxiv.org/abs/hep-th/9305032}{{\tt hep-th/9305032}}].

\bibitem{Kaul:1998ye}
R.~K. Kaul, \emph{{Chern-Simons theory, knot invariants, vertex models and
  three manifold invariants}},  in \emph{{Workshop on Frontiers in Field
  Theory, Quantum Gravity and String Theory}}, pp.~45--63, 4, 1998.
\newblock \href{http://arxiv.org/abs/hep-th/9804122}{{\tt hep-th/9804122}}.

\bibitem{Leigh:2021trp}
R.~G. Leigh and P.-C. Pai, \emph{{Complexity for link complement states in
  Chern-Simons theory}},
  \href{http://dx.doi.org/10.1103/PhysRevD.104.065005}{\emph{Phys. Rev. D} {\bf
  104} (2021) 065005}, [\href{http://arxiv.org/abs/2101.03443}{{\tt
  2101.03443}}].

\bibitem{Camilo:2019bbl}
G.~Camilo, D.~Melnikov, F.~Novaes and A.~Prudenziati, \emph{{Circuit Complexity
  of Knot States in Chern-Simons theory}},
  \href{http://dx.doi.org/10.1007/JHEP07(2019)163}{\emph{JHEP} {\bf 07} (2019)
  163}, [\href{http://arxiv.org/abs/1903.10609}{{\tt 1903.10609}}].

\bibitem{Fliss:2020yrd}
J.~R. Fliss, \emph{{Knots, links, and long-range magic}},
  \href{http://dx.doi.org/10.1007/JHEP04(2021)090}{\emph{JHEP} {\bf 04} (2021)
  090}, [\href{http://arxiv.org/abs/2011.01962}{{\tt 2011.01962}}].

\bibitem{francesco1997conformal}
P.~Di~Francesco, P.~Mathieu and D.~Sénéchal, \emph{{Conformal Field Theory}}.
\newblock Graduate Texts in Contemporary Physics. Springer, New York, 1997.

\bibitem{GEPNER1986493}
D.~Gepner and E.~Witten, \emph{{String Theory on Group Manifolds}},
  \href{http://dx.doi.org/10.1016/0550-3213(86)90051-9}{\emph{Nucl. Phys. B}
  {\bf 278} (1986) 493--549}.

\bibitem{nawata2012super}
S.~Nawata, P.~Ramadevi, X.~Sun et~al., \emph{{Super-A-polynomials for twist
  knots}}, \href{http://dx.doi.org/10.1007/JHEP11(2012)157}{\emph{Journal of
  High Energy Physics} {\bf 2012} (2012) 1--39}.

\bibitem{CJP-twist}
G.~Masbaum, \emph{Skein-theoretical derivation of some formulas of habiro},
  \href{http://dx.doi.org/10.2140/agt.2003.3.537}{\emph{Algebraic \& Geometric
  Topology} {\bf 3} (2003) 537--556},
  [\href{http://arxiv.org/abs/math/0306345}{{\tt math/0306345}}].

\bibitem{Habiro}
K.~Habiro, \emph{{On the colored Jones polynomial of some simple links}},
  {\emph{RIMS Kokyuroku} {\bf 1172} (2000) 34--43}.

\bibitem{hikami2004difference}
K.~Hikami, \emph{{Difference equation of the colored Jones polynomial for torus
  knot}}, \href{http://dx.doi.org/10.1142/S0129167X04002372}{\emph{Int. J.
  Math.} {\bf 15} (2004) 959--965},
  [\href{http://arxiv.org/abs/math/0403224}{{\tt math/0403224}}].

\bibitem{Fuji:2012pi}
H.~Fuji, S.~Gukov, M.~Stosic and P.~Sulkowski, \emph{{3d analogs of
  Argyres-Douglas theories and knot homologies}},
  \href{http://dx.doi.org/10.1007/JHEP01(2013)175}{\emph{JHEP} {\bf 01} (2013)
  175}, [\href{http://arxiv.org/abs/1209.1416}{{\tt 1209.1416}}].

\bibitem{Gukov:2015gmm}
S.~Gukov, S.~Nawata, I.~Saberi, M.~Sto\v{s}i\'c and P.~Su\l{}kowski,
  \emph{{Sequencing BPS Spectra}},
  \href{http://dx.doi.org/10.1007/JHEP03(2016)004}{\emph{JHEP} {\bf 03} (2016)
  004}, [\href{http://arxiv.org/abs/1512.07883}{{\tt 1512.07883}}].

\bibitem{Dwivedi_2020}
S.~Dwivedi, V.~K. Singh and A.~Roy, \emph{{Semiclassical limit of topological
  Rényi entropy in 3d Chern-Simons theory}},
  \href{http://dx.doi.org/10.1007/JHEP12(2020)132}{\emph{JHEP} {\bf 2020}
  (2020) 1--72}, [\href{http://arxiv.org/abs/2007.07033}{{\tt 2007.07033}}].

\bibitem{Stevan_2010}
S.~Stevan, \emph{Chern–simons invariants of torus links},
  \href{http://dx.doi.org/10.1007/s00023-010-0058-z}{\emph{Annales Henri
  Poincaré} {\bf 11} (Nov., 2010) 1201–1224}.

\bibitem{Brini_2012}
A.~Brini, M.~Mariño and B.~Eynard, \emph{Torus knots and mirror symmetry},
  \href{http://dx.doi.org/10.1007/s00023-012-0171-2}{\emph{Annales Henri
  Poincaré} {\bf 13} (Mar., 2012) 1873–1910}.

\bibitem{Perelomov:1971bd}
A.~M. Perelomov, \emph{{Coherent states for arbitrary lie groups}},
  \href{http://dx.doi.org/10.1007/BF01645091}{\emph{Commun. Math. Phys.} {\bf
  26} (1972) 222--236}.

\bibitem{mermin_1990}
N.~D. Mermin, \emph{{Quantum mysteries revisited}},
  \href{http://dx.doi.org/10.1119/1.16503}{\emph{American Journal of Physics}
  {\bf 58} (08, 1990) 731--734}.

\bibitem{D_r_2000}
W.~Dür, G.~Vidal and J.~I. Cirac, \emph{Three qubits can be entangled in two
  inequivalent ways},
  \href{http://dx.doi.org/10.1103/physreva.62.062314}{\emph{Physical Review A}
  {\bf 62} (Nov., 2000) }.

\bibitem{murakami2016periodicity}
H.~Murakami, \emph{{Periodicity properties of the colored Jones polynomial}},
  {\emph{arXiv preprint} (2016) }, [\href{http://arxiv.org/abs/1606.00125}{{\tt
  1606.00125}}].

\bibitem{Jones:1985dw}
V.~F.~R. Jones, \emph{{A polynomial invariant for knots via von Neumann
  algebras}},
  \href{http://dx.doi.org/10.1090/S0273-0979-1985-15304-2}{\emph{Bull. Am.
  Math. Soc.} {\bf 12} (1985) 103--111}.

\bibitem{kauffman1990invariant}
L.~H. Kauffman, \emph{An invariant of regular isotopy},
  \href{http://dx.doi.org/10.1090/S0002-9947-1990-0958895-7}{\emph{Transactions
  of the American Mathematical Society} {\bf 318} (1990) 417--471}.

\bibitem{ramadevi1993chirality}
P.~Ramadevi, T.~R. Govindarajan and R.~K. Kaul, \emph{{Chirality of knots 9(42)
  and 10(71) and Chern-Simons theory}},
  \href{http://dx.doi.org/10.1142/S0217732394003026}{\emph{Mod. Phys. Lett. A}
  {\bf 9} (1994) 3205--3218}, [\href{http://arxiv.org/abs/hep-th/9401095}{{\tt
  hep-th/9401095}}].

\bibitem{Dwivedi2018}
S.~Dwivedi, V.~K. Singh, S.~Dhara, P.~Ramadevi, Y.~Zhou and L.~K. Joshi,
  \emph{{Entanglement on linked boundaries in Chern-Simons theory with generic
  gauge groups}}, \href{http://dx.doi.org/10.1007/JHEP02(2018)163}{\emph{JHEP}
  {\bf 2018} (2018) 163}, [\href{http://arxiv.org/abs/1711.06474}{{\tt
  1711.06474}}].

\bibitem{Nielsen:2005mkt}
M.~A. Nielsen, \emph{{A geometric approach to quantum circuit lower bounds}},
  \href{http://dx.doi.org/10.26421/QIC6.3-2}{\emph{Quant. Inf. Comput.} {\bf 6}
  (2006) 213--262}, [\href{http://arxiv.org/abs/quant-ph/0502070}{{\tt
  quant-ph/0502070}}].

\bibitem{antonelli2003handbook}
P.~Antonelli, \emph{Handbook of Finsler geometry. 1 (2003)}.
\newblock Handbook of Finsler Geometry. Kluwer Academic Publishers, 2003.

\bibitem{strydom2021svd}
T.~Strydom, G.~V. Dalla~Riva and T.~Poisot, \emph{{SVD entropy reveals the high
  complexity of ecological networks}},
  \href{http://dx.doi.org/10.3389/fevo.2021.623141}{\emph{Frontiers in Ecology
  and Evolution} {\bf 9} (2021) }.

\bibitem{2-qubit}
B.~Schumacher, \emph{Quantum coding},
  \href{http://dx.doi.org/10.1103/PhysRevA.51.2738}{\emph{Phys. Rev. A} {\bf
  51} (Apr, 1995) 2738--2747}.

\bibitem{paris2014asymptotic}
R.~B. Paris, \emph{An asymptotic expansion for the generalised quadratic gauss
  sum revisited},
  \href{http://dx.doi.org/dx.doi.org/10.7153/jca-05-02}{\emph{Journal of
  Classical Analysis} {\bf 5} (2014) 15--24},
  [\href{http://arxiv.org/abs/1403.7973}{{\tt 1403.7973}}].

\bibitem{2653689}
achille hui, ``Sum of $1/(1 - x)$ over roots of unity.''
  \url{https://math.stackexchange.com/q/2653689}.

\bibitem{2467077}
pisco, ``Closed form $\int_0^\infty\left(\frac{\tanh
  x}{x^2}-\frac{1}{xe^{2x}}\right)dx=12\log a-\frac{4}{3}\log 2$.''
  \url{https://math.stackexchange.com/q/2467077}.

\bibitem{1577823}
R.~Variable, ``Is there a special value for $\frac{\zeta'(2)}{\zeta(2)}$?.''
  \url{https://math.stackexchange.com/q/1577823}.

\end{thebibliography}\endgroup
\bibliographystyle{JHEP}

\end{document}